\documentclass[twocolumn]{aastex63}

\usepackage{graphicx}	% Including figure files
\usepackage{subfigure}  %
\usepackage{booktabs}
\usepackage{longtable}
\usepackage{supertabular}
\usepackage{color}
\usepackage{array}
\usepackage{verbatim}
\usepackage{lineno}
\shorttitle{ Inner Accretion Disk of EXO~1846--031}
\shortauthors{Ren et al.}
\graphicspath{{./}{figures/}}
\begin{document}

\title{{{\it Insight-HXMT} Study of the Inner Accretion Disk in the Black Hole Candidate EXO 1846--031}}

\author{X.Q. Ren}
%\altaffiliation{E-mail: renxiaoqin@ihep.ac.cn}
\affiliation{Key Laboratory of Particle Astrophysics, Institute of High Energy Physics, Chinese Academy of Sciences, Beijing 100049, China}
\affiliation{University of Chinese Academy of Sciences, Chinese Academy of Sciences, Beijing 100049, People's Republic of China}

\author{Yanan Wang}
\altaffiliation{E-mail: Y.Wang@soton.ac.uk}
\affiliation{Physics \& Astronomy, University of Southampton, Southampton, Hampshire SO17 1BJ, UK}

\author{S. N. Zhang}
\altaffiliation{E-mail: zhangsn@ihep.ac.cn}
\affiliation{Key Laboratory of Particle Astrophysics, Institute of High Energy Physics, Chinese Academy of Sciences, Beijing 100049, China}

\author{R. Soria}
\affiliation{College of Astronomy and Space Sciences, University of the Chinese Academy of Sciences, Beijing 100049, China}
\affiliation{Sydney Institute for Astronomy, School of Physics A28, The University of Sydney, Sydney, NSW 2006, Australia}

\author{ L. Tao}
\affiliation{Key Laboratory of Particle Astrophysics, Institute of High Energy Physics, Chinese Academy of Sciences, Beijing 100049, China}

\author{L. Ji}
\affiliation{School of Physics \& Astronomy, Sun Yat-Sen University, Zhuhai, 519082, China}

\author{Y. J. Yang}
\affiliation{Department of Physics, The University of Hong Kong, Pokfulam Road, Hong Kong}

\author{J. L. Qu}
\affiliation{Key Laboratory of Particle Astrophysics, Institute of High Energy Physics, Chinese Academy of Sciences, Beijing 100049, China}

\author{S. Zhang}
\affiliation{Key Laboratory of Particle Astrophysics, Institute of High Energy Physics, Chinese Academy of Sciences, Beijing 100049, China}

\author{L. M. Song}
\affiliation{Key Laboratory of Particle Astrophysics, Institute of High Energy Physics, Chinese Academy of Sciences, Beijing 100049, China}

\author{M. Y. Ge}
\affiliation{Key Laboratory of Particle Astrophysics, Institute of High Energy Physics, Chinese Academy of Sciences, Beijing 100049, China}

\author{Y. Huang}
\affiliation{Key Laboratory of Particle Astrophysics, Institute of High Energy Physics, Chinese Academy of Sciences, Beijing 100049, China}

\author{X. B. Li}
\affiliation{Key Laboratory of Particle Astrophysics, Institute of High Energy Physics, Chinese Academy of Sciences, Beijing 100049, China}

\author{J. Y. Liao}
\affiliation{Key Laboratory of Particle Astrophysics, Institute of High Energy Physics, Chinese Academy of Sciences, Beijing 100049, China}

\author{H. X. Liu}
\affiliation{Key Laboratory of Particle Astrophysics, Institute of High Energy Physics, Chinese Academy of Sciences, Beijing 100049, China}
\affiliation{University of Chinese Academy of Sciences, Chinese Academy of Sciences, Beijing 100049, People's Republic of China}

\author{R. C. Ma}
\affiliation{Key Laboratory of Particle Astrophysics, Institute of High Energy Physics, Chinese Academy of Sciences, Beijing 100049, China}
\affiliation{University of Chinese Academy of Sciences, Chinese Academy of Sciences, Beijing 100049, People's Republic of China}

\author{Y. L. Tuo}
\affiliation{Key Laboratory of Particle Astrophysics, Institute of High Energy Physics, Chinese Academy of Sciences, Beijing 100049, China}

\author{P. J. Wang}
\affiliation{Key Laboratory of Particle Astrophysics, Institute of High Energy Physics, Chinese Academy of Sciences, Beijing 100049, China}
\affiliation{University of Chinese Academy of Sciences, Chinese Academy of Sciences, Beijing 100049, People's Republic of China}

\author{W. Zhang}
\affiliation{Key Laboratory of Particle Astrophysics, Institute of High Energy Physics, Chinese Academy of Sciences, Beijing 100049, China}
\affiliation{University of Chinese Academy of Sciences, Chinese Academy of Sciences, Beijing 100049, People's Republic of China}

\author{D. K. Zhou}
\affiliation{Key Laboratory of Particle Astrophysics, Institute of High Energy Physics, Chinese Academy of Sciences, Beijing 100049, China}
\affiliation{University of Chinese Academy of Sciences, Chinese Academy of Sciences, Beijing 100049, People's Republic of China}



\begin{abstract}

We study the spectral evolution of the black hole candidate EXO~1846$-$031 during its 2019 outburst, in the 1--150~keV band, with the {\it {Hard X-ray Modulation Telescope}}. The continuum spectrum is well modelled with an absorbed disk-blackbody plus cutoff power-law, in the hard, intermediate and soft states. In addition, we detect an $\approx$6.6~keV Fe emission line in the hard intermediate state. Throughout the soft intermediate and soft states, the fitted inner disk radius remains almost constant; we suggest that it has settled at the innermost stable circular orbit (ISCO). However, in the hard and hard intermediate states, the apparent inner radius was unphysically small (smaller than ISCO), even after accounting for the Compton scattering of some of the disk photons by the corona in the fit. We argue that this is the result of a high hardening factor, $f_{\rm col}\approx2.0-2.7$, in the early phases of outburst evolution, well above the canonical value of 1.7 suitable to a steady disk. We suggest that the inner disk radius was close to ISCO already in the low/hard state. Furthermore, we propose that this high value of hardening factor in the relatively hard state is probably caused by the additional illuminating of the coronal irradiation onto the disk. Additionally, we estimate the spin parameter with the continuum-fitting method, over a range of plausible black hole masses and distances. We compare our results with the spin measured with the reflection-fitting method and find that the inconsistency of the two results is partly caused by the different choices of $f_{\rm col}$.

\end{abstract}



\section{Introduction} \label{1}

A  black hole (BH) low-mass X-ray binary is composed of a stellar-mass BH and a companion star with a mass similar to or lower than the solar mass. 
% In such a system, the BH accretes materials from the companion star via Roche-lobe overflow, which finally forms an optically thick and geometrically thin accretion disk \citep{1973A&A....24..337S}. Driven by mass accretion rate \citep{1974PASJ...26..429O,1984AcA....34..161S,2001NewAR..45..449L}, 
Such a system normally experiences a long period of quiescence followed by a short-lived outburst lasting for months to years \citep{1996ARA&A..34..607T,1997ApJ...491..312C,2000ApJ...537..448T}. During an outburst, its luminosity varies by several orders of magnitude \citep{2006ARA&A..44...49R,2013ApJ...769...16R,2015MNRAS.446.4098P,2019MNRAS.486.2705A}. The variability is usually detected on timescales as short as milliseconds \citep{2006csxs.book...39V,2014SSRv..183...43B, 2016AN....337..398M}. In addition, complex spectral features \citep{2006ARA&A..44...49R, 2018MNRAS.480.4443T} are also seen during outbursts.
%together with short timescales of variability as fast as milliseconds \citep{2006csxs.book...39V,2014SSRv..183...43B, 2016AN....337..398M} and complex spectral features \citep{2006ARA&A..44...49R, 2018MNRAS.480.4443T}.

A typical outburst is usually divided into several different spectral states \citep{2004MNRAS.355.1105F,2005Ap&SS.300..107H,2007A&ARv..15....1D} based on their spectral and timing behavior. In the initial phase of the outburst (low-hard state, LHS), the luminosity is low and the spectrum is dominated by a non-thermal hard component with a power-law form.
%, which is widely attributed to inverse-Compton scattering of disk photons in a hot corona. 
%The spectrum can be described by a \texttt{power-law} with or without a high-energy cutoff \citep{1997ApJ...489..865E} or a more physical model, e.g., \texttt{nthcomp} \citep{1996MNRAS.283..193Z,1999MNRAS.309..561Z} and \texttt{simpl/simplcut} \citep{2009PASP..121.1279S,2016HEAD...1510906P,2017ApJ...836..119S}. 
In the standard scenario, the accretion disk is truncated far from the innermost stable circular orbit (ISCO) \citep{1997ApJ...489..865E} and the inner region is filled with a hot, geometrically thick, radiatively inefficient flow \citep{1995ApJ...452..710N}. In an alternative scenario, a relatively cool inner disk (in addition to a hot corona and an outer, truncated disk) may also exist near ISCO \citep{2007ApJ...671..695L,2006ApJ...652L.113M}.
Radio observations in the LHS show the presence of a compact jet launched from the innermost region \citep{2004MNRAS.355.1105F}.
%the inner region is filled with a hot inner flow or a launching site of jets. 
At this stage, strong noise components dominate the power-density spectra (PDS) and a low-frequency quasi-periodic oscillation (LFQPO) starts to show up \citep{2006ARA&A..44...49R}.
As the accretion rate increases, the source enters the hard intermediate state (HIMS) and soft intermediate state (SIMS) and the spectrum gradually softens. The compact jet disappears between HIMS and SIMS. As the accretion rate increases further, the source enters the high soft state (HSS) and the accretion disk reaches ISCO (or, in the alternative scenario, the outer disk joins the inner disk, removing the coronal gap). In the HSS, the X-ray spectrum is dominated by the disk component \citep{1984PASJ...36..741M,1986ApJ...308..635M}.
%When the accretion disk further moves inwards and eventually reaches the innermost stable circular orbit (ISCO), the source enters the high soft state (HSS); the spectrum is accordingly dominated by the disk component \citep{1984PASJ...36..741M,1986ApJ...308..635M}.
LFQPOs nearly disappear at this stage although occasionally there are some faint signs of oscillations \citep{2012MNRAS.427..595M,2016ASSL..440...61B}. Later on, the source decays in flux, and passes through the lower intensity intermediate and hard states before returning back to quiescence. For a canonical BH transient, the evolution of a full outburst presents a counterclockwise q-shaped loop on the hardness-intensity diagram \citep{2001ApJS..132..377H,2005A&A...440..207B,2020MNRAS.499..851Z}.
%Thus a  BH transient (BHT) shows a counterclockwise Q-shaped loop \citep{2005A&A...440..207B,2020MNRAS.499..851Z} on the hardness-intensity diagram (HID)  when a typical outburst is completed.

% While such a simplified truncated disk model can broadly describe the common disk and corona behavior in most BHTs. However, it does not explain some of the more complex observational phenomena very well especially when the source was in the harder state, and alternative models have been proposed to explain some of the unconventional behaviors of the disk-corona. \citet{2021arXiv210211635D} proposed a stratified accretion flow structure to explain the hard-state disk-corona behavior of MAXI~J1820+070. A simple model for a relativistic adiabatic expanding jet is explored by \citet{2001A&A...372L..25M} in BHT XTE~J1118+480. \citet{2006A&A...447..813F} proposed a magnetized accretion-ejection model that describes a new picture of a BH X-ray binary with multi-flow configuration in its central region. These models are based on the typical scenario but differ slightly from it, suggesting that there are differences between  different sources and that detailed modeling is required to explore these aspects.

The study of the intrinsic properties of an accretion disk (e.g., disk temperature and inner disk radius) helps us to better understand the physical phenomena occurring during an outburst; however, this inference relies on an accurate measurement of the disk emission. The observed disk spectrum must be corrected for the fraction of emitted disk photons that have been upscattered into the power-law component \citep{1995ApJ...445..780S,2016ApJ...822...60P}. To do so, \citet{2009PASP..121.1279S} developed a self-consistent disk-corona model, named  \texttt{simpl}, which accounts for the fraction of Compton-scattered disk photons in the computation of the total disk emission \citep{2005ApJ...619..446Y}. This model provides a more accurate normalization of the disk emission, and therefore of the apparent inner disk radius and color temperature \citep{2017ApJ...836..119S,2020ApJ...890...53S,2021ApJ...906...69Z}. 
%While for the latter, the ignorance of the hardening factor ($f_{\rm col}$) change may also lead to a bias of the disk emission measurement. The numerical simulations conducted by \citet{1995ApJ...445..780S} gave a value of $f_{\rm col}$ that fluctuates around 1.7. 
The true physical radius and effective temperature are then related to the apparent (fitted) radius and temperature via a hardening factor $f_{\rm col}$ \citep{1998PASJ...50..667K}. Numerical simulations \citep{1995ApJ...445..780S, 2005ApJ...621..372D, 2006ApJ...636L.113S} suggest a canonical value of $f_{\rm col} \sim 1.6-1.8$. However, some studies suggest that the value of $f_{\rm col}$ varies during the outburst evolution (e.g., \citealt{ 2000MNRAS.313..193M,2019ApJ...874...23D,2021MNRAS.tmp..941G}), especially in the LHS and HIMS (e.g., \citealt{2011MNRAS.411..337D,2013MNRAS.431.3510S}).

The accurate estimation of the disk emission is also important for the measurement of the spin parameter, especially for the continuum-fitting method  \citep{1997ApJ...479..381Z,2014SSRv..183..295M}. The other main technique to measure the spin parameter is the reflection-fitting method \citep{1989MNRAS.238..729F,2014SSRv..183..277R}.
However, the results obtained from the two techniques do not always agree with each other when the spin is non-extremal, i.e., $a_{*}< 0.9$ \citep{2009ApJ...697..900M, 2011MNRAS.416..941S}.
\citet{2021MNRAS.500.3640S} found that the discrepancies on spin measurements can be brought into agreement when we account for the uncertainties on $f_{\rm col}$.

%While such a simplified truncated disk model can broadly describe the common disk and corona behavior in most BHTs. However, it does not explain some of the more complex observational phenomena very well, and alternative models  \citep{2003MNRAS.342..355Z,2005A&A...430..761M,2006A&A...447..813F,2021arXiv210211635D} have been proposed to explain some of the unconventional behavior of the disk-corona. The disk behaviors of EXO~1846--031, the BH candidate studied in this work, which re-active in 2019, are slightly different from the typical scenario. This suggests that some aspects about the BHTs during the outburst are still worthy of detailed modeling.

To constrain the outburst evolution of disk parameters, coronal parameters and hardening factor, a broadband X-ray spectral coverage is required, so that thermal component and comptonized component can be fitted simultaneously and self-consistently. This is why our team has been monitoring BH outbursts with the {\it Hard X-ray Modulation Telescope} ({\it Insight-HXMT}), over the $\approx$1--150 keV band. One of the X-ray transients we have studied (the subject of this work) is EXO~1846--031, located at R.A. $=$ 18$^{h}$49$^{m}$16$^{s}$.99, Dec.~$=$\,$-$03$^{\circ}$03$^{\prime}$55$^{\prime\prime}$.4, with a 90\% error radius of $\approx$2$^{\prime\prime}$ \citep{2019ATel12969....1M}.

The first outburst of EXO~1846--031 was detected by \textit{EXOSAT} on 1985 April 3 \citep{1993A&A...279..179P}, and lasted for several months. After that, the source stayed in hibernation for over 30 years. Based on its spectrum associated with an ultra-soft component and a hard power-law tail, \citet{1993A&A...279..179P} suggested EXO~1846--031 to be a BH candidate. The recent outburst was first reported by \textit{MAXI}/GSC \citep{2019ATel12968....1N} on 2019 July 23. %Follow-up observations have been performed in radio by \textit{VLA} \citep{2019ATel12977....1M} and \textit{MeerKAT} \citep{2019ATel12992....1W} and in X-ray by \textit{NuSTAR} \citep{2019ATel13012....1M}, \textit{NICER} \citep{2019ATel12976....1B} and \textit{Swift}. 
The outburst was followed in the radio bands with the \textit{VLA} \citep{2019ATel12977....1M} and \textit{MeerKAT} \citep{2019ATel12992....1W}, and in the X-ray bands with \textit{NuSTAR} \citep{2019ATel13012....1M}, \textit{NICER} \citep{2019ATel12976....1B} and \textit{Swift}.
The Hard X-ray Modulation Telescope ({\it Insight-HXMT}) was also triggered by Target of Opportunity (ToO) observations \citep{2019ATel13036....1Y,2019ATel13037....1Y} of the source from 2019 August 2 to October 25. \citet{2020ApJ...900...78D} found an obvious reflection feature in the \textit{NuSTAR} spectrum. They reported an inclination angle of $(73\pm1)^\circ$ of the accretion disk and a high spin value of $0.997^{+0.001}_{-0.002}$. \citet{2021ApJ...906...11W} investigated the spectral properties in the HIMS and the HSS with \textit{NuSTAR} and {\it Insight-HXMT}, respectively. They suggested that the reflection component observed in the two spectral states originate from different sources of illumination; they also found that disk wind and jet probably co-existed in the HIMS in this source. \citet{2021RAA....21...70L} performed a detailed timing analysis with the observations of {\it Insight-HXMT}, \textit{NICER} and \textit{MAXI} and found LFQPOs in the HIMS.

% In addition to the reflection-fitting method \citep{1989MNRAS.238..729F,2014SSRv..183..277R} used by \citet{2020ApJ...900...78D}, the another method of measuring spin is the continuum-fitting method \citep{1997ApJ...479..381Z,2014SSRv..183..295M}. %The key assumption of these two methods is that the inner radius of the accretion disk reaches the ISCO of the BH, which usually occurs in the spectrum state is dominated by the disk component \citep{1997ApJ...479..381Z, 2014SSRv..183..295M}.

%EXO~1846--031 stays at the HSS for a long period in the 2019 outburst, whose spectrum is dominated by the soft thermal component. This fact makes EXO~1846--031 as a good candidate to measure its spin with the continuum-fitting method. 
In this paper, we report on the broad-band spectral evolution of EXO~1846--031 using the {\it Insight-HXMT} observations. In particular, we will show the effect of Compton scattering and of a variable hardening factor on our measurement of the disk parameters. Additionally, we measure the spin parameter with the continuum-fitting method and compare it with the one derived from the reflection-fitting method. The paper is organized as follows: in Section \ref{2}, we introduce the observations and data processing methods. The results are presented in Section \ref{3}. The discussion and conclusion follow in Sections \ref{4} and \ref{5}, respectively.

\section{Observations and Data Reduction}\label{2}

\subsection{Observations}\label{2.1}
Following the MAXI/GSC discovery of a new outburst of EXO~1846--031, we triggered {\it Insight-HXMT} tartget-of-opportunity observations, which covered 85 days, from 2019 August 2 to October 25 (Table~\ref{tab:obsinf}).

{\it Insight-HXMT} \citep{2014SPIE.9144E..21Z,2020SCPMA..63x9502Z} is the first Chinese X-ray astronomy satellite, launched on 2017 June 15. It has a 550-km low-Earth-orbit with an inclination of $43^\circ$. It contains three slat-collimated instruments, sensitive to different energy ranges: the High Energy (HE) \citep{2020SCPMA..63x9503L}, Medium Energy (ME) \citep{2020SCPMA..63x9504C}, and Low Energy (LE) \citep{2020SCPMA..63x9505C} X-ray Telescopes. 
HE, ME and LE are sensitive to the 20.0--250.0~keV, 5.0--30.0~keV and 1.0--15.0~keV bands, with detection areas of 5100~${\rm cm}^{2}$, 952~${\rm cm}^{2}$ and 384~${\rm cm}^{2}$, respectively. The corresponding time resolutions are 4~${\mu}$s, 240~${\mu}$s and 1~ms.
% HE is sensitive in 20.0--250.0~keV and its time resolution is 4~$\mu$s. The total detection area of HE is 5100~${\rm cm}^{2}$. ME is sensitive in 5.0--30.0~keV energy band with the total detection area of 952~${\rm cm}^{2}$. LE is sensitive in the energy range of 1.0--15.0~keV and its total detection area is 384~${\rm cm}^{2}$. 

\subsection{Data Reduction}\label{2.2}
We used the {\it Insight-HXMT} Data Analysis software ({\sc hxmtdas}) v2.02\footnote[1]{http://hxmt.org/software.jhtml} to analyze all the data. The filtering criteria for the good time intervals were: (1) an offset angle from the pointing direction  $<\ 0.04^\circ$; (2) a pointing direction above earth $>\ 10^\circ$; (3) a value of the geomagnetic cutoff rigidity $>$ 8~GeV; (4) a rejection of data within 300~s of the South Atlantic Anomaly passage. A detailed explanation of the {\it Insight-HXMT} data reduction is published on its website\footnote[2]{http://hxmt.org/SoftDoc.jhtml}.
Backgrounds rates were estimated with the tools: {\it lebkgmap}, {\it mebkgmap}, {\it hebkgmap} 
%LEBKGMAP, MEBKGMAP and HEBKGMAP 
(version 2.0.9), based on the standard {\it Insight-HXMT} background models \citep{2020JHEAp..27...24L,2020JHEAp..27...44G,2020JHEAp..27...14L}. We obtained the response files with the tasks {\it lerspgen}, {\it merspgen} and {\it herspgen}.

We rebinned the spectra to a minimum of 30 counts per bin using the {\sc ftools} \citep{1995ASPC...77..367B} task {\it grppha}. We modelled them with {\sc xspec} Version 12.10.1\footnote[3]{https://heasarc.gsfc.nasa.gov/docs/xanadu/Xspec/} \citep{1996ASPC..101...17A}. We used the $\chi^{2}$ statistics, and added a systematic uncertainty of 1.5\%.
The energy bands adopted for our spectral analysis are: 1.0--10.0~keV (LE), 10.0--30.0~keV (ME) and 30.0--150.0~keV (HE).
We jointly fitted the spectra of the three instruments and included a multiplicative constant to account for the relative flux calibration uncertainties \citep{2020JHEAp..27...64L}. All spectral models include a photoelectric absorption, modelled with {\it TBabs} and `wilm' abundances \citep{2000ApJ...542..914W}. 
The results of our spectral analysis are presented in the Section~\ref{3.2}. We only report results from observations with an exposure time longer than 200~s. All uncertainties are quoted at the 90\% confidence level.

\section{Results}\label{3}

\subsection{Light Curves and Hardness Ratio}\label{3.1}

\begin{figure*}
\includegraphics[width=1\linewidth]{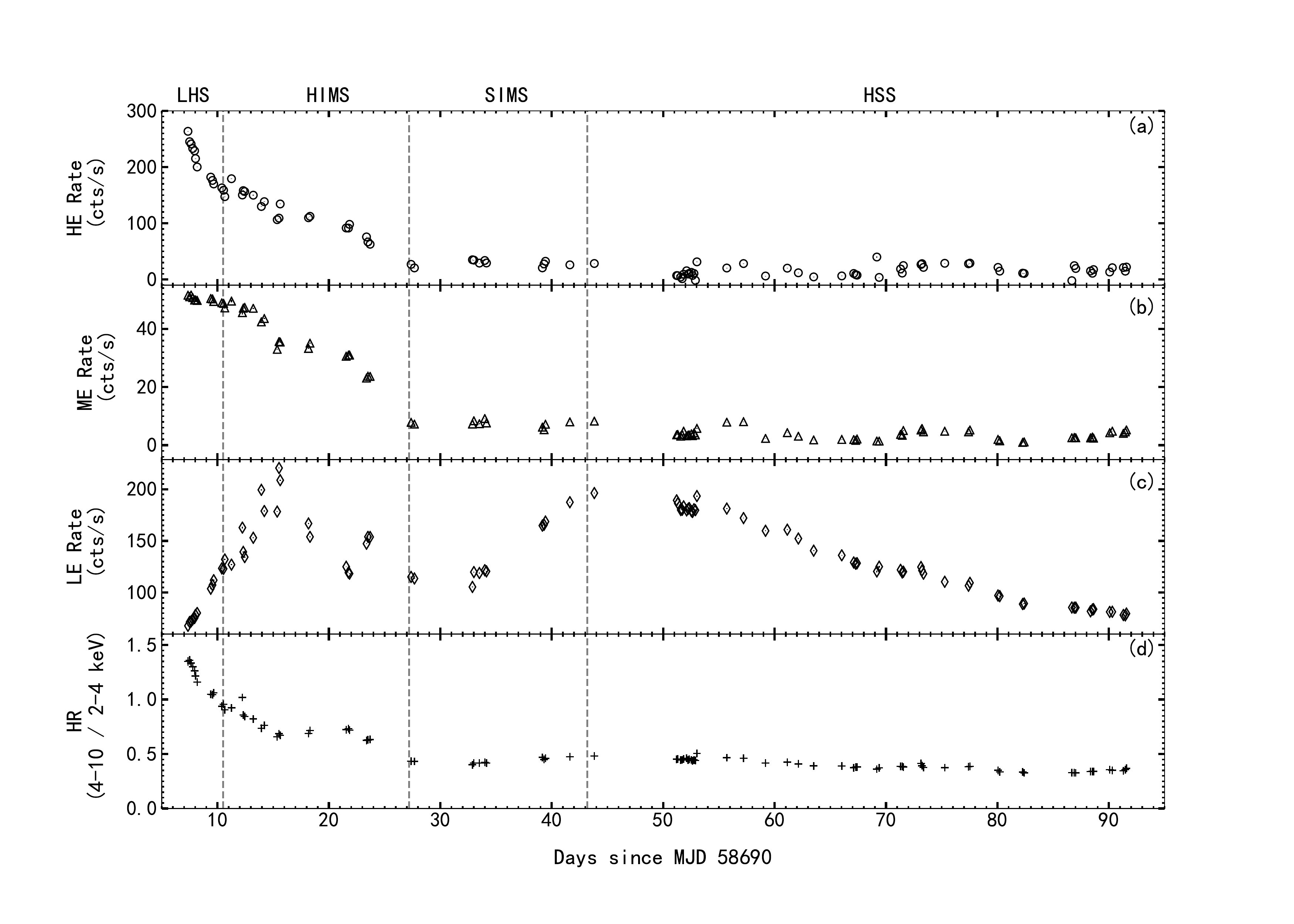}
\vspace{-1cm}
\caption{Top three panels: light curves measured with the {\it Insight-HXMT} detectors: (a) HE (25--150~keV); (b) ME (10--20~keV); (c) LE (1--10~keV). Bottom panel (d): LE hardness ratio (4--10~keV over 2--4 keV count rates). Each data point represents one {\it Insight-HXMT} exposure. The spectral state classification follows \citet{2021RAA....21...70L}, except for the definition of LHS. State transitions are marked with dashed lines.
\vspace{0.3cm}
\label{fig:light curve}}
\end{figure*}

Figure~\ref{fig:light curve} shows the long-term light curves in different energy bands and the evolution of the LE hardness ratio (defined as the 4--10~keV count rate divided by the 2--4~keV rate). The HE and ME count rates show a similar evolution: first a decrease with time and then a relatively low, stable level (Figure ~\ref{fig:light curve}). Instead, the LE light curve shows a more complex evolution, with two peaks. The LE count rate rapidly increased after MJD 58697.35, reached a peak value of $220.5\pm0.4$~cts s$^{-1}$ on MJD 58705.53, and then decreased gradually. Then, it increased again during the SIMS and finally decreased to a relatively low value ($79.7\pm0.3$ cts~s$^{-1}$) during the HSS. We do not know the exact reason for the formation of these two peaks. From the LE hardness ratio we see that the spectrum gradually softened with time (bottom panel of Figure~\ref{fig:light curve}). 
%\textbf{The first peak may be due to an accretion instability process during the transition state.} 

\citet{2021RAA....21...70L} subdivided the outburst into four spectral states based on the relative changes in the hardness-intensity diagram and the fractional rms integrated over the $2^{-5}$--32~Hz band. In the following spectral analysis, we will adopt their spectral state classification, except for their definition of the LHS. 
Their preliminary classification was based on hardness ratios and included only the first three {\it Insight-HXMT} observations in the LHS. Here, we used spectral analysis to obtain a more accurate classification of this state. We adopted the canonical definition of LHS as the epochs when the photon index $\Gamma \sim$ 1.5--1.7 and the disk flux fraction $f_{\rm dbb} \leq$ 20\% \citep{2006ARA&A..44...49R, 2010LNP...794...53B}. With this definition, the LHS interval includes the first eight {\it Insight-HXMT} observations (Section~\ref{3.2.1}).
%\textbf{They only roughly classified the first three observations as LHS based on the absence of significant changes in hardness, and they also mentioned that a more precise classification needs to be combined with the results of spectrum analysis.} 
%We thus identify the LHS interval as the first eight {\it Insight-HXMT} observations, based on the spectral features \citep{2006ARA&A..44...49R, 2010LNP...794...53B}; the photon index $\Gamma \sim$ 1.5--1.7 and the disk flux fraction $f_{\rm dbb} \leq$ 20\% (Section~\ref{3.2.1}).

\subsection{Broad-band Spectral Models}\label{3.2}

\begin{figure*}
\centering
\vspace{-0.5cm}
\subfigure{
\includegraphics[width=8cm,height=9cm]{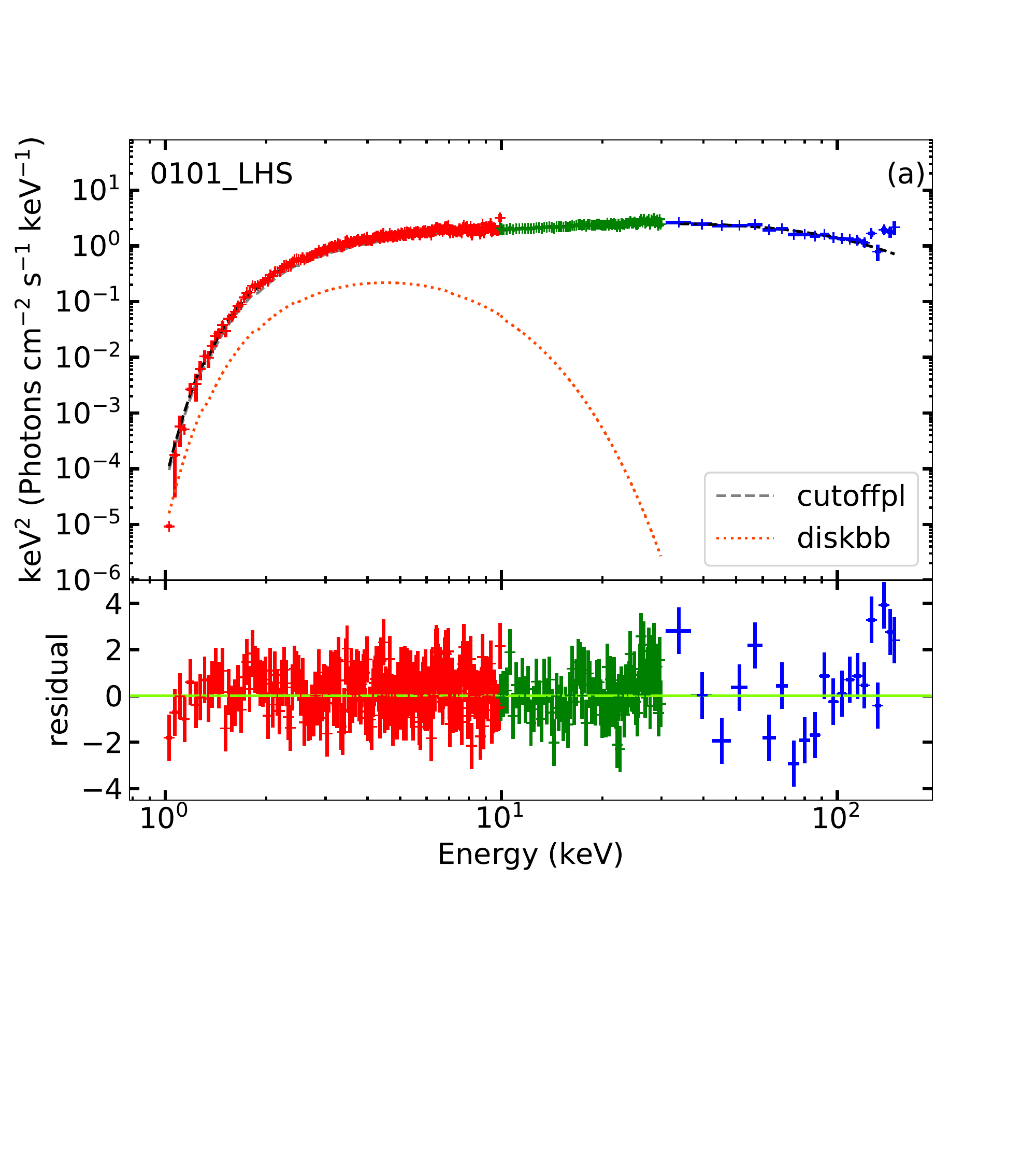}
}\vspace{-1.3cm}
\quad
\subfigure{
\includegraphics[width=8cm,height=9cm]{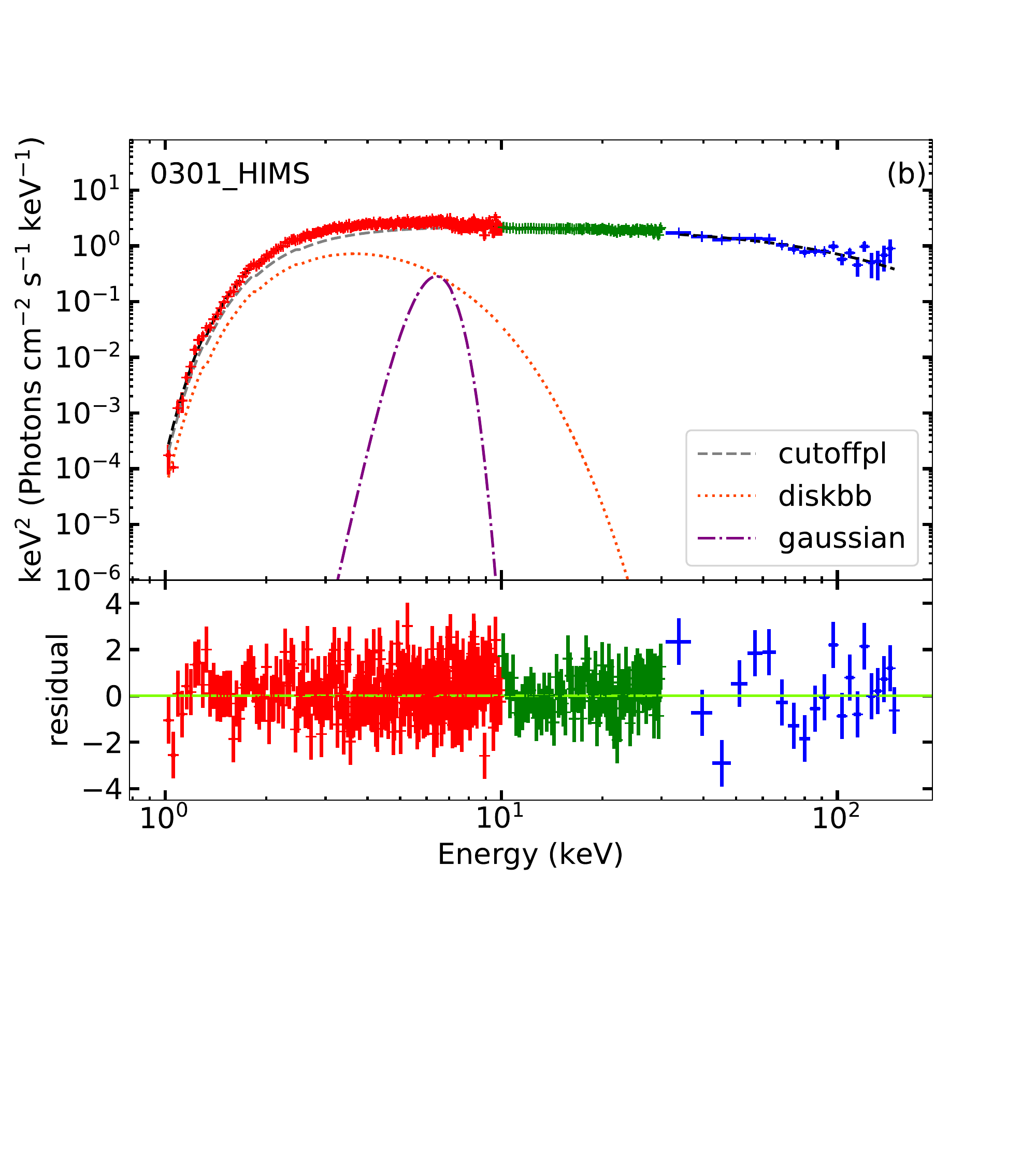}
}\vspace{-1.3cm}
\quad
\subfigure{
\includegraphics[width=8cm,height=9cm]{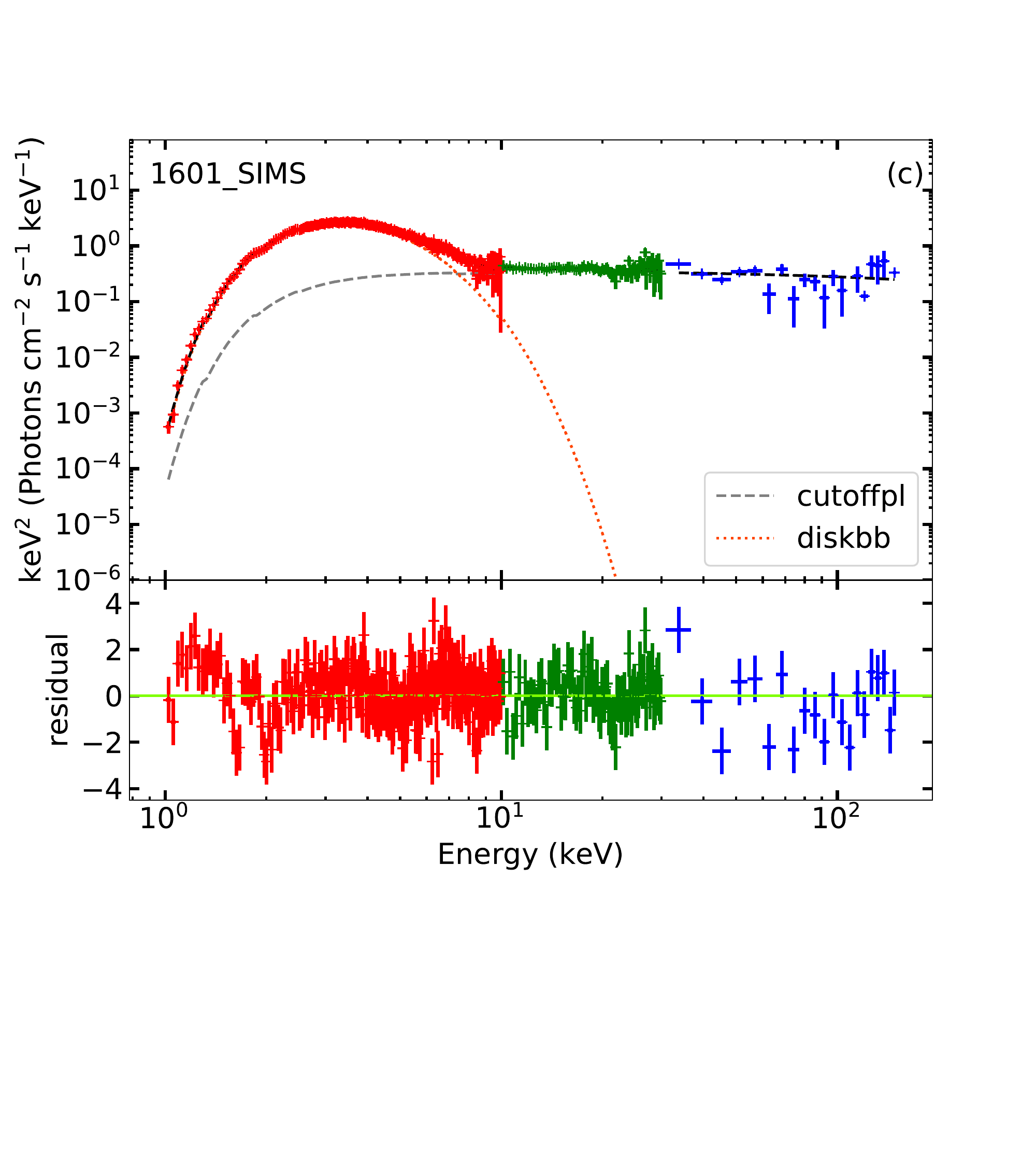}
}\vspace{-1.3cm}
\quad
\subfigure{
\includegraphics[width=8cm,height=9cm]{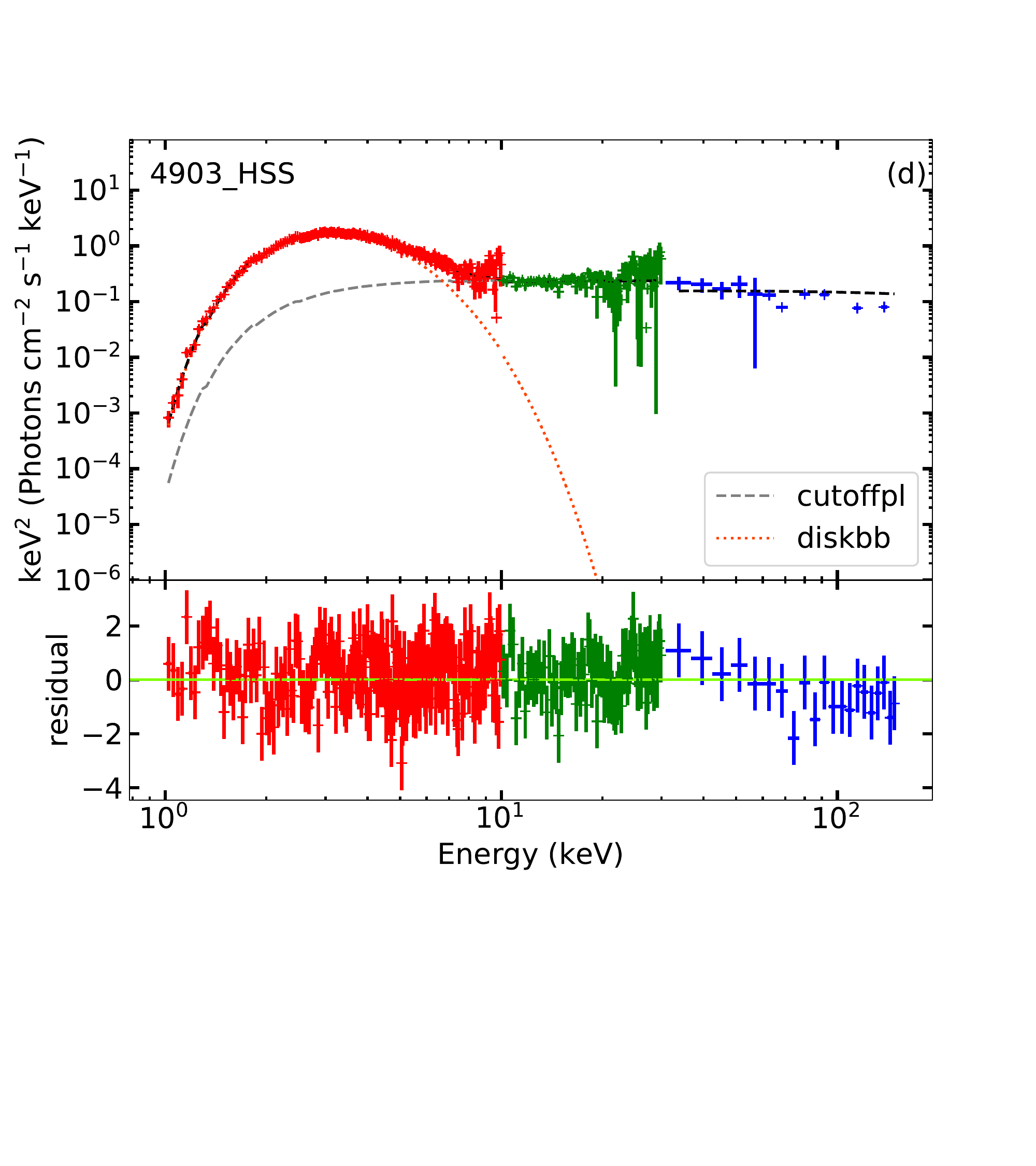}
}\vspace{-1.3cm}
\subfigure{
\includegraphics[width=8cm,height=9cm]{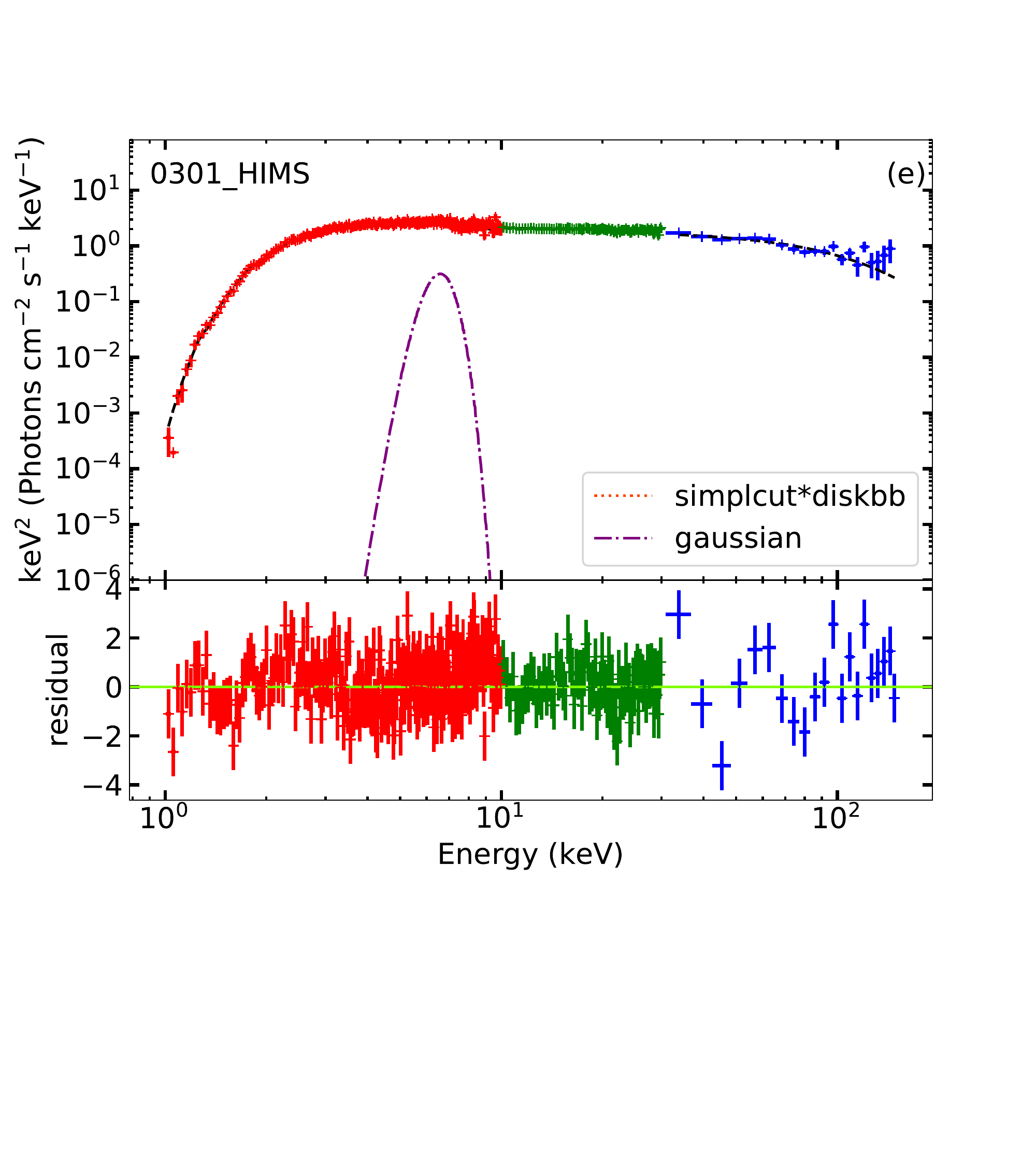}
}\vspace{-1.1cm}
\quad
\subfigure{
\includegraphics[width=8cm,height=9cm]{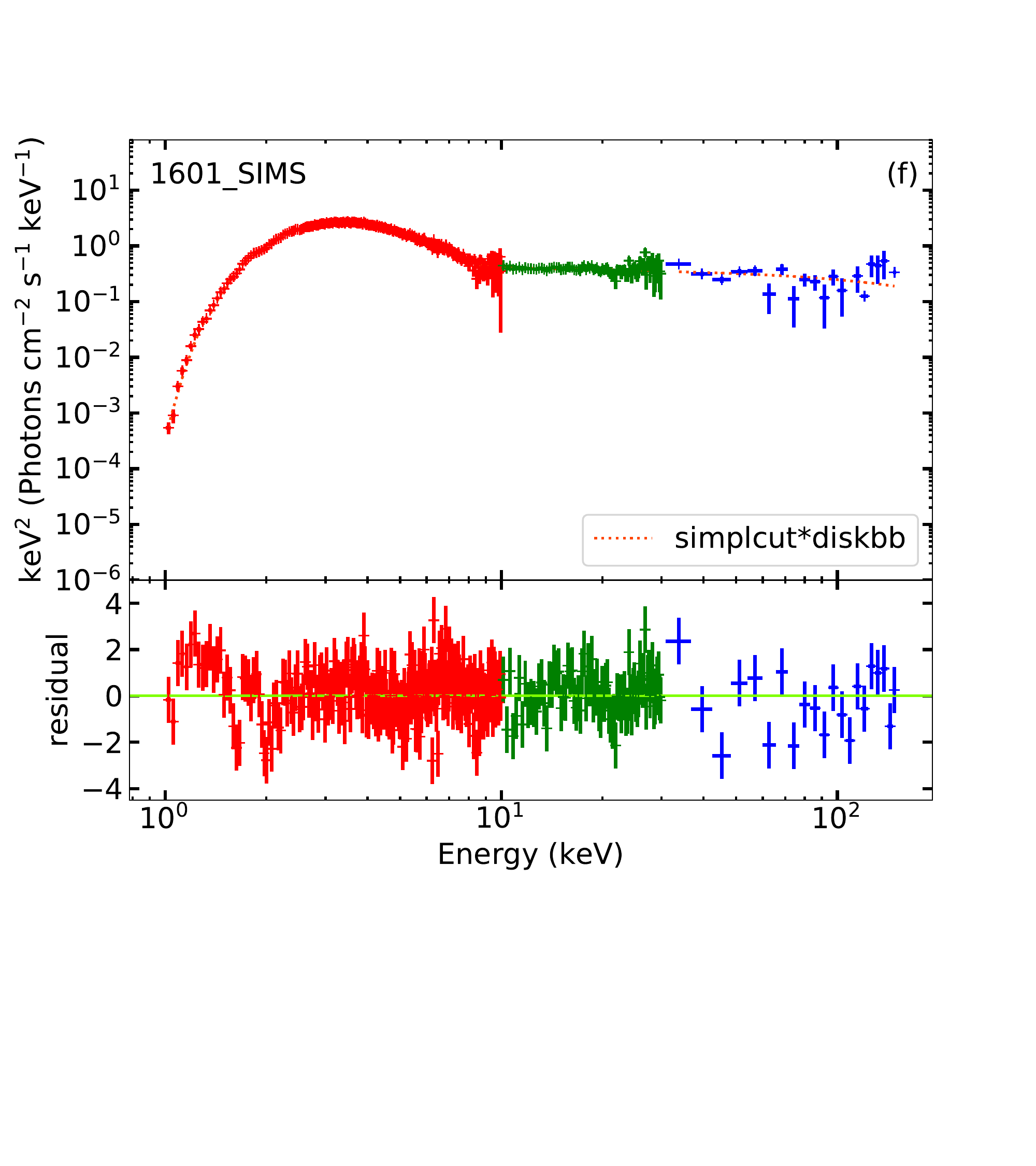}
}\vspace{-1.1cm}
\caption{Unfolded spectra, model components and fit residuals at different epochs (with the corresponding observation IDs labelled inside each panel). Red, green and blue data points correspond to the LE, ME and HE spectra, respectively.}
%Illustration of the unfolded spectra and components of different models. The red, green and blue points are corresponding to the LE, ME and HE spectra, respectively. The observation ID associated with the spectral state and the applied model are marked on each panel.}
\label{fig:fitting results}
\end{figure*}

%To study the spectral evolution of EXO~1846--031, we performed a detailed spectral analysis with the {\it Insight-HXMT} observations in the 1--150~keV energy band. We only included the observations with an exposure time longer than 200~s (Table~\ref{tab:fitting}).

\subsubsection{\texttt{cutoffpl} + \texttt{diskbb} model}\label{3.2.1}
The continuum X-ray emission of BH X-ray binaries generally consists of a thermal/soft and a non-thermal/hard component. The former represents the disk emission and is usually fitted with a multi-color disk-blackbody component (\texttt{diskbb} in {\sc xspec}) \citep{1984PASJ...36..741M,1986ApJ...308..635M}.
The latter represents the coronal emission, {\it i.e.}, it comes from the Comptonization of disk photons in a hot electron cloud \citep{1975ApJ...199L.153E,1980A&A....86..121S}. It is usually fitted with a power-law component, with or without a high-energy cutoff (\texttt{cutoffpl} or \texttt{powerlaw}). 
% These two components play different roles in different spectral states. The hard state and soft state spectra are dominated by the \texttt{cutoffpl} and \texttt{diskbb} components respectively. These two components play comparable roles in the two intermediate states. 
We tried different combinations of the \texttt{cutoffpl} and \texttt{diskbb} components in each spectral state (alone or in combination). Based on the F-test probability ({\it ftest} command in {\sc xspec}), we found that both components are always present throughout the outburst, with a probability $>$99.9\%. 
%We included both components when the F-test probability ({\sc ftest} command in {\sc xspec}) of a single component was lower than 0.001.
%: one \texttt{cutoffpl}, one \texttt{diskbb} and the combination of the two components. We used the {\sc ftest} command in {\sc xspec} to test the reasonableness of the two-component model compared to the one-component model. The combination model was only used when the F-test probability was lower than 0.001.
%We find that all the spectra require a combination model. 
%\textbf{\textcolor[rgb]{1,0,0}{In addition, a Fe emission line is seen in the spectra. In order to test the significance of the emission line, we use the {\it simftest} routine in {\sc XSPEC} with 10,000 simulations for the observation with the longest exposure time in each spectral state. The probabilities of the line to be required by each state from LHS to HSS, are P = 98.1\%, $\sim$100\%, 99.3\%, and 57.7\%, corresponding to a significance of 2.35$\sigma$, 5$\sigma$, 2.7$\sigma$, and 0.8$\sigma$, respectively.  Hence we only added an additional Gaussian component in the HIMS. Additionally, we are aware that if combining more data, the line significance would increase, like what has done by \citet{2021ApJ...906...11W}. As we are aim to closely monitor the changes in the hardening factor, we prefer not to combine the data.}}

In some epochs, the fit is improved with the addition of a Gaussian emission line (the energy of the line is limited to between 6.4 and 7 keV) around 6.4--6.6 keV, corresponding to Fe K$\alpha$ emission. In order to test the significance of this emission line, we followed the approach described by several previous works \citep{2015ApJ...799..123B, 2015MNRAS.447.2274B, 2017ApJ...838..133S, 2020ApJ...899L..19G, 2021ApJ...906...11W} and  used the {\it simftest}\footnote[4]{https://heasarc.gsfc.nasa.gov/xanadu/xspec/manual/node126.html} routine in {\sc XSPEC} with 10,000 simulations for the observation with the longest exposure time in each spectral state. The probabilities of the line to be required in each state, from LHS to HIMS, SIMS and HSS, are P = 98.1\%, $>$99.9\%, 99.3\%, and 57.7\%, corresponding to a significance of 2.35$\sigma$, $>$5$\sigma$, 2.7$\sigma$, and 0.8$\sigma$, respectively. Additionally, we also calculated the ratio between the normalization of the \texttt{gaussian} component and its $1 \sigma$ error, and we only included the \texttt{gaussian} when the ratio is larger than 3, which is true only at HIMS. The line significance and the ratio increases if we combine multiple spectra in each state, as shown in \citet{2021ApJ...906...11W}. However, here we are monitoring the time evolution of the spectra, therefore we prefer not to combine them. Hence, we only included the Fe line in our fits to the HIMS spectra. The mean value of the best-fitting central energy of the line in the 18 HIMS epochs (Tables \ref{tab:fitting} and \ref{tab:fitting_1sc1d}) is $E_{\rm gau} = 6.55$ keV, with a 1$\sigma$ scatter of 0.15 keV. Thus, the line is consistent with both neutral (6.4 keV) and He-like (6.7 keV) iron.

%\textcolor[rgb]{0,0,1}{\citet{2002ApJ...571..545P} pointed out that {\it ftest} is not appropriate for testing the significance of a spectral line and provided a workaround with a similar behavior that described by {\it simftest}\footnote[4]{https://heasarc.gsfc.nasa.gov/xanadu/xspec/manual/node126.html} command in {\sc XSPEC} to detect such a line. Moreover, several previous works \citep{2015ApJ...799..123B, 2015MNRAS.447.2274B, 2017ApJ...838..133S, 2020ApJ...899L..19G, 2021ApJ...906...11W} have uesd {\it simftest} to detect the significance of a line.}

To check the robustness of our results, we refitted the observation with the longest exposure time of each spectral state with the reflection model \texttt{relxill} \citep{2013MNRAS.430.1694D, 2014ApJ...782...76G} plus \texttt{diskbb}. We obtained best-fitting values of $T_{\rm in}$ and $N_{\rm dbb}$ similar to those derived from the simpler model; the latter is therefore preferred. We also fitted some LHS and HIMS data with the high density reflection model \texttt{relxillD} to check whether the disk component was really required. For the LHS and HIMS spectra, the \texttt{relxillD} component alone gives $\chi^{2}$/dof = 1451.90/1300 and 1262.59/1326, respectively. Adding a \texttt{diskbb} component improves the two fits by $\Delta \chi^{2}$ = 78.82 and 29.97, respectively, for the loss of 2 dof. The F-test probabilities of $1.85 \times 10^{-16}$ and $1.24 \times 10^{-7}$ indicate that the \texttt{diskbb} component is significantly required by the LHS and HIMS data. In addition, we compared the best-fitting parameters of \texttt{diskbb} component obtained from model \texttt{relxillD+diskbb} and the simpler model. The best-fitting values of $T_{\rm in}$ and $N_{\rm dbb}$ all vary slightly,  $\lesssim 4\%$  and $\lesssim 5\%$, respectively. Besides, the evolution trends of $T_{\rm in}$ and $N_{\rm dbb}$ from the LHS to the HIMS are all consistent with those derived from the simpler model, thus our main results will not be affected.
%For the LHS and HIMS, the best-fitting values of $T_{\rm in}$ obtained from the two models are very similar, $1.59^{+0.16}_{-0.15}$ vs $1.65^{+0.13}_{-0.16}$ and $1.17^{+0.13}_{-0.11}$ vs $1.19^{+0.05}_{-0.05}$, respectively. $N_{\rm dbb}$ varies ($\lesssim 5\%$) slightly and its evolution trend from the LHS to the HIMS is consistent, thus our main results will not be affected.}
%This has actually been confirmed by \citet{2020ApJ...900...78D} with \textit{NuSTAR} data. Since the focus of this work is on the evolution continuum parameters, thus we prefer to adopt the simpler model.

%\textbf{\textcolor[rgb]{1,0,0}{To further check the presence of Fe emission line in the LHS and HSS, we try to combine some LHS data, and the significance of the line increase to 3$\sigma$. However, since the LHS count rate is significantly different in different observations, and we are aiming to monitor the changes in the hardening factor closely (Section~\ref{3.3}), we prefer not to combine the data. Similarly, we also try to combine the HSS data and see the existence of weak iron lines, but this has already been reported in \citet{2021ApJ...906...11W}. The focus of our work is to study the evolution of continuum parameters, adding or not adding a \texttt{gaussian} component would not affect our results except for the HIMS.}}

We show a representative sample of unfolded spectra (one for each state), together with the corresponding models, in Figure~\ref{fig:fitting results}.

\begin{figure*}
\vspace{-1.5cm}
\includegraphics[width=1\linewidth]{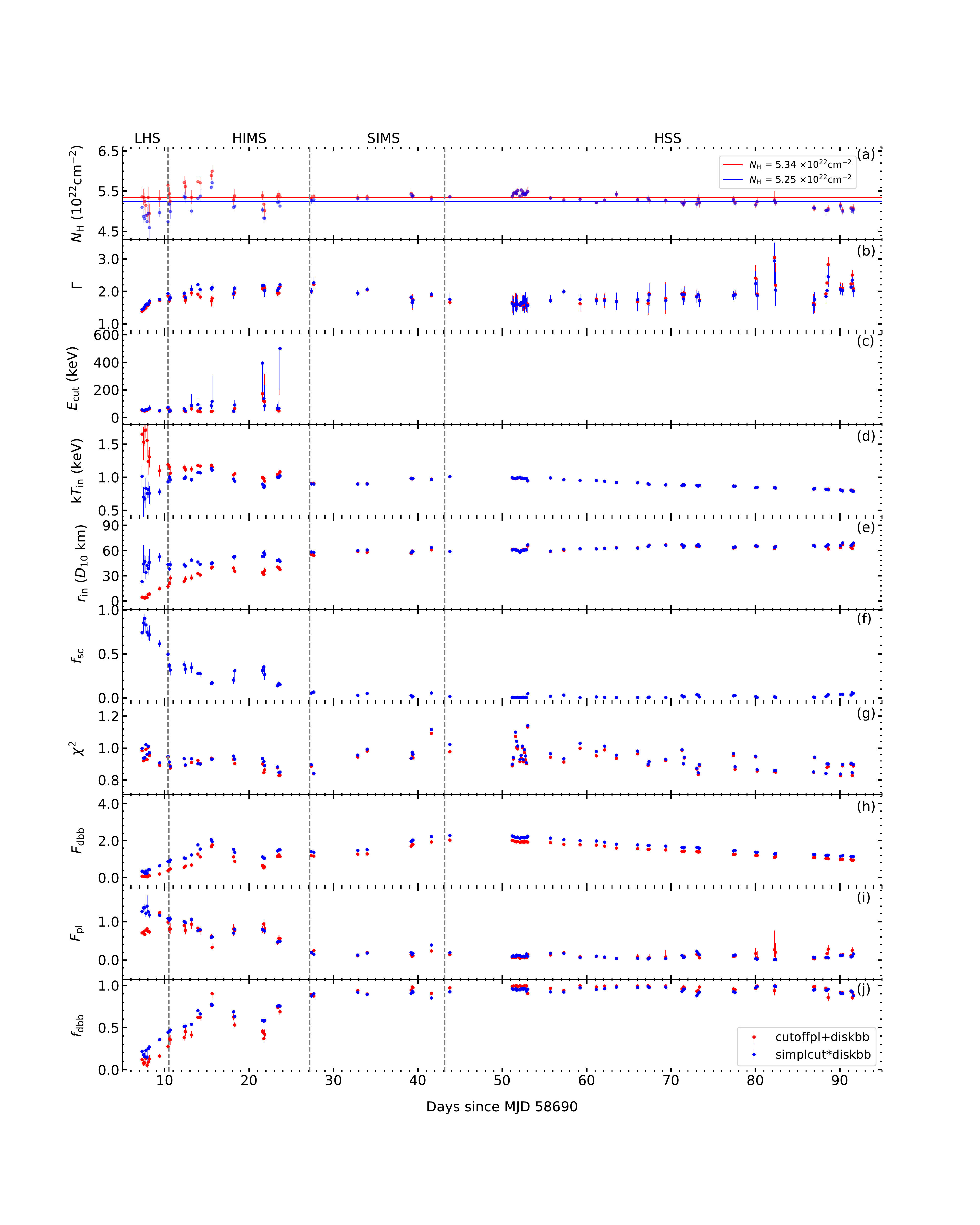}
\vspace{-1.5cm}
\caption{Best-fitting parameters of the {\it Insight-HXMT} spectra. Red and blue dots indicate the parameter values derived from the models \texttt{cutoffpl+diskbb} and \texttt{simplcut*diskbb}, respectively. Panel (a) shows the best-fitting values of the X-ray absorption column density, $N_{\rm H}$, when it is left as a free parameter; it also shows the two constant values of $N_{\rm H}$ that we have adopted for further analysis, that is $N_{\rm H} = 5.34 \times 10^{22}$ cm$^{-2}$ and $N_{\rm H} = 5.25 \times 10^{22}$ cm$^{-2}$ for \texttt{cutoffpl+diskbb} and \texttt{simplcut*diskbb}, respectively. The parameter values plotted in  panels (b)--(j) are the best-fitting values for the models with fixed $N_{\rm H}$.
$\Gamma$ is the photon index of the power-law or Comptonized component. $E_{\rm cut}$ is the exponential cutoff energy. It was unconstrained in the SIMS and HSS (no statistical improvement over a simple power-law), and was fixed at 500~keV for fitting purposes in {\sc xspec}; we omitted those values in panel (c). $T_{\rm in}$ is the peak colour temperature of the disk.  $r_{\rm in}$ is the apparent inner disk radius, defined as $r_{\rm in}$ = $\sqrt{N_{\rm dbb}/\cos~\theta} \times D_{\rm 10}$, where $D_{10}$ is the source distance in units of~10 kpc, 
%and takes the value of $D_{10} = 0.7$ \citep{1993A&A...279..179P} for this plot, 
and $\theta$ is the inclination angle of the accretion disk.  We adopted $\theta=73^{\circ}$ \citep{2020ApJ...900...78D}. $f_{\rm sc}$ is the scattered fraction, {\it i.e.}, the proportion of disk photons scattered by the corona. $F_{\rm dbb}$ and $F_{\rm pl}$ stand for the unabsorbed bolometric flux of the disk component and power-law component, in units of $10^{-8}\ {\rm erg\ cm^{-2}\ s^{-1}}$. $f_{\rm dbb}$ is the ratio between the disk flux and the total flux.
\label{fig: para_all}}
\end{figure*}

We set out to determine the evolution of the main physical parameters over the course of the outburst. As a first step of our analysis, we fitted each spectrum with a free absorbing column density parameter $N_{\rm H}$ (red datapoints in Figure~\ref{fig: para_all}a). The best-fitting value of $N_{\rm H}$ is stable in the SIMS and HSS, but there is some scatter in the LHS and HIMS. There is some degeneracy between the best-fitting values of $N_{\rm H}$ and $T_{\rm in}$, which makes it difficult to determine the intrinsic evolution of the disk parameters. To reduce this hindrance, we froze $N_{\rm H}$ at the mean value ($N_{\rm H} \approx 5.34 \times 10^{22}$ cm$^{-2}$, red line in Figure~\ref{fig: para_all}a) derived from the SIMS and HSS fits, and refitted all the spectra with fixed $N_{\rm H}$.

The cutoff energy of the (dominant) power-law component is well constrained for most of the spectra in the LHS and HIMS (Figure 3). Instead, in the SIMS and HSS, the power-law component weakens: the low-energy spectrum becomes dominated by the \texttt{diskbb} component, and the high-energy tail above $\approx$60--70~keV becomes dominated by background counts. For this reason, if we let the cutoff energy vary freely in the fit, most of the times $E_{\rm cut}$ will peg at its upper limit, meaning that a free cutoff energy does not provide a significant improvement to the fit.
To better constrain $E_{\rm cut}$ in the HSS, we also tried a simultaneous fit of several spectra near the peak of the outburst, to increase the signal to noise. The cutoff energy obtained with this method is formally constrained between 46 and 100~keV. However, the corresponding photon index is only $1.17\pm{0.08}$, even lower than in the LHS and HIMS, which we regard as unphysical. This contradicts the expectations from the observed evolution (strong softening) of the hardness ratio from LHS to HSS. We suspect that the main reason for this unphysical fitting result is the strong degeneracy between the photon index and the cutoff energy. The two quantities cannot be well constrained simultaneously after the evolution of the source into the SIMS, and the presence or absence of a cutoff makes no substantial difference to the rest of the spectrum. To keep the model self-consistent throughout the outburst evolution, we fixed $E_{\rm cut}$ at 500~keV in the SIMS and HSS. The photon index then increases to $\approx 2$, which is more consistent with the typical value of photon index in the soft state. In addition, other parameters such as $T_{\rm in}$ ($\approx 1$~keV), the fraction of \texttt{diskbb} component ($\geq 90\%$) and the fraction of \texttt{cutoffpl} component ($\leq 10\%$) at this time all in the canonical range of soft state.

We show the best-fitting parameters (Table~\ref{tab:fitting}) of the \texttt{cutoffpl} + \texttt{diskbb} model (with the assumed constant $N_{\rm H}$), as well as the unabsorbed bolometric disk flux, power-law flux, and the ratio between disk flux and the total flux, as red dots in Fig~\ref{fig: para_all}. The four parameters of interest are the power-law photon index, the cutoff energy (only in the LHS and HIMS), the inner disk radius and the peak disk temperature.
% The X-ray absorption column density ($N_{\rm H}$), the photon index ($\Gamma$), the exponential roll-off energy ($E_{\rm cut}$), the inner disk temperature ($T_{\rm in}$), the apparent radius of inner disk ($r_{\rm in}$) obtained with the normalization of \texttt{diskbb} component $r_{\rm in}\ =\ N_{\rm dbb}^{1/2}\ *\ cos(\theta)^{-1/2}\ *\ D_{10}(7)$, where $D_{10}$ is the source distance in units of 10 kpc and $\theta$ is the inclination angle of the disk; in our calculation we adopt the source distance $D$ to be 7~kpc \citep{1993A&A...279..179P} and the angle of inclination to be $73^{\circ}$ \citep{2020ApJ...900...78D}),
% %the square root of the normalization of the \texttt{diskbb} model ($\sqrt{N_{\rm dbb}} = r_{\rm in} * D_{10}^{-1} * \sqrt{\rm cos(\theta)}$, where $r_{\rm in}$ is the apparent radius of inner disk, $D_{10}$ is the source distance in units of 10 kpc, and $\theta$ is the inclination angle of the disk),
% the characterizing the goodness of fit reduced $\chi^{2}$,  the unabsorbed flux of the \texttt{cutoffpl} ($Flux_{\rm pl}$) and \texttt{disk} components ($F_{\rm dbb}$) in the 0--150 keV band (for a more accurate calculation of disk flux), and the evolution of the ratios of the \texttt{diskbb} component flux to the total flux ($f_{\rm dbb})$ are shown in Fig~\ref{fig: para_all} (red dots). 

The photon index $\Gamma$ gradually increased from $\approx$1.4 to $\approx$2.2 as the source evolved from the LHS to the HIMS, and then remained almost constant within uncertainties in the SIMS and HSS.
%and settled at $\approx$2.0 after the transition to the HSS. 
% There is a slight fluctuation in $\Gamma$, but it stays roughly around 2.0 when the source transitioned to the SIMS and HSS (hereafter we refer to SIMS and HSS together as soft state, SS). 

$E_{\rm cut}$ fluctuated between $\approx$40 and $\approx$90~keV when the source was in the LHS and the HIMS, except for the last few observations of the HIMS. As mentioned above, it was unconstrained in the SIMS and HSS, and was fixed at 500 keV for fitting purposes in {\sc xspec}; we did not plot such data points as shown in Figure~\ref{fig: para_all}c.

The inner-disk color temperature $T_{\rm in}$ unexpectedly decreased during the LHS. It remained approximately constant around 1 keV throughout the HIMS. Finally, it gradually decreased (as expected) during the outburst decline in the HSS.

The apparent inner radius of the accretion disk, $r_{\rm in}$ ($\propto \sqrt{N_{\rm dbb}}$ where $N_{\rm dbb}$ is the \texttt{diskbb} normalization), unexpectedly increased during the LHS and HIMS. This is seemingly inconsistent with the canonical scenario that the inner disk moves towards ISCO during those states \citep{2006ARA&A..44...49R,2011MNRAS.415..292M,2015ApJ...811...51T}. We will suggest a possible explanation in Section~\ref{3.3}. After reaching its peak in the SIMS, $r_{\rm in}$ remained mostly constant throughout the HSS, apart from a slight increasing trend (only by $\sim$10\%) near the end of that state. We interpret the evolution of $r_{\rm in}$ as evidence that the disk had reached a steady state (Shakura-Sunyaev solution) and the inner disk radius had settled at ISCO at the beginning of the HSS.

%We calculate the apparent inner radius of the accretion disk as $r_{\rm in}$ = $\sqrt{N_{\rm dbb}/\cos~\theta}$ * $D_{\rm 10}$, where $N_{\rm dbb}$ is the normalization of the \texttt{diskbb} component, $D_{10}$ is the source distance in units of 10~kpc and $\theta$ is the inclination angle of the disk. Here, we adopt an inclination angle of $73^{\circ}$ from \citet{2020ApJ...900...78D}.
%The apparent inner radius $r_{\rm in}$ shows an increasing trend in the LHS and HIMS. After reaching its peak in the SIMS, $r_{\rm in}$ remains constant. We interpret this as evidence that the inner disk radius settled at ISCO. The overall evolution trend of $r_{\rm in}$ is inconsistent with previous studies \citep{2006ARA&A..44...49R,2011MNRAS.415..292M,2015ApJ...811...51T}. We will investigate the behavior of the disk radius further in the next Section.

The unabsorbed disk flux $F_{\rm dbb}$ increased rapidly and then decreased slowly in the LHS and HIMS, then starts to increase again in the SIMS and decrease slightly in the HSS. The power-law flux $F_{\rm pl}$ increased slowly in the LHS and then started to decrease after reached HIMS.
%and total flux $F_{\rm total}$ evolved in a similar way to each other (Figure~\ref{fig: para_all}h and Figure~\ref{fig: para_all}i).
% The flux of the \texttt{diskbb} component $F_{\rm dbb}$ increases rapidly and then decreases slowly in the HIMS, then starts to increase again in the SIMS and decreases slightly in the HSS. The total flux $F_{\rm total}$ has the similar evolution trend with $F_{\rm dbb}$. 
The ratio between the disk flux and the total flux, $f_{\rm dbb}$, increased significantly in the LHS and HIMS, reached its peak in the SIMS and remained almost constant ($\geq 90\%$) after that. 
%\textbf{The value of $f_{\rm dbb}$ increases from 20\% to 60\% in the HIMS; it gradually increases in the SS first and then remains almost at 90\%. (do not understand what you want to address!)}

\subsubsection{\texttt{simplcut} * \texttt{diskbb} model}\label{3.2.2}
Before we try to attribute any physical meaning to the unusually small value of $r_{\rm in}$ in the LHS and HIMS, we need to be aware that a phenomenological \texttt{cutoffpl} + \texttt{diskbb} model does not self-consistently account for the photons that are removed from the disk component and upscattered into the power-law component. As a result, the normalization of the disk emission (hence, $r_{\rm in}$) is usually underestimated by such model when a corona is present \citep{2005ApJ...619..446Y}. To get around this problem, we re-fitted the spectra with the self-consistent Comptonization model \texttt{simplcut} \citep{2017ApJ...836..119S} applied to a disk-blackbody seed spectrum.

When fitting the \texttt{simplcutx*diskbb} model, we adopted the same method described in Section 3.2.1 
%as fitting \texttt{cutoffpl} + \texttt{diskbb} model 
to determine the value of $N_{\rm H}$. We fixed it at the mean value during the SIMS and HSS ($5.25 \times 10^{22}$ cm$^{-2}$, blue line in Figure~\ref{fig: para_all}a) for all the subsequent spectral modelling.

Panels (e) and (f) of Figure~\ref{fig:fitting results} show two representative unfolded spectra and residuals based on the \texttt{simplcut*diskbb} model (one for the HIMS, and one for the SIMS).

%The broad-band spectra from the LHS to the HSS  are well fitted by the \texttt{cutoffpl+diskbb} model. However, the derived inner radius of the accretion disk in the LHS and HIMS is unrealistically smaller than that in the HSS when the inner disk radius has probably reached the ISCO. To see if this peculiar behavior of $r_{\rm in}$ is caused by the inaccurate measurement of the disk emission, we considered two potential factors.

%The current model does not properly accounts for the disk photons that are Compton-scattered by the corona. This will probably underestimate the normalization of the \texttt{diskbb} component and thus result in a smaller inner radius when the source is in a relatively hard state \citep{2005ApJ...619..446Y}.  To test this possible issue, we redo the modelling with the \texttt{simpl} development model \texttt{simplcut} \citep{2017ApJ...836..119S}, a self-consistent Comptonization model applied to a disk-blackbody seed spectrum.
%To test this, we replace the \texttt{cutoffpl} component with the \texttt{simpl} development model \texttt{simplcut} in which a Comptonized accretion disk are considered \citep{2017ApJ...836..119S}. 
% Accordingly, we also add a \texttt{gaussian} model to fit the extra emission line structure in the HIMS. 
%Panels (e) and (f) of Figure~\ref{fig:fitting results} show two representative unfolded spectra based on the \texttt{simplcut*diskbb} model.

The fitting statistics of the two models is equally good: the reduced $\chi^{2}$ only varies by up to 2\%.
The best-fitting parameter values (Table~\ref{tab:fitting_1sc1d}) of the \texttt{simplcut*diskbb} model are shown as blue dots in Figure~\ref{fig: para_all}.
Compared to the \texttt{cutoffpl} model, the simplcut model has one more parameter: $f_{\rm sc}$, defined as the fraction of seed photons scattered into the power-law distribution. Its value decreases rapidly from $\approx$0.91 in the LHS to $\approx$0.13 in the HIMS, and then remains constant in the SIMS and HSS, at $\approx$0.05.
%Its evolution is consistent with previous results, e.g., GX~339--4 \citep{2020ApJ...890...53S}, XTE~J1550–564 \citep{2020ApJ...892...47C} and 4U~1630--47 \citep{2021ApJ...909..146C}.
As before, the high-energy cutoff is unconstrained for the observations in the SIMS and HSS. 
Most of the other parameters are consistent between the two models. In particular, we obtain the same best-fitting values of $r_{\rm in}$ and $T_{\rm in}$ for the SIMS and HSS observations. Instead, in the LHS and HIMS, the apparent radius derived from  \texttt{simplcut*diskbb} is higher than the value initially obtained from  \texttt{cutoffpl} + \texttt{diskbb}, and the color temperature is lower. We argue that the self-consistent Comptonization model provides more plausible values of those two quantities. Although the anomalous behavior of $r_{\rm in}$ and $T_{\rm in}$ is now reduced, it is not entirely removed. Even in the Comptonization model, $r_{\rm in}$ is still smaller in the LHS and HIMS than in the HSS, and the disk temperature is surprisingly higher ($\approx$1 keV) at those same epochs. This suggests that there is a physical reason for this strange behaviour. We propose that it is caused by changes in the hardening factor.
%\textbf{In the \texttt{simplcut*diskbb} model,  $r_{\rm in}$ is almost constant from an overall point of view except for a small fluctuation during the initial phase of the LHS. This means that the position of the accretion disk may not have changed and remain at ISCO throughout the entire outburst. If the above proposed scenario is realistic, the disk flux must be proportional to the fourth power of the effective temperature. Therefore the role of the hardening factor is to be considered.}

\subsection{Evolution of the hardening factor}\label{3.3}

\begin{figure}
	\vspace{-0.3cm}
	\includegraphics[width=1\linewidth]{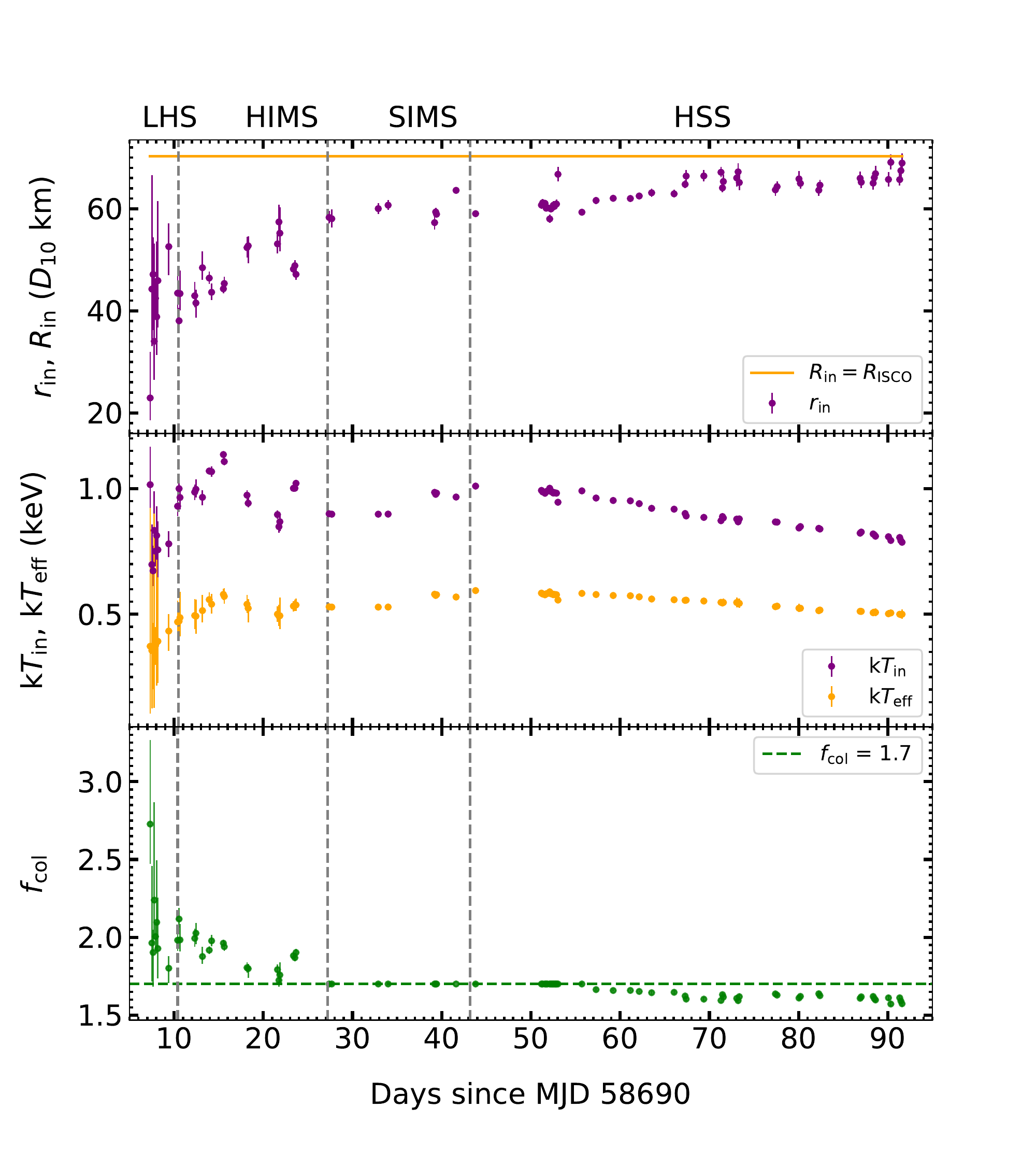}
	\vspace{-0.4cm}
    \caption{Evolution of $T_{\rm eff}$, $R_{\rm in}$ and $f_{\rm col}$. $f_{\rm col}$ is calculated from $R_{\rm in}\ =\ r_{\rm in} \, \xi \, f_{\rm col}^{2}$ under the assumption that $R_{\rm in}=R_{\rm ISCO}$.
    Then, $T_{\rm eff}$ is calculated from $T_{\rm eff}\ =\ T_{\rm in}\ / \ f_{\rm col} $. 
    %where $R_{\rm ISCO}$ is derived from $R_{\rm ISCO}$ = $r^{\prime}_{\rm in} * \xi * f_{\rm col}^{2} \approx 42$~km. 
    Purple dots represent $T_{\rm in}$ and $r_{\rm in}$ obtained from the \texttt{simplcut*diskbb} model, orange dots and the orange line indicate $T_{\rm eff}$ and $R_{\rm in} = R_{\rm ISCO}$, respectively. 
    If $R_{\rm in} > R_{\rm ISCO}$ in the LHS and HIMS, the values of $T_{\rm eff}$ plotted in the middle panel are the upper limits of the true effective temperature, and the values of $f_{\rm col}$ in the bottom panel are the lower limits of the hardening factor.}
%    \textbf{It is worth noting that in the LHS and HIMS, $f_{\rm col}$ and $T_{\rm eff}$ are plotted here as the lower and upper limits as we do not know for sure if $R_{\rm in}$ is at $R_{\rm ISCO}$.}}
    \label{fig:Teff_Rin}
\end{figure}

The hardening factor (also called color correction factor) $f_{\rm col}$ accounts for the deviation of the disk emission from a pure blackbody spectrum \citep{1995ApJ...445..780S}. The peak effective temperature $T_{\rm eff}$ is related to the fitted color temperature $T_{\rm in}$ by the relation $T_{\rm eff} = T_{\rm in}/f_{\rm col}$. Conversely, the relation between physical inner radius, $R_{\rm in}$, and apparent inner radius, $r_{\rm in}$, is:
\begin{equation}
%R_{\rm in}\! =\!  r_{\rm in}\! *\! \xi\! *\! f_{\rm col}^{2}\! =\! \sqrt{N_{\rm dbb}/\cos~\theta}\! *\! \xi\! *\! f_{\rm col}^{2}\! *\! D_{\rm 10}, 
R_{\rm in} = r_{\rm in} \, \xi \, f_{\rm col}^{2} = \sqrt{N_{\rm dbb}/\cos~\theta} \, \xi \, f_{\rm col}^{2} \, D_{\rm 10}, 
\end{equation}
\citep{1998PASJ...50..667K}, where $\xi=0.412$ is the general relativity correction factor, $N_{\rm dbb}$ is \texttt{diskbb} normalization, $\theta$ is the inclination angle of the accretion disk and $D_{\rm 10}$ is the distance in units of 10 kpc. 
%\textcolor{red}{The logical steps in the following par were wrong. I outline here what I think should be the right steps. Please consider also changing the plots in Fig 4 accordingly.}\\
We can plausible assume that $R_{\rm in} \approx R_{\rm ISCO}$ and $f_{\rm col} \approx 1.7$ (''canonical'' value from \citealt{1995ApJ...445..780S}) in the SIMS and at least near the peak of the HSS, before the start of the decline ({\it i.e.}, MJD 58717--58745). We adopt an inclination angle of $73^{\circ}$ \citep{2020ApJ...900...78D}. Instead, the distance is completely unconstrained, in the absence of a detected optical counterpart\footnote[5]{The value of 7 kpc introduced by \cite{1993A&A...279..179P} is only a guess based on a now out-of-date analogy with the luminosity of other BH transients.}. The fitted value of $r_{\rm in}$ in the first epoch of the HSS is $59.0\pm0.5$ $D_{\rm 10}$ km, and this value is indeed substantially unchanged throughout MJD 58717--58745. Then, from Eq.(1), $R_{\rm ISCO} = 70.3\pm0.6\,D_{\rm 10}$ km. In the LHS and HIMS, the physical inner radius $R_{\rm in}$ must be at least as large as $R_{\rm ISCO}$ (it may be equal if there is already a condensed thin disk in the innermost region). Thus, in our proposed scenario, we can determine the minimum value of $f_{\rm col}$ required to make $R_{\rm in} \geq R_{\rm ISCO}$ in the early epochs. We obtain $f_{\rm col} \gtrsim 2.7$ in the first few days of LHS observations. In the last part of the LHS and HIMS, $f_{\rm col}$ decreases to $\approx$1.7--2.4. Conversely, in the declining phase of the HSS, a canonical value $f_{\rm col} = 1.7$ implies $R_{\rm in} \geq R_{\rm ISCO}$. This can be interpreted in two alternative ways: either the inner disk was starting to recede from ISCO after MJD 58745, or it was still at ISCO but the hardening factor declined to $f_{\rm col} \approx 1.6$. Finally, we derived the effective temperature $T_{\rm eff} = T_{\rm in}/f_{\rm col}$. We find that with our estimated hardening factors, $T_{\rm eff}$ increased monotonically during the LHS (as expected) instead of showing an unphysical decline in the first part of the outburst. We show the evolution of radii, temperatures and $f_{\rm col}$ in Figure~\ref{fig:Teff_Rin}.

\section{Discussion}\label{4}

\subsection{The inner disk radius in the initial hard state}\label{4.1}
The early stages of transient BH outbursts are characterized by the coupled evolution of a corona and of the underlying disk. For EXO 1846$-$031, the evolution of the coronal parameters follows the canonical expectations. For example, the scattering fraction $f_{\rm sc}$ follows the same trend seen in recent outbursts of well-known BH transients such as GX~339--4 \citep{2020ApJ...890...53S}, XTE~J1550–564 \citep{2020ApJ...892...47C} and 4U~1630--47 \citep{2021ApJ...909..146C}. Instead, the evolution of the disk parameters may have subtle differences.

Our spectral modelling shows (Section~\ref{3.2}) that if the hardening factor $f_{\rm col}$ is kept constant from LHS to HSS, the inner radius of the disk is smaller in the LHS than in the HSS. We reject this possibility as unphysical: the inner disk radius in the LHS must be at least as large as in the HSS. If we assume that the inner disk radius is constant throughout the entire outburst, $f_{\rm col}$ must be decreasing from $\approx$2.7 at the beginning of the outburst, towards the canonical value of $\approx$1.6 in the HSS. With this assumption, we preserve the scaling law $F_{\rm dbb}~\propto T_{\rm eff}^4$ throughout the outburst, characteristic of a radiatively efficient disk with constant inner radius. Finally, if we also allow for the possibility that $R_{\rm in} > R_{\rm ISCO}$ in the LHS, the hardening factor must be even higher at the beginning of the outburst ($f_{\rm col} > 2.7$).

In previous studies of BH transients, a “canonical” value of $f_{\rm col} \approx 1.7$ (from \citealt{1995ApJ...445..780S}) was usually adopted for the soft state, to account for a “diluted” disk spectrum. However, there are several observations of BH outbursts that show convincing evidence of a variable hardening factor, especially in the initial hard state. For example in GX~339-4, \citet{2013MNRAS.431.3510S} proposed that a variable hardening factor is an alternative to disk truncation, to explain changes in the disk spectrum. Other sources with possible evidence of variable hardening factor are 4U~1957$+$11 \citep{2014ApJ...794...85M}, MAXI~J1820$+$070 \citep{2021MNRAS.tmp..941G}, and MAXI~J1348$-$630 \citep{2022arXiv220111919Z}.  From simulations of disk spectra, 
\citet{2000MNRAS.313..193M} also suggested a variable hardening factor $f_{\rm col} \approx 1.7$--3, arguing that $f_{\rm col}$ increases when the disk emission is relatively less dominant. This result is supported by a global study of disk properties by \citet{2011MNRAS.411..337D}. They found that for almost all BH transients, $f_{\rm col}$ is relatively stable in the disk-dominated state, but increase from 1.6 to 2.6 as the disk fraction decreases. 
The variation of $f_{\rm col}$ may be caused by changes in the accretion energy dissipation \citep{2000MNRAS.313..193M}, in the vertical disk structure \citep{2005ApJ...621..372D}, in the magnetic energy density \citep{2006ApJ...645.1402B} and in the disk density and optical depth \citep{2008xng..conf...48S}.  
%\textcolor[rgb]{1,0,0}{\textbf{Besides, \textcolor[rgb]{1,0,1}{Zhang et al. (submitted) found that MAXI~J1348--630 in the initial hard state also has a similar trend of disk parameter evolution to our case}, i.e., the inner radius is also smaller than the constant value in the HSS. \textcolor[rgb]{1,0,1}{They interpreted this anomalous behavior in the initial hard state as a result of the increase in the disk density (corresponding

%Furthermore, in the study of the disk evolutionary behavior of the typical BHT GX~339--4, \citet{2013MNRAS.431.3510S} proposed that a variable hardening factor could be an alternative to disk truncation for explaining the changes in the disk spectrum. 
% \textcolor[rgb]{1,0,0}{\textbf{This scenario that the accretion disk already extends to the ISCO in the beginning of the hard state would have important implications for the interpretation of the observational phenomena happening in this stage. For instance, the mechanism of jet launching and the relative fraction of jet power \citep{2004MNRAS.355.1105F, 2021MNRAS.504..444C}; the process in which the corona disappears and the jet gets quenched during the hard-to-soft transition \citep{1999ApJ...519L.165F, 2021ApJ...910L...3W}; and the origin of LFQPOs \citep{2009MNRAS.397L.101I, 2016AN....337..398M}.}} 
The presence or absence of a full disk has important implications also for the interpretation of QPOs.
LFQPOs with a centroid frequency between 0.7 and 8 Hz have been detected in the HIMS of EXO~1846--031 by \cite{2021RAA....21...70L} from a timing analysis of the same {\it Insight-HXMT} data used in this work. If there was already a thin disk at or very close to ISCO, as we have suggested here, this geometry disfavours QPO models ({\it e.g.}, \citealt{1998ApJ...492L..59S,  2009MNRAS.397L.101I,2016A&A...591A..36V}) that require significant changes in the inner disk edge. Instead, jet precession would be a viable model. The presence of a jet in EXO~1846$-$031 was confirmed by VLA  \citep{2019ATel12977....1M} and MeerKAT \citep{2019ATel12992....1W} detections in the HIMS. By comparison, the BH transient MAXI~J1820$+$070 is another source in which LFQPOs have been detected in the LHS (e.g., \citealt{2020ApJ...896...33W,2021NatAs...5...94M}) and the disk is thought to be non-truncated based on the reflection-fitting method \citep{2019MNRAS.490.1350B}. A likely interpretation for the origin of the QPOs is jet precession \citep{2021NatAs...5...94M}.

\subsection{Additional contributions to the hardening factor in the LHS and HIMS}\label{4.2}

\begin{figure}
	\vspace{-0.3cm}
	\includegraphics[width=1\linewidth]{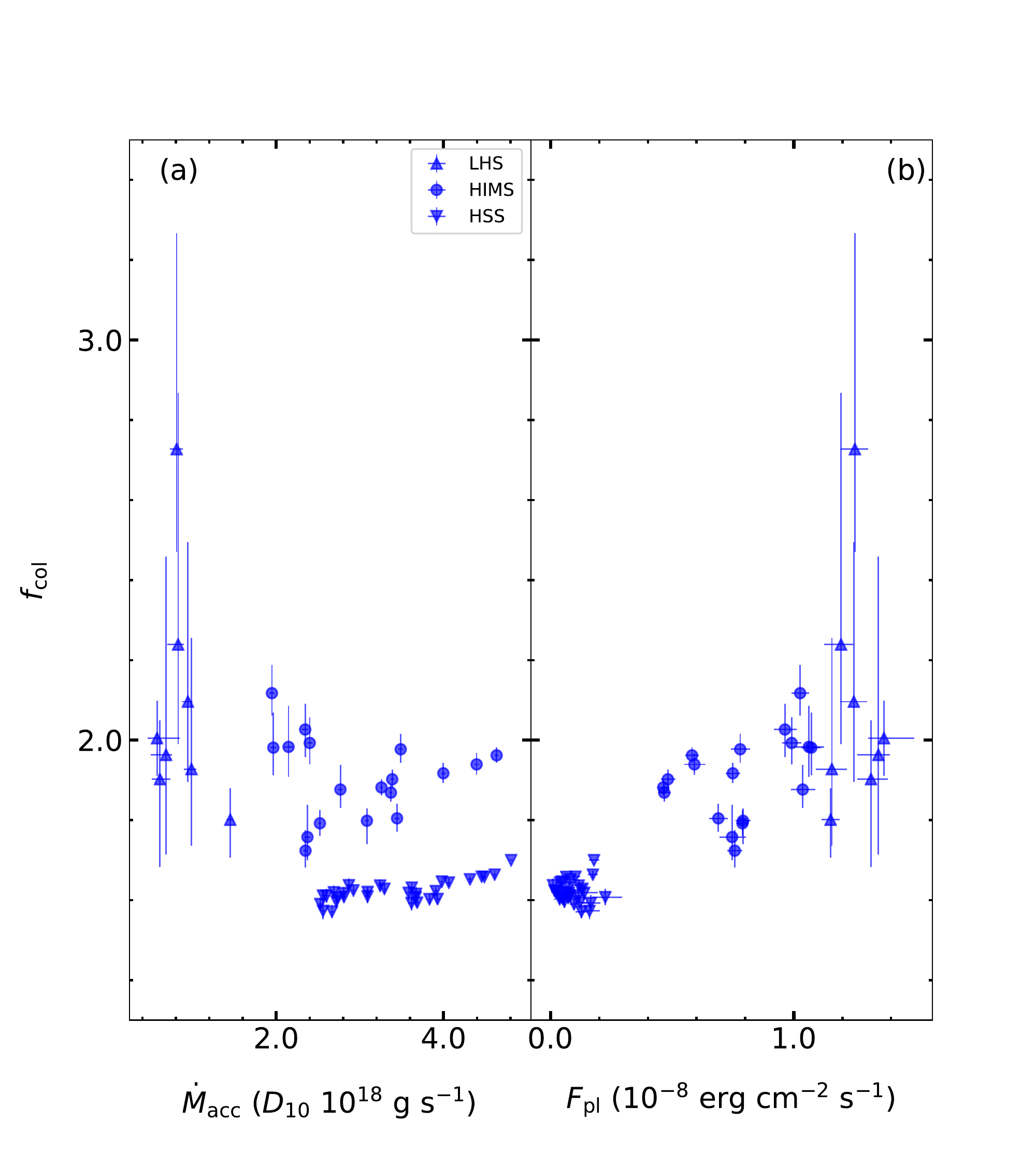}
	\vspace{-0.4cm}
    \caption{$f_{\rm col}$ as a function of mass accretion rate $\dot M_{\rm acc}$  (a) and power-law flux $F_{\rm pl}$ (b) respectively. The mass accretion rate is roughly estimated as $\dot{M}_{\rm acc}=L/(\eta c^2)$ where the efficiency $\eta=0.1$ is assumed. The blue upward triangles, circles and downward triangles represent the LHS, HIMS and HSS respectively.}
    \label{fig:fcol_M}
\end{figure}

\citet{2008ApJ...683..389D} reported a detailed investigation of $f_{\rm col}$ in a disk-dominated state. Their work shows that $f_{\rm col}$ increases as the mass accretion rate $\dot{M}_{\rm acc}$ increases, which links the variation of $f_{\rm col}$ to the mass accretion rate. To examine if this relationship is still valid in a hard state, we plot $f_{\rm col}$ as a function of $\dot{M}_{\rm acc}$ in the LHS, HIMS and most of the HSS in Figure~\ref{fig:fcol_M}a. Data points between MJD 58717 and 58745 are excluded here since we have assumed a constant $f_{\rm col}$ in this period (we explained this in Section~\ref{3.3}). It is clear that $f_{\rm col}$ and $\dot{M}_{\rm acc}$ indeed show a positive correlation in the HSS, which is in good agreement with the results of \citet{2008ApJ...683..389D}. However, the values of $f_{\rm col}$ in the LHS and HIMS where $\dot{M}_{\rm acc}$ are relatively low, is significantly higher than that in the HSS. We thus suspect that there should be additional contributions to the $f_{\rm col}$ in the LHS and HIMS.

We show the relationship between $f_{\rm col}$ and power-law flux $F_{\rm pl}$ in Figure~\ref{fig:fcol_M}b, in which $f_{\rm col}$ increases as $F_{\rm pl}$ increases throughout the outburst. Apparently, $F_{\rm pl}$ gradually decreases as the source evolves from the LHS to the HIMS and becomes negligible in the HSS. This implies that the high value of $f_{\rm col}$ in the LHS and the HIMS is probably associated with the hard emission. Compared to HSS, the additional power-law emission in the hard state will heat up the disk surface, leading to an increase in the disk temperature, which results in a significant increase in $f_{\rm col}$. Similar results that $f_{\rm col}$ increases as the disk fraction decreases have been reported by \citet{2000MNRAS.313..193M} and \citet{2011MNRAS.411..337D}. We plot $f_{\rm col}$ versus $F_{\rm pl}$ in Figure~\ref{fig:fcol_M}b, which is to emphasize the contribution of the power-law flux when the source was in the LHS and HIMS.

\subsection{Constraints on system parameters}\label{4.3}
Using the reflection-fitting method, \citet{2020ApJ...900...78D} obtained a spin parameter as high as $a_* = 0.997^{+0.001}_{-0.002}$. They applied 12 reflection models to the \textit{NuSTAR} spectra to test this value, with different assumptions on coronal geometry, disk density and emission spectrum, and obtained consistent spin values. However, \citet{2021ApJ...906...11W} studied the same \textit{NuSTAR} data with two reflection models and argued that the spin cannot be constrained. Here, we try to constrain the spin parameter with the continuum-fitting method, and test whether the result is consistent with the extreme value claimed by \citet{2020ApJ...900...78D}.
%whereas found the spin parameter cannot be well constrained. Here we used another technique, the continuum-fitting method, to estimate the spin parameter of EXO~1846--031 and to compare the results of these two classical techniques. 

% Since neither the spin nor the mass parameters of EXO~1846--031 have been accurately measured; EXO~1846--031 is at HSS for a long period, and the inner disk radius reaches $R_{\rm ISCO}$ at this point, this is well suited to the estimation of the spin parameter.
% Thus we make estimates of the spin parameter and the mass of the BH in EXO~1846--031. 
As discussed in Section~\ref{3.3}, the inner disk radius reached the ISCO in the HSS, with $R_{\rm ISCO} \approx 70\,D_{10}$~km (for $f_{\rm col} =1.7$ and $\theta = 73^{\circ}$).
%\begin{equation}
%R_{\rm ISCO}=\ R_{\rm in}\ =\ \sqrt{N_{\rm dbb}/\cos~\theta}\ *\ \xi\ *\ f_{\rm col}^{2}\ *\ D_{\rm 10}.
%\end{equation}
%where $N_{\rm dbb}$ is the \texttt{diskbb} normalization, $\theta$ is the inclination angle of the accretion disk, $\xi$ is the general relativity correction factor, $f_{\rm col}$ is the hardening factor and $D_{\rm 10}$ is the distance to the source in units of 10~kpc.
We apply the relationship between the ISCO radius and the spin parameter (\citealt{1997ApJ...482L.155Z}; see also \citealt{1972ApJ...178..347B}):
%, we could also write $R_{\rm ISCO}$ as a function of $a_{*}$:
\begin{equation}
\footnotesize
R_{\rm ISCO}=\left\{
\begin{array}{ll}
(3\! +\! A_{2}\! +\! \sqrt{(3\! -\! A_{1})(3\! +\! A_{1}\! +\! 2A_{2})}) \times r_{\rm g}, {a_{*} \leq 0}\\
(3\! +\! A_{2}\! -\! \sqrt{(3\! -\! A_{1})(3\! +\! A_{1}\! +\! 2A_{2})}) \times r_{\rm g}, {a_{*} > 0},\\
\end{array}
\right.
\end{equation}
%\begin{equation}
%R_{\rm ISCO}=r_{\rm g}\ \times \{3 + A_{2} \pm [(3 - A_{1})(3 + A_{1} + 2A_{2})]^{1/2}\},
%\end{equation}
where $r_{\rm g}=GM_{\rm BH}/c^2$, $A_{1}=1 + (1 - a_{*}^{2})^{1/3} [(1 + a_{*})^{1/3} + (1 - a_{*})^{1/3}]$ and $A_{2} = (3a_{*}^{2} + A_{1}^{2})^{1/2}$. 
Thus, from Equation 2, we can derive $a_{*}$ as a function of $M_{\rm BH}$ and distance.
%Combining Equations 1 and 2, we know that the value of $a_{*}$ is determined by $M_{\rm BH}$, $\theta$, $N_{\rm dbb}$, $f_{\rm col}$ and $D_{\rm 10}$. Here, we adopt $N_{\rm dbb}=1000$ from the fit to the first observation in the HSS where $f_{\rm col} =1.7$. 
Since neither the BH mass nor the distance is known, we determined $a_{*}$ over a grid of values spanning a mass range of 5--15~$M_{\odot}$ (based on the distribution of kinematic masses in Galactic BH transients; \citealt{Ozel2010, 2016A&A...587A..61C}) and a distance $D$ of 2--12~kpc (based on the distribution of Galactic BH distances in \citealt{2016ApJS..222...15T}). 
%We adopted the inclination angle of $73^{\circ}$ from \cite{2020ApJ...900...78D}. 

\begin{figure}
	\vspace{0.3cm}
	\includegraphics[width=1\linewidth]{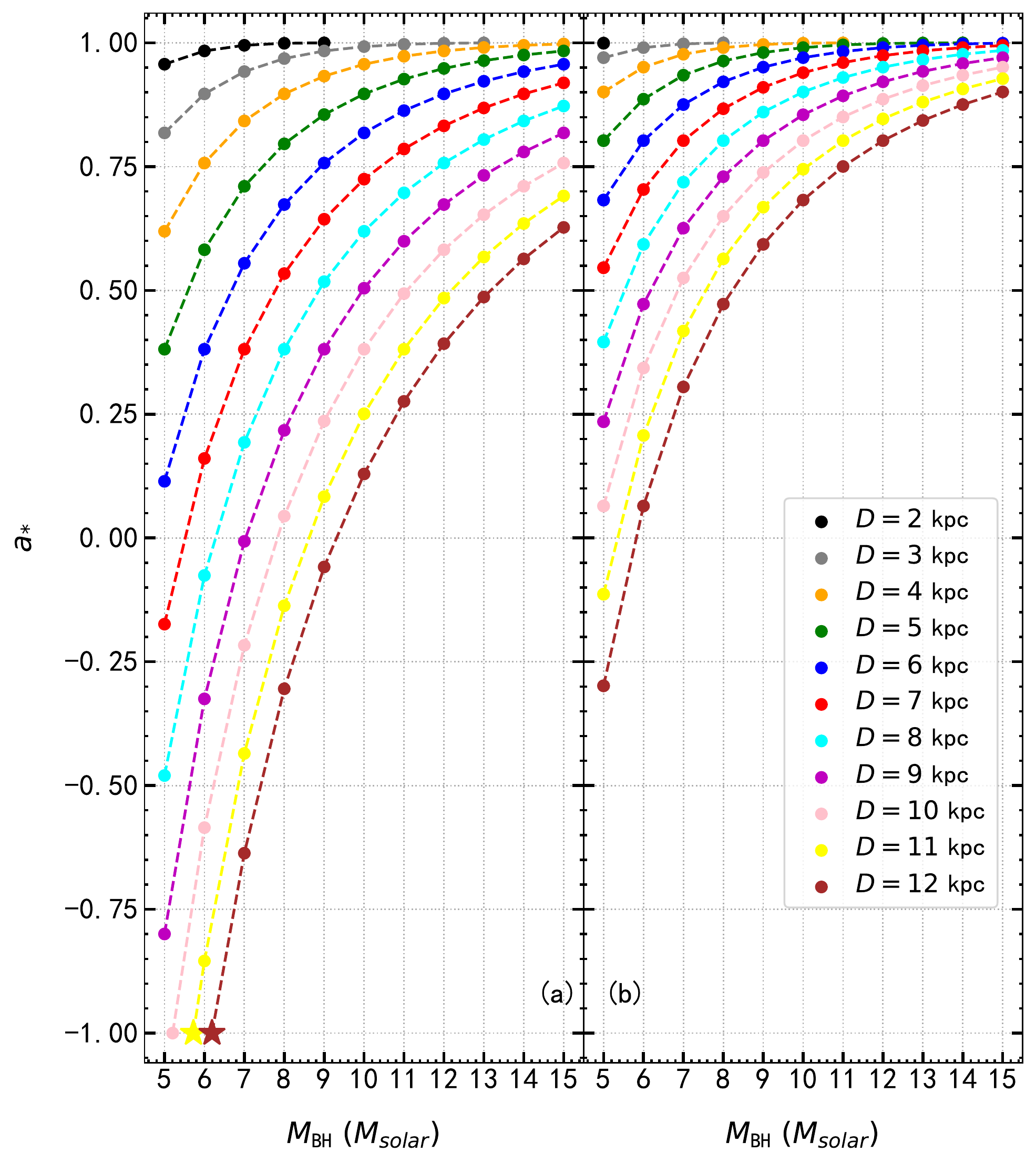}\vspace{0.1cm}
    \caption{Spin parameter as a function of BH mass and distance to EXO 1846$–$031 when the inclination is assumed to be $73^{\circ}$ (a) or $40^{\circ}$ (b). Stars mark the lowest acceptable limit of the BH mass for a given distance. 
    }
    \label{fig:amd}
\end{figure}

The values of the spin parameter versus BH mass for different distances are shown in Figure~\ref{fig:amd}a. The extreme value of $\emph{a}_* \approx 0.997$ suggested by \citet{2020ApJ...900...78D} is consistent only with a small distance ($D \approx 2$--4~kpc) and relatively high mass ($M_{\rm BH} > 7 M_{\odot}$). 
%\approx  7.3,\ 10.9,\ 14.6~M_{\odot}$.
%For an inclination of $40^{\circ}$, $D_{\rm 10}$ = 2~kpc and $M_{\rm bh} \geq 12~M_{\odot}$ are required while for the case of $73^{\circ}$, it even requests a range beyond our assumption, i.e., $D_{\rm 10} < 2$~kpc and $M_{\rm bh} \geq 15 M_{\odot}$.
%D = 2~kpc and $M_{\rm BH} \geq 15 M_{\odot}$ for a spin of $a_{*}\ \approx\ 0.98$. 
However, this combination of mass and distance would imply a peak Eddington ratio $\lambda \equiv L_{\rm disk}/L_{\rm Edd} \approx 0.02$--0.04
%1.2\%,\  1.8\%,\ 2.4\%$ 
in the HSS. This is much lower than the typical X-ray luminosities observed in the disk-dominated states of other Galactic BHs \citep{2016ApJS..222...15T}. 
We also calculate the spin parameter when the inclination is assumed to be $40^{\circ}$ \citep{2021ApJ...906...11W}, as shown in Figure~\ref{fig:amd}b. In this case,  the extreme value of $\emph{a}_* \approx 0.997$ is consistent with the combination of distance ($D \approx 2$--6~kpc) and mass ($M_{\rm BH} \approx 5$--14 $M_{\odot}$). The corresponding  peak Eddington ratio would then be $\approx$0.1 but is still lower than expected and observed in most other systems \citep{2011RAA....11..434T, 2015ApJ...805...87Y}.
For a higher Eddington ratio, the spin parameter has to be non-extremal.

%It is also much lower than expected from the best-fitting value of the colour temperature $kT_{\rm in} \approx 1.0$ keV at the peak of the outburst (Figure~\ref{fig:Teff_Rin} and Table~\ref{tab:fitting_1sc1d}), a typical value of stellar-mass BHs at an Eddington ratio $\gtrsim$0.3. More precisely, if we assume a standard Shakura-Sunyaev disk, simple scaling relations ({\it e.g.}, summarized in \citealt{2007Ap&SS.311..213S}) suggest that at $kT_{\rm in} \approx 1.0$ keV, the bolometric disk luminosity $L_{\rm {disk}} \approx 5 \times 10^{38} \, (M_{\rm BH}/10M_{\odot})^2 \approx 0.4 \, (M_{\rm BH}/10M_{\odot})\, L_{\rm Edd}$. Since the outburst duration, the sequence of state transitions, and the peak disk temperature of EXO~1846$-$031 appear perfectly consistent with most other BH transients, there is no reason to believe that its peak luminosity was only at 0.04 Eddington. For more plausible Eddington ratios, the spin parameter has to be non-extremal.

Previous work has already reported the inconsistency of the spin measurements obtained for the same systems with the reflection-fitting and continuum-fitting methods. %\citealt{2006ApJ...636L.113S,2009MNRAS.395.1257R,2014SSRv..183..277R}.
%4U 1543--47 \citep{2006ApJ...636L.113S,2014SSRv..183..277R}, GRO J1665--40 \citep{2006ApJ...636L.113S,2009MNRAS.395.1257R,2014SSRv..183..277R} and XTE~J1550--564 \citep{2011MNRAS.416..941S}. 
For example, for GRO J1655$-$40, the reflection-fitting method provides a higher spin than the continuum-fitting method \citep{2006ApJ...636L.113S,2009MNRAS.395.1257R,2014SSRv..183..277R}, similar to what we found here for EXO~1846$-$031.
Allowing for uncertainties in $f_{\rm col}$ at a level of $\pm$0.2--0.3 can alleviate the discrepancy \citep{2021MNRAS.500.3640S}; this effect is more pronounced for non-extremal spins. 
For example, if we assume a smaller value of the hardening factor, $f_{\rm col}=1.4$, we can reach $a_{*}=0.997$ for $D \approx 5$--6~kpc and $M_{\rm BH} \gtrsim 12 M_{\odot}$ at $\theta = 73^{\circ}$, or $D \approx 4$--9 kpc and $M_{\rm BH} \approx 6$--14 $M_{\odot}$ at $\theta = 40^{\circ}$.
%For the two cases of the inclination is assumed to be $73^{\circ}$ and $40^{\circ}$, if we assume a smaller value of the hardening factor, $f_{\rm col}=1.4$, we can reach $a_{*}=0.997$ for $D \approx 5$--6~kpc and $M_{\rm BH} \gtrsim 12 M_{\odot}$ or D $\approx 4$--9 kpc and $M_{\rm BH} \approx 6$--14 $M_{\odot}$. 
The corresponding peak Eddington ratio would then be $\approx$0.09 and $\approx$0.22, respectively.

%\textbf{\textcolor[rgb]{1,0,0}{It is clear that different values of inclination will result in different spin parameters according to Equations 1 and 2. Therefore, we also calculate the spin parameter when the inclination is assumed to be $40^{\circ}$ \citep{2021ApJ...906...11W}, as shown in Figure~\ref{fig:amd}b. In this case,  the extreme value of $\emph{a}_* \approx 0.997$ is consistent with the combination of distance ($D \approx 2$--6~kpc) and mass ($M_{\rm BH} \approx 5$--14 $M_{\odot}$). The corresponding  peak Eddington ratio would then be $\approx$0.1. Similarly, if we assume $f_{\rm col} = 1.4$, we can reach $\emph{a}_* \approx 0.997$ for D $\approx 4$--9 kpc and $M_{\rm BH} \approx 6$--14 $M_{\odot}$. Then the Eddington ratio increases to 0.22.}}

In general, for the same continuum parameters, lower values of $f_{\rm col}$ imply higher Eddington ratios. We conclude that, in principle, the inconsistency of the spin parameter measured with the continuum-fitting and reflection-fitting methods can be somewhat mitigated by the choice of $f_{\rm col}$. However, without knowing its mass and distance, we cannot draw a further conclusion about the spin parameter of the BH in EXO~1846--031.

\section{Conclusions}\label{5}

We have presented a broad-band (1--150~keV band) spectral analysis of the 2019 outburst of the BH candidate EXO~1846$-$031, based on {\it Insight-HXMT} observations. The source exhibited state transitions from the LHS to the HIMS, SIMS and HSS. The broad-band spectra can be well modelled with  \texttt{diskbb} plus \texttt{cutoffpl} components. Most of the best-fitting parameter values are consistent with those inferred in other ``canonical'' BH transients. In particular, the almost constant \texttt{diskbb} normalization in the SIMS and HSS indicates that the accretion disk reached ISCO in the two states, and suggests a value of $R_{\rm ISCO} \approx 70\,D_{10}$ km.
However, the apparent inner radius in the LHS and HIMS is unphysically small (in some epochs, smaller than that in the HSS), even after we take into account the fraction of disk photons upscattered into the Comptonization component. In order to ensure that the true radius is at least never smaller than ISCO in the hard state, we need to allow for a variable hardening factor, decreasing as the outburst progresses and the disk becomes brighter and more dominant. If we normalize the hardening factor to its canonical value $f_{\rm col} =  1.7$ at the SIMS and the beginning of the HSS, we find that $f_{\rm col} \gtrsim 2.7$ at the beginning of the LHS. Conversely, the last phase of the HSS is consistent with a small decrease of $f_{\rm col}$ below the canonical value. 

Furthermore, we find the value of the hardening factor in the relatively hard states is significantly higher than that in the HSS. We suggest that the coronal irradiation onto the disk provides additional contributions to the hardening factor in the LHS and HIMS.
%We propose that this could be caused by the additional illuminating of the hard emission onto the disk surface in the LHS and HIMS.
%and a canonical hardening factor $f_{\rm col}$ =  1.7, 
%assuming the inner disk radius in the LHS and HIMS also located at the ISCO, 
%we obtained a variable hardening factor decreases from $\approx$2.8 to $\approx$1.6 as the disk component becomes more dominant. 
%The $F_{\rm dbb} \propto T^{4}_{\rm eff}$ relation holds for all the spectral states justifies our assumption. 

Using our continuum-fitting parameters, we also tested previous claims of an extreme spin for this BH, inferred from reflection-fitting. We derived the range of acceptable spin values as a function of a plausible range of BH masses and distances based on two inclination angles. The inconsistency of the spin measured with the continuum-fitting and reflection-fitting methods is partly due to the different choices of $f_{\rm col}$.
%We concluded that, even with the current large uncertainty in those two parameters and in $f_{\rm col}$, an extreme spin is highly unlikely and would make EXO~1846$-$031 a complete outlier in terms of peak outburst luminosity in the HSS. 
%\textcolor{red}{CHECK IF YOU AGREE WITH THIS LAST STATEMENT.}

%Therefore, the accretion disk of EXO~1846--031 is probably not truncated during the outburst. Additionally, using the continuum-fitting method, we obtained the spin parameter of the system as a function of mass and distance. The inconsistency of the spin measured with continuum-fitting and reflection-fitting methods is influenced by the choice of the hardening factor.

In conclusion, our study has shown that $f_{\rm col}$ plays an important role for the intrinsic emission and physical parameters of the accretion disk, and for our understanding of the outburst evolution. It also influences the measurement of the spin parameter in the continuum-fitting method. Therefore, we argue that $f_{\rm col}$ should never simply be assumed as a constant value, especially when the disk emission evolves over a broad range of luminosities \citep{2013MNRAS.431.3510S}. More specifically in the case of EXO~1846$-$031, there is empirical evidence that the accretion disk was close to ISCO already in the initial hard state (unless $f_{\rm col} \gtrsim 2.7$). A similar non-canonical interpretation was proposed for several other transient BH candidates \citep{2008MNRAS.387.1489R, 2010MNRAS.402..836R, 2015ApJ...808....9P, 2018ApJ...864...25G, 2019ApJ...885...48G, 2019MNRAS.490.1350B, 2019Natur.565..198K}.
%\textbf{This seems to be more in favor of the non-canonical argument that the disk in accreting BH binaries is not truncated in the bright hard state \citep{2008MNRAS.387.1489R, 2010MNRAS.402..836R, 2015ApJ...808....9P, 2018ApJ...864...25G, 2019ApJ...885...48G, 2019MNRAS.490.1350B, 2019Natur.565..198K}. The question of whether the disk in the accretion BH binaries is truncated in the bright hard state is hotly debated and will be left for future studies.}
%Thus, EXO~1846$–$031 joins a growing list of BH transients likely to have a full disk already at the upper end of the LHS, before the suppression of the jet and the removal of the corona. This temporary coexistence of disk and jet/corona poses intriguing constraints to theoretical models of coronal structure, jet launching mechanism, disk condensation/evaporation, origin of LFQPOs, etc. 
Simultaneous broad-band investigations of the early outburst evolution in BH X-ray binaries, such as those made possible by {\it Insight-HXMT}, hold the key for the physical understanding of state transitions.

\acknowledgments
This work made use of data from the {\it Insight-HXMT} mission, a project funded by China National Space Administration (CNSA) and the Chinese Academy of Sciences (CAS). This work is supported by the National Key R\&D Program of China (2021YFA0718500) and the National Natural Science Foundation of China under grants U1838201, 11473027, U1838202, 11733009, U1838104, U1938101, U1838115, U2038101, U1938103 and NSF grant 12073029. Y.~W. acknowledges support from the Royal Society Newton Funds. J.~L. thanks Guangdong Major Project of Basic and Applied Basic Research (Grant No. 2019B030302001). R.~S. thanks NSF grant 12073029.\\


\vspace{3mm}
\facilities{{\it Insight-HXMT}}




\clearpage
\bibliography{sample63}{}

\begin{thebibliography}{}
\expandafter\ifx\csname natexlab\endcsname\relax\def\natexlab#1{#1}\fi
\providecommand{\url}[1]{\href{#1}{#1}}
\providecommand{\dodoi}[1]{doi:~\href{http://doi.org/#1}{\nolinkurl{#1}}}
\providecommand{\doeprint}[1]{\href{http://ascl.net/#1}{\nolinkurl{http://ascl.net/#1}}}
\providecommand{\doarXiv}[1]{\href{https://arxiv.org/abs/#1}{\nolinkurl{https://arxiv.org/abs/#1}}}

\bibitem[{{Aneesha} {et~al.}(2019){Aneesha}, {Mandal}, \&
  {Sreehari}}]{2019MNRAS.486.2705A}
{Aneesha}, U., {Mandal}, S., \& {Sreehari}, H. 2019, \mnras, 486, 2705,
  \dodoi{10.1093/mnras/stz1000}

\bibitem[{{Arnaud}(1996)}]{1996ASPC..101...17A}
{Arnaud}, K.~A. 1996, in Astronomical Society of the Pacific Conference Series,
  Vol. 101, Astronomical Data Analysis Software and Systems V, ed. G.~H.
  {Jacoby} \& J.~{Barnes}, 17

\bibitem[{{Bardeen} {et~al.}(1972){Bardeen}, {Press}, \&
  {Teukolsky}}]{1972ApJ...178..347B}
{Bardeen}, J.~M., {Press}, W.~H., \& {Teukolsky}, S.~A. 1972, \apj, 178, 347,
  \dodoi{10.1086/151796}

\bibitem[{{Barri{\`e}re} {et~al.}(2015){Barri{\`e}re}, {Krivonos}, {Tomsick},
  {Bachetti}, {Boggs}, {Chakrabarty}, {Christensen}, {Craig}, {Hailey},
  {Harrison}, {Hong}, {Mori}, {Stern}, \& {Zhang}}]{2015ApJ...799..123B}
{Barri{\`e}re}, N.~M., {Krivonos}, R., {Tomsick}, J.~A., {et~al.} 2015, \apj,
  799, 123, \dodoi{10.1088/0004-637X/799/2/123}

\bibitem[{{Belloni} {et~al.}(2005){Belloni}, {Homan}, {Casella}, {van der
  Klis}, {Nespoli}, {Lewin}, {Miller}, \& {M{\'e}ndez}}]{2005A&A...440..207B}
{Belloni}, T., {Homan}, J., {Casella}, P., {et~al.} 2005, \aap, 440, 207,
  \dodoi{10.1051/0004-6361:20042457}

\bibitem[{{Belloni}(2010)}]{2010LNP...794...53B}
{Belloni}, T.~M. 2010, {States and Transitions in Black Hole Binaries}, ed.
  T.~{Belloni}, Vol. 794, 53, \dodoi{10.1007/978-3-540-76937-8\_3}

\bibitem[{{Belloni} \& {Motta}(2016)}]{2016ASSL..440...61B}
{Belloni}, T.~M., \& {Motta}, S.~E. 2016, {Transient Black Hole Binaries}, ed.
  C.~{Bambi}, Vol. 440, 61, \dodoi{10.1007/978-3-662-52859-4\_2}

\bibitem[{{Belloni} \& {Stella}(2014)}]{2014SSRv..183...43B}
{Belloni}, T.~M., \& {Stella}, L. 2014, \ssr, 183, 43,
  \dodoi{10.1007/s11214-014-0076-0}

\bibitem[{{Bhalerao} {et~al.}(2015){Bhalerao}, {Romano}, {Tomsick},
  {Natalucci}, {Smith}, {Bellm}, {Boggs}, {Chakrabarty}, {Christensen},
  {Craig}, {Fuerst}, {Hailey}, {Harrison}, {Krivonos}, {Lu}, {Madsen}, {Stern},
  {Younes}, \& {Zhang}}]{2015MNRAS.447.2274B}
{Bhalerao}, V., {Romano}, P., {Tomsick}, J., {et~al.} 2015, \mnras, 447, 2274,
  \dodoi{10.1093/mnras/stu2495}

\bibitem[{{Blackburn}(1995)}]{1995ASPC...77..367B}
{Blackburn}, J.~K. 1995, in Astronomical Society of the Pacific Conference
  Series, Vol.~77, Astronomical Data Analysis Software and Systems IV, ed.
  R.~A. {Shaw}, H.~E. {Payne}, \& J.~J.~E. {Hayes}, 367

\bibitem[{{Blaes} {et~al.}(2006){Blaes}, {Davis}, {Hirose}, {Krolik}, \&
  {Stone}}]{2006ApJ...645.1402B}
{Blaes}, O.~M., {Davis}, S.~W., {Hirose}, S., {Krolik}, J.~H., \& {Stone},
  J.~M. 2006, \apj, 645, 1402, \dodoi{10.1086/503741}

\bibitem[{{Buisson} {et~al.}(2019){Buisson}, {Fabian}, {Barret}, {F{\"u}rst},
  {Gandhi}, {Garc{\'\i}a}, {Kara}, {Madsen}, {Miller}, {Parker}, {Shaw},
  {Tomsick}, \& {Walton}}]{2019MNRAS.490.1350B}
{Buisson}, D.~J.~K., {Fabian}, A.~C., {Barret}, D., {et~al.} 2019, \mnras, 490,
  1350, \dodoi{10.1093/mnras/stz2681}

\bibitem[{{Bult} {et~al.}(2019){Bult}, {Gendreau}, {Arzoumanian}, {Strohmayer},
  {Ray}, {Guillot}, {Iwakiri}, {Homan}, {Altamirano}, {Jaisawal}, \&
  {Miller}}]{2019ATel12976....1B}
{Bult}, P.~M., {Gendreau}, K.~C., {Arzoumanian}, Z., {et~al.} 2019, The
  Astronomer's Telegram, 12976, 1

\bibitem[{{Cao} {et~al.}(2020){Cao}, {Jiang}, {Meng}, {Zhang}, {Luo}, {Yang},
  {Zhang}, {Gu}, {Sun}, {Liu}, {Yang}, {Li}, {Tan}, {Liu}, {Du}, {Lu}, {Xu},
  {Guan}, {Zhang}, {Wang}, {Li}, {Zhang}, {Wen}, {Qu}, {Song}, {Li}, {Ge},
  {Zhou}, {Xiong}, {Zhang}, {Zhang}, {Cheng}, {Zhang}, {Li}, {Liang}, {Gao},
  {Yang}, {Liu}, {Liu}, {Yang}, \& {Zhang}}]{2020SCPMA..63x9504C}
{Cao}, X., {Jiang}, W., {Meng}, B., {et~al.} 2020, Science China Physics,
  Mechanics, and Astronomy, 63, 249504, \dodoi{10.1007/s11433-019-1506-1}

\bibitem[{{Chen} {et~al.}(1997){Chen}, {Shrader}, \&
  {Livio}}]{1997ApJ...491..312C}
{Chen}, W., {Shrader}, C.~R., \& {Livio}, M. 1997, \apj, 491, 312,
  \dodoi{10.1086/304921}

\bibitem[{{Chen} {et~al.}(2020){Chen}, {Cui}, {Li}, {Wang}, {Xu}, {Lu}, {Wang},
  {Chen}, {Han}, {Hu}, {Zhang}, {Huo}, {Yang}, {Li}, {Lu}, {Zhang}, {Li},
  {Zhang}, {Xiong}, {Zhang}, {Xue}, {Zhao}, {Zhu}, {Zhu}, {Liu}, {Yang}, \&
  {Zhang}}]{2020SCPMA..63x9505C}
{Chen}, Y., {Cui}, W., {Li}, W., {et~al.} 2020, Science China Physics,
  Mechanics, and Astronomy, 63, 249505, \dodoi{10.1007/s11433-019-1469-5}

\bibitem[{{Connors} {et~al.}(2020){Connors}, {Garc{\'\i}a}, {Dauser},
  {Grinberg}, {Steiner}, {Sridhar}, {Wilms}, {Tomsick}, {Harrison}, \&
  {Licklederer}}]{2020ApJ...892...47C}
{Connors}, R. M.~T., {Garc{\'\i}a}, J.~A., {Dauser}, T., {et~al.} 2020, \apj,
  892, 47, \dodoi{10.3847/1538-4357/ab7afc}

\bibitem[{{Connors} {et~al.}(2021){Connors}, {Garc{\'\i}a}, {Tomsick}, {Hare},
  {Dauser}, {Grinberg}, {Steiner}, {Mastroserio}, {Sridhar}, {Fabian}, {Jiang},
  {Parker}, {Harrison}, \& {Kallman}}]{2021ApJ...909..146C}
{Connors}, R. M.~T., {Garc{\'\i}a}, J.~A., {Tomsick}, J., {et~al.} 2021, \apj,
  909, 146, \dodoi{10.3847/1538-4357/abdd2c}

\bibitem[{{Corral-Santana} {et~al.}(2016){Corral-Santana}, {Casares},
  {Mu{\~n}oz-Darias}, {Bauer}, {Mart{\'\i}nez-Pais}, \&
  {Russell}}]{2016A&A...587A..61C}
{Corral-Santana}, J.~M., {Casares}, J., {Mu{\~n}oz-Darias}, T., {et~al.} 2016,
  \aap, 587, A61, \dodoi{10.1051/0004-6361/201527130}

\bibitem[{{Dauser} {et~al.}(2013){Dauser}, {Garcia}, {Wilms}, {B{\"o}ck},
  {Brenneman}, {Falanga}, {Fukumura}, \& {Reynolds}}]{2013MNRAS.430.1694D}
{Dauser}, T., {Garcia}, J., {Wilms}, J., {et~al.} 2013, \mnras, 430, 1694,
  \dodoi{10.1093/mnras/sts710}

\bibitem[{{Davis} {et~al.}(2005){Davis}, {Blaes}, {Hubeny}, \&
  {Turner}}]{2005ApJ...621..372D}
{Davis}, S.~W., {Blaes}, O.~M., {Hubeny}, I., \& {Turner}, N.~J. 2005, \apj,
  621, 372, \dodoi{10.1086/427278}

\bibitem[{{Davis} \& {El-Abd}(2019)}]{2019ApJ...874...23D}
{Davis}, S.~W., \& {El-Abd}, S. 2019, \apj, 874, 23,
  \dodoi{10.3847/1538-4357/ab05c5}

\bibitem[{{Done} \& {Davis}(2008)}]{2008ApJ...683..389D}
{Done}, C., \& {Davis}, S.~W. 2008, \apj, 683, 389, \dodoi{10.1086/589647}

\bibitem[{{Done} {et~al.}(2007){Done}, {Gierli{\'n}ski}, \&
  {Kubota}}]{2007A&ARv..15....1D}
{Done}, C., {Gierli{\'n}ski}, M., \& {Kubota}, A. 2007, \aapr, 15, 1,
  \dodoi{10.1007/s00159-007-0006-1}

\bibitem[{{Draghis} {et~al.}(2020){Draghis}, {Miller}, {Cackett}, {Kammoun},
  {Reynolds}, {Tomsick}, \& {Zoghbi}}]{2020ApJ...900...78D}
{Draghis}, P.~A., {Miller}, J.~M., {Cackett}, E.~M., {et~al.} 2020, \apj, 900,
  78, \dodoi{10.3847/1538-4357/aba2ec}

\bibitem[{{Dunn} {et~al.}(2011){Dunn}, {Fender}, {K{\"o}rding}, {Belloni}, \&
  {Merloni}}]{2011MNRAS.411..337D}
{Dunn}, R.~J.~H., {Fender}, R.~P., {K{\"o}rding}, E.~G., {Belloni}, T., \&
  {Merloni}, A. 2011, \mnras, 411, 337,
  \dodoi{10.1111/j.1365-2966.2010.17687.x}

\bibitem[{{Eardley} {et~al.}(1975){Eardley}, {Lightman}, \&
  {Shapiro}}]{1975ApJ...199L.153E}
{Eardley}, D.~M., {Lightman}, A.~P., \& {Shapiro}, S.~L. 1975, \apjl, 199,
  L153, \dodoi{10.1086/181871}

\bibitem[{{Esin} {et~al.}(1997){Esin}, {McClintock}, \&
  {Narayan}}]{1997ApJ...489..865E}
{Esin}, A.~A., {McClintock}, J.~E., \& {Narayan}, R. 1997, \apj, 489, 865,
  \dodoi{10.1086/304829}

\bibitem[{{Fabian} {et~al.}(1989){Fabian}, {Rees}, {Stella}, \&
  {White}}]{1989MNRAS.238..729F}
{Fabian}, A.~C., {Rees}, M.~J., {Stella}, L., \& {White}, N.~E. 1989, \mnras,
  238, 729, \dodoi{10.1093/mnras/238.3.729}

\bibitem[{{Fender} {et~al.}(2004){Fender}, {Belloni}, \&
  {Gallo}}]{2004MNRAS.355.1105F}
{Fender}, R.~P., {Belloni}, T.~M., \& {Gallo}, E. 2004, \mnras, 355, 1105,
  \dodoi{10.1111/j.1365-2966.2004.08384.x}

\bibitem[{{Garc{\'\i}a} {et~al.}(2014){Garc{\'\i}a}, {Dauser}, {Lohfink},
  {Kallman}, {Steiner}, {McClintock}, {Brenneman}, {Wilms}, {Eikmann},
  {Reynolds}, \& {Tombesi}}]{2014ApJ...782...76G}
{Garc{\'\i}a}, J., {Dauser}, T., {Lohfink}, A., {et~al.} 2014, \apj, 782, 76,
  \dodoi{10.1088/0004-637X/782/2/76}

\bibitem[{{Garc{\'\i}a} {et~al.}(2018){Garc{\'\i}a}, {Steiner}, {Grinberg},
  {Dauser}, {Connors}, {McClintock}, {Remillard}, {Wilms}, {Harrison}, \&
  {Tomsick}}]{2018ApJ...864...25G}
{Garc{\'\i}a}, J.~A., {Steiner}, J.~F., {Grinberg}, V., {et~al.} 2018, \apj,
  864, 25, \dodoi{10.3847/1538-4357/aad231}

\bibitem[{{Garc{\'\i}a} {et~al.}(2019){Garc{\'\i}a}, {Tomsick}, {Sridhar},
  {Grinberg}, {Connors}, {Wang}, {Steiner}, {Dauser}, {Walton}, {Xu},
  {Harrison}, {Foster}, {Grefenstette}, {Madsen}, \&
  {Fabian}}]{2019ApJ...885...48G}
{Garc{\'\i}a}, J.~A., {Tomsick}, J.~A., {Sridhar}, N., {et~al.} 2019, \apj,
  885, 48, \dodoi{10.3847/1538-4357/ab384f}

\bibitem[{{Ge} {et~al.}(2020){Ge}, {Ji}, {Zhang}, {Santangelo}, {Liu},
  {Doroshenko}, {Staubert}, {Qu}, {Zhang}, {Lu}, {Song}, {Li}, {Tao}, {Xu},
  {Cao}, {Chen}, {Bu}, {Cai}, {Chang}, {Chen}, {Chen}, {Chen}, {Chen}, {Chen},
  {Cui}, {Cui}, {Deng}, {Dong}, {Du}, {Fu}, {Gao}, {Gao}, {Gao}, {Gu}, {Guan},
  {Guo}, {Han}, {Huang}, {Huo}, {Jia}, {Jiang}, {Jiang}, {Jin}, {Jin}, {Kong},
  {Li}, {Li}, {Li}, {Li}, {Li}, {Li}, {Li}, {Li}, {Li}, {Li}, {Liang}, {Liao},
  {Liu}, {Liu}, {Liu}, {Liu}, {Liu}, {Lu}, {Lu}, {Luo}, {Luo}, {Ma}, {Meng},
  {Nang}, {Nie}, {Ou}, {Sai}, {Shang}, {Song}, {Sun}, {Tan}, {Tuo}, {Wang},
  {Wang}, {Wang}, {Wang}, {Wang}, {Wang}, {Wang}, {Wen}, {Wu}, {Wu}, {Wu},
  {Xiao}, {Xiao}, {Xiong}, {Xu}, {Yang}, {Yang}, {Yang}, {Yang}, {Yi}, {Yin},
  {You}, {Zhang}, {Zhang}, {Zhang}, {Zhang}, {Zhang}, {Zhang}, {Zhang},
  {Zhang}, {Zhang}, {Zhang}, {Zhang}, {Zhang}, {Zhang}, {Zhang}, {Zhang},
  {Zhang}, {Zhao}, {Zhao}, {Zheng}, {Zheng}, {Zhou}, {Zhou}, {Zhuang}, {Zhu},
  \& {Zhu}}]{2020ApJ...899L..19G}
{Ge}, M.~Y., {Ji}, L., {Zhang}, S.~N., {et~al.} 2020, \apjl, 899, L19,
  \dodoi{10.3847/2041-8213/abac05}

\bibitem[{{Guan} {et~al.}(2021){Guan}, {Tao}, {Qu}, {Zhang}, {Zhang}, {Zhang},
  {Ma}, {Ge}, {Song}, {Lu}, {Li}, {Xu}, {Chen}, {Cao}, {Liu}, {Zhang}, {Wang},
  {Chen}, {Bu}, {Cai}, {Chang}, {Chen}, {Chen}, {Chen}, {Cui}, {Du}, {Gao},
  {Gao}, {Gu}, {Guo}, {Han}, {Huang}, {Huo}, {Jia}, {Jiang}, {Jin}, {Kong},
  {Li}, {Li}, {Li}, {Li}, {Li}, {Li}, {Li}, {Li}, {Liang}, {Liao}, {Liu},
  {Liu}, {Liu}, {Liu}, {Lu}, {Luo}, {Luo}, {Ma}, {Meng}, {Nang}, {Nie}, {Ou},
  {Ren}, {Sai}, {Song}, {Sun}, {Tan}, {Wang}, {Wang}, {Wang}, {Wang}, {Wang},
  {Wen}, {Wu}, {Wu}, {Wu}, {Xiao}, {Xiao}, {Xiong}, {Yang}, {Yang}, {Yang},
  {Yang}, {Yi}, {Yin}, {You}, {Zhang}, {Zhang}, {Zhang}, {Zhang}, {Zhang},
  {Zhang}, {Zhang}, {Zhao}, {Zhao}, {Zheng}, {Zheng}, \&
  {Zhou}}]{2021MNRAS.tmp..941G}
{Guan}, J., {Tao}, L., {Qu}, J.~L., {et~al.} 2021, \mnras,
  \dodoi{10.1093/mnras/stab945}

\bibitem[{{Guo} {et~al.}(2020){Guo}, {Liao}, {Zhang}, {Zhang}, {Tan}, {Song},
  {Lu}, {Cao}, {Chang}, {Chen}, {Du}, {Ge}, {Gu}, {Jiang}, {Jin}, {Li}, {Li},
  {Li}, {Liu}, {Liu}, {Lu}, {Luo}, {Meng}, {Sun}, {Yang}, {Yang}, {You},
  {Zhang}, {Zhao}, \& {Zhang}}]{2020JHEAp..27...44G}
{Guo}, C.-C., {Liao}, J.-Y., {Zhang}, S., {et~al.} 2020, Journal of High Energy
  Astrophysics, 27, 44, \dodoi{10.1016/j.jheap.2020.02.008}

\bibitem[{{Homan} \& {Belloni}(2005)}]{2005Ap&SS.300..107H}
{Homan}, J., \& {Belloni}, T. 2005, \apss, 300, 107,
  \dodoi{10.1007/s10509-005-1197-4}

\bibitem[{{Homan} {et~al.}(2001){Homan}, {Wijnands}, {van der Klis}, {Belloni},
  {van Paradijs}, {Klein-Wolt}, {Fender}, \&
  {M{\'e}ndez}}]{2001ApJS..132..377H}
{Homan}, J., {Wijnands}, R., {van der Klis}, M., {et~al.} 2001, \apjs, 132,
  377, \dodoi{10.1086/318954}

\bibitem[{{Ingram} {et~al.}(2009){Ingram}, {Done}, \&
  {Fragile}}]{2009MNRAS.397L.101I}
{Ingram}, A., {Done}, C., \& {Fragile}, P.~C. 2009, \mnras, 397, L101,
  \dodoi{10.1111/j.1745-3933.2009.00693.x}

\bibitem[{{Kara} {et~al.}(2019){Kara}, {Steiner}, {Fabian}, {Cackett},
  {Uttley}, {Remillard}, {Gendreau}, {Arzoumanian}, {Altamirano}, {Eikenberry},
  {Enoto}, {Homan}, {Neilsen}, \& {Stevens}}]{2019Natur.565..198K}
{Kara}, E., {Steiner}, J.~F., {Fabian}, A.~C., {et~al.} 2019, \nat, 565, 198,
  \dodoi{10.1038/s41586-018-0803-x}

\bibitem[{{Kubota} {et~al.}(1998){Kubota}, {Tanaka}, {Makishima}, {Ueda},
  {Dotani}, {Inoue}, \& {Yamaoka}}]{1998PASJ...50..667K}
{Kubota}, A., {Tanaka}, Y., {Makishima}, K., {et~al.} 1998, \pasj, 50, 667,
  \dodoi{10.1093/pasj/50.6.667}

\bibitem[{{Li} {et~al.}(2020){Li}, {Li}, {Tan}, {Yang}, {Ge}, {Zhang}, {Tuo},
  {Wu}, {Liao}, {Zhang}, {Song}, {Zhang}, {Qu}, {Zhang}, {Lu}, {Xu}, {Liu},
  {Cao}, {Chen}, {Nie}, {Zhao}, \& {Li}}]{2020JHEAp..27...64L}
{Li}, X., {Li}, X., {Tan}, Y., {et~al.} 2020, Journal of High Energy
  Astrophysics, 27, 64, \dodoi{10.1016/j.jheap.2020.02.009}

\bibitem[{{Liao} {et~al.}(2020{\natexlab{a}}){Liao}, {Zhang}, {Chen}, {Zhang},
  {Jin}, {Chang}, {Chen}, {Ge}, {Guo}, {Li}, {Li}, {Lu}, {Lu}, {Nie}, {Song},
  {Yang}, {You}, {Zhao}, \& {Zhang}}]{2020JHEAp..27...24L}
{Liao}, J.-Y., {Zhang}, S., {Chen}, Y., {et~al.} 2020{\natexlab{a}}, Journal of
  High Energy Astrophysics, 27, 24, \dodoi{10.1016/j.jheap.2020.02.010}

\bibitem[{{Liao} {et~al.}(2020{\natexlab{b}}){Liao}, {Zhang}, {Lu}, {Zhang},
  {Li}, {Chang}, {Chen}, {Ge}, {Guo}, {Huang}, {Jin}, {Li}, {Li}, {Li}, {Liu},
  {Lu}, {Nie}, {Song}, {Wang}, {You}, {Zhang}, {Zhao}, \&
  {Zhang}}]{2020JHEAp..27...14L}
{Liao}, J.-Y., {Zhang}, S., {Lu}, X.-F., {et~al.} 2020{\natexlab{b}}, Journal
  of High Energy Astrophysics, 27, 14, \dodoi{10.1016/j.jheap.2020.04.002}

\bibitem[{{Liu} {et~al.}(2007){Liu}, {Taam}, {Meyer-Hofmeister}, \&
  {Meyer}}]{2007ApJ...671..695L}
{Liu}, B.~F., {Taam}, R.~E., {Meyer-Hofmeister}, E., \& {Meyer}, F. 2007, \apj,
  671, 695, \dodoi{10.1086/522619}

\bibitem[{{Liu} {et~al.}(2020){Liu}, {Zhang}, {Li}, {Lu}, {Chang}, {Li},
  {Zhang}, {Jin}, {Yu}, {Zhang}, {Fu}, {Chen}, {Ji}, {Xu}, {Deng}, {Shang},
  {Liu}, {Lu}, {Zhang}, {Dong}, {Li}, {Wu}, {Li}, {Wang}, {Wu}, {Zhang},
  {Zhang}, {Xiong}, {Liu}, {Zhang}, {Liu}, {Yang}, \&
  {Zhang}}]{2020SCPMA..63x9503L}
{Liu}, C., {Zhang}, Y., {Li}, X., {et~al.} 2020, Science China Physics,
  Mechanics, and Astronomy, 63, 249503, \dodoi{10.1007/s11433-019-1486-x}

\bibitem[{{Liu} {et~al.}(2021){Liu}, {Huang}, {Xiao}, {Bu}, {Qu}, {Zhang},
  {Zhang}, {Jia}, {Lu}, {Ma}, {Tao}, {Zhang}, {Chen}, {Song}, {Li}, {Xu},
  {Cao}, {Chen}, {Liu}, {Cai}, {Chang}, {Chen}, {Chen}, {Chen}, {Chen}, {Cui},
  {Cui}, {Deng}, {Dong}, {Du}, {Fu}, {Gao}, {Gao}, {Gao}, {Ge}, {Gu}, {Guan},
  {Guo}, {Han}, {Huo}, {Jiang}, {Jiang}, {Jin}, {Jin}, {Kong}, {Li}, {Li},
  {Li}, {Li}, {Li}, {Li}, {Li}, {Li}, {Li}, {Li}, {Liang}, {Liao}, {Liu},
  {Liu}, {Liu}, {Liu}, {Liu}, {Lu}, {Lu}, {Luo}, {Luo}, {Meng}, {Nang}, {Nie},
  {Ou}, {Sai}, {Shang}, {Song}, {Sun}, {Tan}, {Tuo}, {Wang}, {Wang}, {Wang},
  {Wang}, {Wang}, {Wang}, {Wen}, {Wu}, {Wu}, {Wu}, {Xiao}, {Xiong}, {Xu},
  {Yang}, {Yang}, {Yang}, {Yang}, {Yi}, {Yin}, {You}, {Zhang}, {Zhang},
  {Zhang}, {Zhang}, {Zhang}, {Zhang}, {Zhang}, {Zhang}, {Zhang}, {Zhang},
  {Zhang}, {Zhang}, {Zhang}, {Zhang}, {Zhang}, {Zhang}, {Zhao}, {Zhao},
  {Zheng}, {Zheng}, {Zhou}, {Zhou}, {Zhu}, {Zhuang}, \&
  {Zhu}}]{2021RAA....21...70L}
{Liu}, H.-X., {Huang}, Y., {Xiao}, G.-C., {et~al.} 2021, Research in Astronomy
  and Astrophysics, 21, 070, \dodoi{10.1088/1674-4527/21/3/70}

\bibitem[{{Ma} {et~al.}(2021){Ma}, {Tao}, {Zhang}, {Zhang}, {Bu}, {Ge}, {Chen},
  {Qu}, {Zhang}, {Lu}, {Song}, {Yang}, {Yuan}, {Cai}, {Cao}, {Chang}, {Chen},
  {Chen}, {Chen}, {Chen}, {Chen}, {Cui}, {Cui}, {Deng}, {Dong}, {Du}, {Fu},
  {Gao}, {Gao}, {Gao}, {Gu}, {Guan}, {Guo}, {Han}, {Huang}, {Huo}, {Ji}, {Jia},
  {Jiang}, {Jiang}, {Jin}, {Jin}, {Kong}, {Li}, {Li}, {Li}, {Li}, {Li}, {Li},
  {Li}, {Li}, {Li}, {Li}, {Li}, {Liang}, {Liao}, {Liu}, {Liu}, {Liu}, {Liu},
  {Liu}, {Liu}, {Lu}, {Lu}, {Luo}, {Luo}, {Meng}, {Nang}, {Nie}, {Ou}, {Sai},
  {Shang}, {Song}, {Sun}, {Tan}, {Tuo}, {Wang}, {Wang}, {Wang}, {Wang}, {Wang},
  {Wang}, {Wen}, {Wu}, {Wu}, {Wu}, {Xiao}, {Xiao}, {Xie}, {Xiong}, {Xu}, {Xu},
  {Yang}, {Yang}, {Yang}, {Yi}, {Yin}, {You}, {Zhang}, {Zhang}, {Zhang},
  {Zhang}, {Zhang}, {Zhang}, {Zhang}, {Zhang}, {Zhang}, {Zhang}, {Zhang},
  {Zhang}, {Zhang}, {Zhang}, {Zhang}, {Zhang}, {Zhao}, {Zhao}, {Zheng}, {Zhou},
  {Zhou}, {Zhu}, {Zhu}, \& {Zhuang}}]{2021NatAs...5...94M}
{Ma}, X., {Tao}, L., {Zhang}, S.-N., {et~al.} 2021, Nature Astronomy, 5, 94,
  \dodoi{10.1038/s41550-020-1192-2}

\bibitem[{{Maitra} {et~al.}(2014){Maitra}, {Miller}, {Reynolds}, {Reis}, \&
  {Nowak}}]{2014ApJ...794...85M}
{Maitra}, D., {Miller}, J.~M., {Reynolds}, M.~T., {Reis}, R., \& {Nowak}, M.
  2014, \apj, 794, 85, \dodoi{10.1088/0004-637X/794/1/85}

\bibitem[{{Makishima} {et~al.}(1986){Makishima}, {Maejima}, {Mitsuda}, {Bradt},
  {Remillard}, {Tuohy}, {Hoshi}, \& {Nakagawa}}]{1986ApJ...308..635M}
{Makishima}, K., {Maejima}, Y., {Mitsuda}, K., {et~al.} 1986, \apj, 308, 635,
  \dodoi{10.1086/164534}

\bibitem[{{McClintock} {et~al.}(2014){McClintock}, {Narayan}, \&
  {Steiner}}]{2014SSRv..183..295M}
{McClintock}, J.~E., {Narayan}, R., \& {Steiner}, J.~F. 2014, \ssr, 183, 295,
  \dodoi{10.1007/s11214-013-0003-9}

\bibitem[{{Mereminskiy} {et~al.}(2019){Mereminskiy}, {Krivonos}, {Medvedev}, \&
  {Grebenev}}]{2019ATel12969....1M}
{Mereminskiy}, I.~A., {Krivonos}, R.~A., {Medvedev}, P.~S., \& {Grebenev},
  S.~A. 2019, The Astronomer's Telegram, 12969, 1

\bibitem[{{Merloni} {et~al.}(2000){Merloni}, {Fabian}, \&
  {Ross}}]{2000MNRAS.313..193M}
{Merloni}, A., {Fabian}, A.~C., \& {Ross}, R.~R. 2000, \mnras, 313, 193,
  \dodoi{10.1046/j.1365-8711.2000.03226.x}

\bibitem[{{Miller} {et~al.}(2006){Miller}, {Homan}, \&
  {Miniutti}}]{2006ApJ...652L.113M}
{Miller}, J.~M., {Homan}, J., \& {Miniutti}, G. 2006, \apjl, 652, L113,
  \dodoi{10.1086/510015}

\bibitem[{{Miller} {et~al.}(2009){Miller}, {Reynolds}, {Fabian}, {Miniutti}, \&
  {Gallo}}]{2009ApJ...697..900M}
{Miller}, J.~M., {Reynolds}, C.~S., {Fabian}, A.~C., {Miniutti}, G., \&
  {Gallo}, L.~C. 2009, \apj, 697, 900, \dodoi{10.1088/0004-637X/697/1/900}

\bibitem[{{Miller} {et~al.}(2019){Miller}, {Zoghbi}, {Gandhi}, \&
  {Paice}}]{2019ATel13012....1M}
{Miller}, J.~M., {Zoghbi}, A., {Gandhi}, P., \& {Paice}, J. 2019, The
  Astronomer's Telegram, 13012, 1

\bibitem[{{Miller-Jones} {et~al.}(2019){Miller-Jones}, {Russell}, {Sivakoff},
  \& {Tetarenko}}]{2019ATel12977....1M}
{Miller-Jones}, J., {Russell}, T., {Sivakoff}, G., \& {Tetarenko}, A. 2019, The
  Astronomer's Telegram, 12977, 1

\bibitem[{{Mitsuda} {et~al.}(1984){Mitsuda}, {Inoue}, {Koyama}, {Makishima},
  {Matsuoka}, {Ogawara}, {Shibazaki}, {Suzuki}, {Tanaka}, \&
  {Hirano}}]{1984PASJ...36..741M}
{Mitsuda}, K., {Inoue}, H., {Koyama}, K., {et~al.} 1984, \pasj, 36, 741

\bibitem[{{Motta} {et~al.}(2012){Motta}, {Homan}, {Mu{\~n}oz Darias},
  {Casella}, {Belloni}, {Hiemstra}, \& {M{\'e}ndez}}]{2012MNRAS.427..595M}
{Motta}, S., {Homan}, J., {Mu{\~n}oz Darias}, T., {et~al.} 2012, \mnras, 427,
  595, \dodoi{10.1111/j.1365-2966.2012.22037.x}

\bibitem[{{Motta}(2016)}]{2016AN....337..398M}
{Motta}, S.~E. 2016, Astronomische Nachrichten, 337, 398,
  \dodoi{10.1002/asna.201612320}

\bibitem[{{Mu{\~n}oz-Darias} {et~al.}(2011){Mu{\~n}oz-Darias}, {Motta},
  {Stiele}, \& {Belloni}}]{2011MNRAS.415..292M}
{Mu{\~n}oz-Darias}, T., {Motta}, S., {Stiele}, H., \& {Belloni}, T.~M. 2011,
  \mnras, 415, 292, \dodoi{10.1111/j.1365-2966.2011.18702.x}

\bibitem[{{Narayan} \& {Yi}(1995)}]{1995ApJ...452..710N}
{Narayan}, R., \& {Yi}, I. 1995, \apj, 452, 710, \dodoi{10.1086/176343}

\bibitem[{{Negoro} {et~al.}(2019){Negoro}, {Nakajima}, {Sugita}, {Sasaki},
  {Mihara}, {Maruyama}, {Aoki}, {Kobayashi}, {Tamagawa}, {Matsuoka},
  {Sakamoto}, {Serino}, {Nishida}, {Yoshida}, {Tsuboi}, {Kawai}, {Sato},
  {Shidatsu}, {Kawai}, {Sugizaki}, {Oeda}, {Shiraishi}, {Nakahira}, {Sugawara},
  {Ueno}, {Tomida}, {Ishikawa}, {Isobe}, {Shimomukai}, {Tominaga}, {Ueda},
  {Tanimoto}, {Yamada}, {Ogawa}, {Setoguchi}, {Yoshitake}, {Tsunemi},
  {Yoneyama}, {Asakura}, {Ide}, {Yamauchi}, {Iwahori}, {Kurihara}, {Kurogi},
  {Miike}, {Kawamuro}, {Yamaoka}, \& {Kawakubo}}]{2019ATel12968....1N}
{Negoro}, H., {Nakajima}, M., {Sugita}, S., {et~al.} 2019, The Astronomer's
  Telegram, 12968, 1

\bibitem[{{{\"O}zel} {et~al.}(2010){{\"O}zel}, {Psaltis}, {Narayan}, \&
  {McClintock}}]{Ozel2010}
{{\"O}zel}, F., {Psaltis}, D., {Narayan}, R., \& {McClintock}, J.~E. 2010,
  \apj, 725, 1918, \dodoi{10.1088/0004-637X/725/2/1918}

\bibitem[{{Parker} {et~al.}(2015){Parker}, {Tomsick}, {Miller}, {Yamaoka},
  {Lohfink}, {Nowak}, {Fabian}, {Alston}, {Boggs}, {Christensen}, {Craig},
  {F{\"u}rst}, {Gandhi}, {Grefenstette}, {Grinberg}, {Hailey}, {Harrison},
  {Kara}, {King}, {Stern}, {Walton}, {Wilms}, \& {Zhang}}]{2015ApJ...808....9P}
{Parker}, M.~L., {Tomsick}, J.~A., {Miller}, J.~M., {et~al.} 2015, \apj, 808,
  9, \dodoi{10.1088/0004-637X/808/1/9}

\bibitem[{{Parmar} {et~al.}(1993){Parmar}, {Angelini}, {Roche}, \&
  {White}}]{1993A&A...279..179P}
{Parmar}, A.~N., {Angelini}, L., {Roche}, P., \& {White}, N.~E. 1993, \aap,
  279, 179

\bibitem[{{Peris} {et~al.}(2016){Peris}, {Remillard}, {Steiner}, {Vrtilek},
  {Varni{\`e}re}, {Rodriguez}, \& {Pooley}}]{2016ApJ...822...60P}
{Peris}, C.~S., {Remillard}, R.~A., {Steiner}, J.~F., {et~al.} 2016, \apj, 822,
  60, \dodoi{10.3847/0004-637X/822/2/60}

\bibitem[{{Plotkin} {et~al.}(2015){Plotkin}, {Gallo}, {Markoff}, {Homan},
  {Jonker}, {Miller-Jones}, {Russell}, \& {Drappeau}}]{2015MNRAS.446.4098P}
{Plotkin}, R.~M., {Gallo}, E., {Markoff}, S., {et~al.} 2015, \mnras, 446, 4098,
  \dodoi{10.1093/mnras/stu2385}

\bibitem[{{Reis} {et~al.}(2010){Reis}, {Fabian}, \&
  {Miller}}]{2010MNRAS.402..836R}
{Reis}, R.~C., {Fabian}, A.~C., \& {Miller}, J.~M. 2010, \mnras, 402, 836,
  \dodoi{10.1111/j.1365-2966.2009.15976.x}

\bibitem[{{Reis} {et~al.}(2009){Reis}, {Fabian}, {Ross}, \&
  {Miller}}]{2009MNRAS.395.1257R}
{Reis}, R.~C., {Fabian}, A.~C., {Ross}, R.~R., \& {Miller}, J.~M. 2009, \mnras,
  395, 1257, \dodoi{10.1111/j.1365-2966.2009.14622.x}

\bibitem[{{Reis} {et~al.}(2008){Reis}, {Fabian}, {Ross}, {Miniutti}, {Miller},
  \& {Reynolds}}]{2008MNRAS.387.1489R}
{Reis}, R.~C., {Fabian}, A.~C., {Ross}, R.~R., {et~al.} 2008, \mnras, 387,
  1489, \dodoi{10.1111/j.1365-2966.2008.13358.x}

\bibitem[{{Remillard} \& {McClintock}(2006)}]{2006ARA&A..44...49R}
{Remillard}, R.~A., \& {McClintock}, J.~E. 2006, \araa, 44, 49,
  \dodoi{10.1146/annurev.astro.44.051905.092532}

\bibitem[{{Reynolds}(2014)}]{2014SSRv..183..277R}
{Reynolds}, C.~S. 2014, \ssr, 183, 277, \dodoi{10.1007/s11214-013-0006-6}

\bibitem[{{Reynolds} \& {Miller}(2013)}]{2013ApJ...769...16R}
{Reynolds}, M.~T., \& {Miller}, J.~M. 2013, \apj, 769, 16,
  \dodoi{10.1088/0004-637X/769/1/16}

\bibitem[{{Salvesen} \& {Miller}(2021)}]{2021MNRAS.500.3640S}
{Salvesen}, G., \& {Miller}, J.~M. 2021, \mnras, 500, 3640,
  \dodoi{10.1093/mnras/staa3325}

\bibitem[{{Salvesen} {et~al.}(2013){Salvesen}, {Miller}, {Reis}, \&
  {Begelman}}]{2013MNRAS.431.3510S}
{Salvesen}, G., {Miller}, J.~M., {Reis}, R.~C., \& {Begelman}, M.~C. 2013,
  \mnras, 431, 3510, \dodoi{10.1093/mnras/stt436}

\bibitem[{{Shafee} {et~al.}(2006){Shafee}, {McClintock}, {Narayan}, {Davis},
  {Li}, \& {Remillard}}]{2006ApJ...636L.113S}
{Shafee}, R., {McClintock}, J.~E., {Narayan}, R., {et~al.} 2006, \apjl, 636,
  L113, \dodoi{10.1086/498938}

\bibitem[{{Shimura} \& {Takahara}(1995)}]{1995ApJ...445..780S}
{Shimura}, T., \& {Takahara}, F. 1995, \apj, 445, 780, \dodoi{10.1086/175740}

\bibitem[{{Sidoli} {et~al.}(2017){Sidoli}, {Tiengo}, {Paizis}, {Sguera},
  {Lotti}, \& {Natalucci}}]{2017ApJ...838..133S}
{Sidoli}, L., {Tiengo}, A., {Paizis}, A., {et~al.} 2017, \apj, 838, 133,
  \dodoi{10.3847/1538-4357/aa671a}

\bibitem[{{Soria} {et~al.}(2008){Soria}, {Wu}, \&
  {Kunic}}]{2008xng..conf...48S}
{Soria}, R., {Wu}, K., \& {Kunic}, Z. 2008, in X-rays From Nearby Galaxies, ed.
  S.~{Carpano}, M.~{Ehle}, \& W.~{Pietsch}, 48--51.
\newblock \doarXiv{0711.2448}

\bibitem[{{Sridhar} {et~al.}(2020){Sridhar}, {Garc{\'\i}a}, {Steiner},
  {Connors}, {Grinberg}, \& {Harrison}}]{2020ApJ...890...53S}
{Sridhar}, N., {Garc{\'\i}a}, J.~A., {Steiner}, J.~F., {et~al.} 2020, \apj,
  890, 53, \dodoi{10.3847/1538-4357/ab64f5}

\bibitem[{{Steiner} {et~al.}(2017){Steiner}, {Garc{\'\i}a}, {Eikmann},
  {McClintock}, {Brenneman}, {Dauser}, \& {Fabian}}]{2017ApJ...836..119S}
{Steiner}, J.~F., {Garc{\'\i}a}, J.~A., {Eikmann}, W., {et~al.} 2017, \apj,
  836, 119, \dodoi{10.3847/1538-4357/836/1/119}

\bibitem[{{Steiner} {et~al.}(2009){Steiner}, {Narayan}, {McClintock}, \&
  {Ebisawa}}]{2009PASP..121.1279S}
{Steiner}, J.~F., {Narayan}, R., {McClintock}, J.~E., \& {Ebisawa}, K. 2009,
  \pasp, 121, 1279, \dodoi{10.1086/648535}

\bibitem[{{Steiner} {et~al.}(2011){Steiner}, {Reis}, {McClintock}, {Narayan},
  {Remillard}, {Orosz}, {Gou}, {Fabian}, \& {Torres}}]{2011MNRAS.416..941S}
{Steiner}, J.~F., {Reis}, R.~C., {McClintock}, J.~E., {et~al.} 2011, \mnras,
  416, 941, \dodoi{10.1111/j.1365-2966.2011.19089.x}

\bibitem[{{Stella} \& {Vietri}(1998)}]{1998ApJ...492L..59S}
{Stella}, L., \& {Vietri}, M. 1998, \apjl, 492, L59, \dodoi{10.1086/311075}

\bibitem[{{Sunyaev} \& {Titarchuk}(1980)}]{1980A&A....86..121S}
{Sunyaev}, R.~A., \& {Titarchuk}, L.~G. 1980, \aap, 500, 167

\bibitem[{{Tanaka} \& {Shibazaki}(1996)}]{1996ARA&A..34..607T}
{Tanaka}, Y., \& {Shibazaki}, N. 1996, \araa, 34, 607,
  \dodoi{10.1146/annurev.astro.34.1.607}

\bibitem[{{Tang} {et~al.}(2011){Tang}, {Yu}, \& {Yan}}]{2011RAA....11..434T}
{Tang}, J., {Yu}, W.-F., \& {Yan}, Z. 2011, Research in Astronomy and
  Astrophysics, 11, 434, \dodoi{10.1088/1674-4527/11/4/006}

\bibitem[{{Tao} {et~al.}(2015){Tao}, {Tomsick}, {Walton}, {F{\"u}rst},
  {Kennea}, {Miller}, {Boggs}, {Christensen}, {Craig}, {Gandhi},
  {Grefenstette}, {Hailey}, {Harrison}, {Krimm}, {Pottschmidt}, {Stern},
  {Tendulkar}, \& {Zhang}}]{2015ApJ...811...51T}
{Tao}, L., {Tomsick}, J.~A., {Walton}, D.~J., {et~al.} 2015, \apj, 811, 51,
  \dodoi{10.1088/0004-637X/811/1/51}

\bibitem[{{Tao} {et~al.}(2018){Tao}, {Chen}, {G{\"u}ng{\"o}r}, {Huang}, {Lu},
  {Qu}, {Song}, {Zhang}, {Zhang}, \& {Zhang}}]{2018MNRAS.480.4443T}
{Tao}, L., {Chen}, Y., {G{\"u}ng{\"o}r}, C., {et~al.} 2018, \mnras, 480, 4443,
  \dodoi{10.1093/mnras/sty2157}

\bibitem[{{Tetarenko} {et~al.}(2016){Tetarenko}, {Sivakoff}, {Heinke}, \&
  {Gladstone}}]{2016ApJS..222...15T}
{Tetarenko}, B.~E., {Sivakoff}, G.~R., {Heinke}, C.~O., \& {Gladstone}, J.~C.
  2016, \apjs, 222, 15, \dodoi{10.3847/0067-0049/222/2/15}

\bibitem[{{Tomsick} \& {Kaaret}(2000)}]{2000ApJ...537..448T}
{Tomsick}, J.~A., \& {Kaaret}, P. 2000, \apj, 537, 448, \dodoi{10.1086/308999}

\bibitem[{{van der Klis}(2006)}]{2006csxs.book...39V}
{van der Klis}, M. 2006, {Rapid X-ray Variability}, Vol.~39, 39--112

\bibitem[{{Varniere} \& {Vincent}(2016)}]{2016A&A...591A..36V}
{Varniere}, P., \& {Vincent}, F.~H. 2016, \aap, 591, A36,
  \dodoi{10.1051/0004-6361/201527711}

\bibitem[{{Wang} {et~al.}(2020){Wang}, {Ji}, {Zhang}, {M{\'e}ndez}, {Qu},
  {Maggi}, {Ge}, {Qiao}, {Tao}, {Zhang}, {Altamirano}, {Zhang}, {Ma}, {Lu},
  {Li}, {Huang}, {Zheng}, {Chen}, {Chang}, {Tuo}, {G{\"u}ng{\"o}r}, {Song},
  {Xu}, {Cao}, {Chen}, {Liu}, {Bu}, {Cai}, {Chen}, {Chen}, {Chen}, {Chen},
  {Cui}, {Cui}, {Deng}, {Dong}, {Du}, {Fu}, {Gao}, {Gao}, {Gao}, {Gu}, {Guan},
  {Guo}, {Han}, {Huo}, {Jia}, {Jiang}, {Jiang}, {Jin}, {Jin}, {Kong}, {Li},
  {Li}, {Li}, {Li}, {Li}, {Li}, {Li}, {Li}, {Li}, {Li}, {Liang}, {Liao}, {Liu},
  {Liu}, {Liu}, {Liu}, {Lu}, {Lu}, {Luo}, {Luo}, {Meng}, {Nang}, {Nie}, {Ou},
  {Sai}, {Shang}, {Song}, {Sun}, {Tan}, {Wang}, {Wang}, {Wang}, {Wang}, {Wang},
  {Wen}, {Wu}, {Wu}, {Wu}, {Xiao}, {Xiao}, {Xiong}, {Yang}, {Yang}, {Yang},
  {Yang}, {Yi}, {Yin}, {You}, {Zhang}, {Zhang}, {Zhang}, {Zhang}, {Zhang},
  {Zhang}, {Zhang}, {Zhang}, {Zhang}, {Zhang}, {Zhang}, {Zhang}, {Zhang},
  {Zhang}, {Zhang}, {Zhang}, {Zhao}, {Zhao}, {Zhou}, {Zhou}, {Zhuang}, {Zhu},
  {Zhu}, \& {Wang}}]{2020ApJ...896...33W}
{Wang}, Y., {Ji}, L., {Zhang}, S.~N., {et~al.} 2020, \apj, 896, 33,
  \dodoi{10.3847/1538-4357/ab8db4}

\bibitem[{{Wang} {et~al.}(2021){Wang}, {Ji}, {Garc{\'\i}a}, {Dauser},
  {M{\'e}ndez}, {Mao}, {Tao}, {Altamirano}, {Maggi}, {Zhang}, {Ge}, {Zhang},
  {Qu}, {Zhang}, {Ma}, {Lu}, {Li}, {Huang}, {Zheng}, {Chang}, {Tuo}, {Song},
  {Xu}, {Chen}, {Liu}, {Bu}, {Cai}, {Cao}, {Chen}, {Chen}, {Chen}, {Cui}, {Du},
  {Gao}, {Gu}, {Guan}, {Guo}, {Han}, {Huo}, {Jia}, {Jiang}, {Jin}, {Kong},
  {Li}, {Li}, {Li}, {Li}, {Li}, {Li}, {Li}, {Li}, {Liang}, {Liao}, {Liu},
  {Liu}, {Lu}, {Luo}, {Luo}, {Meng}, {Nang}, {Nie}, {Ou}, {Sai}, {Shang},
  {Song}, {Sun}, {Tan}, {Wang}, {Wang}, {Wang}, {Wen}, {Wu}, {Wu}, {Wu},
  {Xiao}, {Xiao}, {Xiong}, {Yang}, {Yang}, {Yi}, {Yin}, {You}, {Zhang},
  {Zhang}, {Zhang}, {Zhang}, {Zhang}, {Zhang}, {Zhao}, {Zhao}, \&
  {Zhou}}]{2021ApJ...906...11W}
{Wang}, Y., {Ji}, L., {Garc{\'\i}a}, J.~A., {et~al.} 2021, \apj, 906, 11,
  \dodoi{10.3847/1538-4357/abc55e}

\bibitem[{{Williams} {et~al.}(2019){Williams}, {Fender}, {Woudt}, \&
  {Miller-Jones}}]{2019ATel12992....1W}
{Williams}, D., {Fender}, R., {Woudt}, P., \& {Miller-Jones}, J. 2019, The
  Astronomer's Telegram, 12992, 1

\bibitem[{{Wilms} {et~al.}(2000){Wilms}, {Allen}, \&
  {McCray}}]{2000ApJ...542..914W}
{Wilms}, J., {Allen}, A., \& {McCray}, R. 2000, \apj, 542, 914,
  \dodoi{10.1086/317016}

\bibitem[{{Yan} \& {Yu}(2015)}]{2015ApJ...805...87Y}
{Yan}, Z., \& {Yu}, W. 2015, \apj, 805, 87, \dodoi{10.1088/0004-637X/805/2/87}

\bibitem[{{Yang} {et~al.}(2019{\natexlab{a}}){Yang}, {Soria}, {Russell},
  {Xiao}, {Qu}, \& {Zhang}}]{2019ATel13036....1Y}
{Yang}, Y.-J., {Soria}, R., {Russell}, D., {et~al.} 2019{\natexlab{a}}, The
  Astronomer's Telegram, 13036, 1

\bibitem[{{Yang} {et~al.}(2019{\natexlab{b}}){Yang}, {Xiao}, {Soria},
  {Russell}, {Qu}, \& {Zhang}}]{2019ATel13037....1Y}
{Yang}, Y.-J., {Xiao}, G., {Soria}, R., {et~al.} 2019{\natexlab{b}}, The
  Astronomer's Telegram, 13037, 1

\bibitem[{{Yao} {et~al.}(2005){Yao}, {Zhang}, {Zhang}, {Feng}, \&
  {Robinson}}]{2005ApJ...619..446Y}
{Yao}, Y., {Zhang}, S.~N., {Zhang}, X., {Feng}, Y., \& {Robinson}, C.~R. 2005,
  \apj, 619, 446, \dodoi{10.1086/426374}

\bibitem[{{Zdziarski} {et~al.}(2021){Zdziarski}, {De Marco}, {Szanecki},
  {Nied{\'z}wiecki}, \& {Markowitz}}]{2021ApJ...906...69Z}
{Zdziarski}, A.~A., {De Marco}, B., {Szanecki}, M., {Nied{\'z}wiecki}, A., \&
  {Markowitz}, A. 2021, \apj, 906, 69, \dodoi{10.3847/1538-4357/abca9c}

\bibitem[{{Zhang} {et~al.}(2020{\natexlab{a}}){Zhang}, {Altamirano},
  {C{\'u}neo}, {Alabarta}, {Enoto}, {Homan}, {Remillard}, {Uttley},
  {Vincentelli}, {Arzoumanian}, {Bult}, {Gendreau}, {Markwardt}, {Sanna},
  {Strohmayer}, {Steiner}, {Basak}, {Neilsen}, \&
  {Tombesi}}]{2020MNRAS.499..851Z}
{Zhang}, L., {Altamirano}, D., {C{\'u}neo}, V.~A., {et~al.} 2020{\natexlab{a}},
  \mnras, 499, 851, \dodoi{10.1093/mnras/staa2842}

\bibitem[{{Zhang} {et~al.}(2014){Zhang}, {Lu}, {Zhang}, \&
  {Li}}]{2014SPIE.9144E..21Z}
{Zhang}, S., {Lu}, F.~J., {Zhang}, S.~N., \& {Li}, T.~P. 2014, in Society of
  Photo-Optical Instrumentation Engineers (SPIE) Conference Series, Vol. 9144,
  Space Telescopes and Instrumentation 2014: Ultraviolet to Gamma Ray, ed.
  T.~{Takahashi}, J.-W.~A. {den Herder}, \& M.~{Bautz}, 914421,
  \dodoi{10.1117/12.2054144}

\bibitem[{{Zhang} {et~al.}(1997{\natexlab{a}}){Zhang}, {Cui}, \&
  {Chen}}]{1997ApJ...482L.155Z}
{Zhang}, S.~N., {Cui}, W., \& {Chen}, W. 1997{\natexlab{a}}, \apjl, 482, L155,
  \dodoi{10.1086/310705}

\bibitem[{{Zhang} {et~al.}(1997{\natexlab{b}}){Zhang}, {Ebisawa}, {Sunyaev},
  {Ueda}, {Harmon}, {Sazonov}, {Fishman}, {Inoue}, {Paciesas}, \&
  {Takahash}}]{1997ApJ...479..381Z}
{Zhang}, S.~N., {Ebisawa}, K., {Sunyaev}, R., {et~al.} 1997{\natexlab{b}},
  \apj, 479, 381, \dodoi{10.1086/303870}

\bibitem[{{Zhang} {et~al.}(2020{\natexlab{b}}){Zhang}, {Li}, {Lu}, {Song},
  {Xu}, {Liu}, {Chen}, {Cao}, {Bu}, {Chang}, {Chen}, {Chen}, {Chen}, {Chen},
  {Chen}, {Cui}, {Cui}, {Deng}, {Dong}, {Du}, {Fu}, {Gao}, {Gao}, {Gao}, {Ge},
  {Gu}, {Guan}, {Gungor}, {Guo}, {Han}, {Hu}, {Huang}, {Huo}, {Jia}, {Jiang},
  {Jiang}, {Jin}, {Jin}, {Li}, {Li}, {Li}, {Li}, {Li}, {Li}, {Li}, {Li}, {Li},
  {Li}, {Li}, {Liang}, {Liao}, {Liu}, {Liu}, {Liu}, {Liu}, {Liu}, {Liu}, {Lu},
  {Lu}, {Luo}, {Ma}, {Meng}, {Nang}, {Nie}, {Ou}, {Qu}, {Sai}, {Shang}, {Shen},
  {Sun}, {Tan}, {Tao}, {Tuo}, {Wang}, {Wang}, {Wang}, {Wang}, {Wang}, {Wang},
  {Wang}, {Wen}, {Wu}, {Wu}, {Wu}, {Xiao}, {Xiong}, {Yan}, {Yang}, {Yang},
  {Yang}, {Yi}, {Yuan}, {Zhang}, {Zhang}, {Zhang}, {Zhang}, {Zhang}, {Zhang},
  {Zhang}, {Zhang}, {Zhang}, {Zhang}, {Zhang}, {Zhang}, {Zhang}, {Zhang},
  {Zhang}, {Zhang}, {Zhang}, {Zhang}, {Zhang}, {Zhang}, {Zhao}, {Zhao},
  {Zheng}, {Zhou}, {Zhu}, {Zhu}, {Zhuang}, \& {Insight-HXMT
  Team}}]{2020SCPMA..63x9502Z}
{Zhang}, S.-N., {Li}, T., {Lu}, F., {et~al.} 2020{\natexlab{b}}, Science China
  Physics, Mechanics, and Astronomy, 63, 249502,
  \dodoi{10.1007/s11433-019-1432-6}

\bibitem[{{Zhang} {et~al.}(2022){Zhang}, {Tao}, {Soria}, {Qu}, {Zhang}, {Weng},
  {zhang}, {Wang}, {Huang}, {Ma}, {Zhang}, {Ge}, {Song}, {Ma}, {Bu}, {Cai},
  {Cao}, {Chang}, {Chen}, {Chen}, {Chen}, {Chen}, {Chen}, {Cui}, {Du}, {Gao},
  {Gao}, {Gu}, {Guan}, {Guo}, {Han}, {Huo}, {Jia}, {Jiang}, {Jin}, {Kong},
  {Li}, {Li}, {Li}, {Li}, {Li}, {Li}, {Li}, {Li}, {Li}, {Liang}, {Liao}, {Liu},
  {Liu}, {Liu}, {Liu}, {Liu}, {Lu}, {Lu}, {Luo}, {Luo}, {Meng}, {Nang}, {Nie},
  {Ou}, {Ren}, {Sai}, {Song}, {Sun}, {Tan}, {Tuo}, {Wang}, {Wang}, {Wang},
  {Wang}, {Wang}, {Wen}, {Wu}, {Wu}, {Wu}, {Xiao}, {Xiao}, {Xiong}, {Chen},
  {Yang}, {Yang}, {Yang}, {Yang}, {Yi}, {Yin}, {Yuan}, {Zhang}, {Zhang},
  {Zhang}, {Zhang}, {Zhang}, {Zhang}, {Zhao}, {Zhao}, {Zheng}, {Zheng}, \&
  {Zhou}}]{2022arXiv220111919Z}
{Zhang}, W., {Tao}, L., {Soria}, R., {et~al.} 2022, arXiv e-prints,
  arXiv:2201.11919.
\newblock \doarXiv{2201.11919}

\end{thebibliography}
\bibliographystyle{aasjournal}

\clearpage
\begin{center}
%\onecolumn
\begin{longtable*}{cccccccccc}
	\caption{{\it Insight-HXMT} observations of EXO~1846--031.}\\
	\label{tab:obsinf} \\
	\toprule
	\toprule
	 ExpID$^{a}$	&	Start Time	&	Start Time 	&	HE Rate$^{b}$	&	ME Rate$^{c}$	&	LE Rate$^{d}$	&	HE Exp$^{e}$ 	&	ME Exp	&	LE Exp &	state	\\
	&	 (day)	& (MJD)	&	(cts $\rm s^{-1}$)	&	(cts $\rm s^{-1}$)	&	(cts $\rm s^{-1}$)	&	(s)	&	(s)	&	 (s)	&		\\
	\hline
	0101	&	2019-08-02T06:14:48	&	58697.35 	&	$263.6\pm0.3$	&	$51.58\pm0.15$	&	$67.8\pm0.3$	&	3001	&	2359	&	718	&	LHS	\\
0102	&	2019-08-02T10:09:02	&	58697.50 	&	$245.2\pm0.3$	&	$51.0\pm0.2$	&	$71.5\pm0.2$	&	3062	&	1964	&	1436	&	LHS	\\
0103	&	2019-08-02T13:19:55	&	58697.63 	&	$241.2\pm0.3$	&	$51.7\pm0.2$	&	$72.1\pm0.3$	&	2816	&	1823	&	762	&	LHS	\\
0104	&	2019-08-02T16:30:48	&	58697.78 	&	$233.1\pm0.3$	&	$50.8\pm0.2$	&	$74.1\pm0.3$	&	1987	&	1703	&	718	&	LHS	\\
0105	&	2019-08-02T19:41:41	&	58697.95 	&	$229.0\pm0.6$	&	$49.9\pm0.2$	&	$75.5\pm0.2$	&	616	&	1637	&	1715	&	LHS	\\
0106	&	2019-08-02T22:52:33	&	58698.03 	&	$215.2\pm0.3$	&	$50.0\pm0.2$	&	$78.5\pm0.4$	&	2896	&	1890	&	563	&	LHS	\\
0107	&	2019-08-03T02:03:26	&	58698.18 	&	$200.2\pm0.4$	&	$49.8\pm0.5$	&	$80.3\pm0.4$	&	447	&	1350	&	656	&	LHS	\\
0201	&	2019-08-04T07:33:08	&	58699.40 	&	$182.1\pm0.2$	&	$50.43\pm0.14$	&	$103.7\pm0.4$	&	3892	&	2751	&	690	&	LHS	\\
%0202	&	2019-08-04T11:27:33	&	58699.56 	&	$176.2\pm0.2$	&	$50.3\pm0.2$	&	$107.3\pm0.3$	&	2865	&	1792	&	1017	&	LHS	\\
%0203	&	2019-08-04T14:38:25	&	58699.66 	&	$170.0\pm0.5$	&	$49.4\pm0.3$	&	$112.0\pm0.7$	&	793	&	496	&	239	&	LHS	\\
0301	&	2019-08-05T07:24:32	&	58700.39 	&	$163.0\pm0.2$	&	$48.97\pm0.14$	&	$123.4\pm0.4$	&	3643	&	2540	&	700	&	HIMS	\\
0302	&	2019-08-05T11:18:58	&	58700.55 	&	$158.8\pm0.2$	&	$48.8\pm0.2$	&	$123.1\pm0.3$	&	2548	&	1500	&	1102	&	HIMS	\\
0303	&	2019-08-05T14:29:49	&	58700.66 	&	$147.4\pm0.3$	&	$47.2\pm0.2$	&	$132.1\pm0.7$	&	1421	&	848	&	299	&	HIMS	\\
0401	&	2019-08-06T02:29:39	&	58701.26 	&	$179.3\pm0.6$	&	$49.6\pm0.4$	&	$127.1\pm0.7$	&	500	&	287	&	299	&	HIMS	\\
0502	&	2019-08-07T05:47:50	&	58702.32 	&	$158.1\pm0.2$	&	$47.1\pm0.2$	&	$139.3\pm0.4$	&	2768	&	1880	&	691	&	HIMS	\\
0503	&	2019-08-07T09:26:08	&	58702.45 	&	$156.5\pm0.4$	&	$47.4\pm0.2$	&	$134.3\pm0.5$	&	1255	&	794	&	539	&	HIMS	\\
0601	&	2019-08-08T02:12:29	&	58703.21 	&	$150.0\pm0.4$	&	$47.1\pm0.2$	&	$153.0\pm0.4$	&	768	&	1311	&	1130	&	HIMS	\\
0701	&	2019-08-08T21:17:40	&	58703.94 	&	$129.9\pm0.2$	&	$42.4\pm0.2$	&	$199.4\pm0.4$	&	2549	&	1564	&	1163	&	HIMS	\\
0703	&	2019-08-09T03:48:43	&	58704.20 	&	$138.5\pm0.3$	&	$43.6\pm0.2$	&	$178.9\pm0.5$	&	1659	&	1314	&	838	&	HIMS	\\
0801	&	2019-08-10T06:41:48	&	58705.36 	&	$106.4\pm0.2$	&	$32.94\pm0.12$	&	$178.3\pm0.3$	&	1883	&	2147	&	1592	&	HIMS	\\
0802	&	2019-08-10T10:35:08	&	58705.53 	&	$109.2\pm0.2$	&	$35.64\pm0.14$	&	$220.5\pm0.4$	&	2742	&	1864	&	1371	&	HIMS	\\
0803	&	2019-08-10T13:45:57	&	58705.63 	&	$134.4\pm0.4$	&	$35.5\pm0.2$	&	$208.7\pm0.6$	&	814	&	959	&	658	&	HIMS	\\
0901	&	2019-08-13T01:30:34	&	58708.17 	&	$109.8\pm0.2$	&	$33.3\pm0.2$	&	$166.8\pm0.4$	&	1836	&	1266	&	835	&	HIMS	\\
0902	&	2019-08-13T05:22:20	&	58708.30 	&	$112.4\pm0.2$	&	$35.08\pm0.15$	&	$153.9\pm0.4$	&	2421	&	1631	&	1197	&	HIMS	\\
1001	&	2019-08-16T12:15:31	&	58711.55 	&	$91.6\pm0.3$	&	$30.6\pm0.1$	&	$125.0\pm0.2$	&	1164	&	2994	&	2765	&	HIMS	\\
1002	&	2019-08-16T16:03:46	&	58711.75 	&	$91.5\pm0.2$	&	$30.95\pm0.13$	&	$119.3\pm0.3$	&	2247	&	1777	&	1736	&	HIMS	\\
1003	&	2019-08-16T19:14:36	&	58711.85 	&	$98.1\pm0.4$	&	$31.0\pm0.2$	&	$118.4\pm0.4$	&	696	&	671	&	599	&	HIMS	\\
1101	&	2019-08-18T07:14:35	&	58713.38 	&	$75.8\pm0.2$	&	$23.1\pm0.1$	&	$147.3\pm0.3$	&	3342	&	2324	&	2072	&	HIMS	\\
1102	&	2019-08-18T11:00:05	&	58713.50 	&	$66.8\pm0.2$	&	$23.7\pm0.1$	&	$153.9\pm0.3$	&	1538	&	2202	&	1975	&	HIMS	\\
1103	&	2019-08-18T14:10:56	&	58713.70 	&	$62.7\pm0.5$	&	$23.65\pm0.14$	&	$153.8\pm0.4$	&	264	&	1255	&	1229	&	HIMS	\\
1301	&	2019-08-22T06:47:11	&	58717.38 	&	$26.9\pm0.1$	&	$7.84\pm0.06$	&	$115.2\pm0.3$	&	2714	&	1867	&	1399	&	SIMS	\\
1303	&	2019-08-22T13:36:56	&	58717.68 	&	$20.6\pm0.1$	&	$7.24\pm0.11$	&	$113.6\pm0.4$	&	917	&	589	&	599	&	SIMS	\\
1501	&	2019-08-27T18:51:16	&	58722.88 	&	$35.0\pm0.1$	&	$7.26\pm0.05$	&	$105.5\pm0.2$	&	3534	&	3039	&	2119	&	SIMS	\\
1601	&	2019-08-28T21:53:27	&	58723.99 	&	$33.79\pm0.08$	&	$9.16\pm0.05$	&	$121.7\pm0.3$	&	4836	&	3419	&	1541	&	SIMS	\\
1701	&	2019-09-03T01:55:34	&	58729.16 	&	$20.7\pm0.1$	&	$6.18\pm0.07$	&	$165.0\pm1.0$	&	2048	&	1370	&	397	&	SIMS	\\
1702	&	2019-09-03T05:25:45	&	58729.31 	&	$27.6\pm0.2$	&	$5.30\pm0.05$	&	$165.7\pm0.4$	&	556	&	2391	&	1257	&	SIMS	\\
1703	&	2019-09-03T08:44:47	&	58729.44 	&	$32.4\pm0.1$	&	$7.19\pm0.05$	&	$168.7\pm0.4$	&	3297	&	2398	&	1217	&	SIMS	\\
1801	&	2019-09-05T12:45:46	&	58731.63 	&	$26.04\pm0.12$	&	$8.01\pm0.04$	&	$187.4\pm0.3$	&	1901	&	3967	&	1533	&	SIMS	\\
1901	&	2019-09-07T17:14:33	&	58733.82 	&	$28.45\pm0.08$	&	$8.27\pm0.06$	&	$196.3\pm0.3$	&	3952	&	2610	&	2040	&	HSS	\\
2201	&	2019-09-15T01:49:09	&	58741.21 	&	$6.98\pm0.06$	&	$3.56\pm0.04$	&	$189.3\pm0.5$	&	1666	&	1839	&	718	&	HSS	\\
2202	&	2019-09-15T05:20:18	&	58741.31 	&	$6.80\pm0.04$	&	$3.68\pm0.04$	&	$186.4\pm0.4$	&	3896	&	2486	&	1436	&	HSS	\\
2204	&	2019-09-15T11:52:26	&	58741.57 	&	$4.30\pm0.04$	&	$3.09\pm0.04$	&	$180.0\pm0.3$	&	2889	&	1967	&	2179	&	HSS	\\
2205	&	2019-09-15T15:03:15	&	58741.70 	&	$1.37\pm0.02$	&	$3.62\pm0.04$	&	$180.1\pm0.3$	&	2876	&	1920	&	2155	&	HSS	\\
2206	&	2019-09-15T18:14:04	&	58741.84 	&	$9.43\pm0.07$	&	$4.82\pm0.06$	&	$183.4\pm0.4$	&	2037	&	1348	&	1162	&	HSS	\\
2208	&	2019-09-16T00:35:41	&	58742.10 	&	$15.5\pm0.2$	&	$3.20\pm0.06$	&	$179.6\pm0.5$	&	480	&	1047	&	838	&	HSS	\\
2209	&	2019-09-16T03:46:30	&	58742.24 	&	$9.65\pm0.05$	&	$3.46\pm0.04$	&	$180.9\pm0.5$	&	3889	&	2584	&	838	&	HSS	\\
2210	&	2019-09-16T06:52:15	&	58742.33 	&	$10.88\pm0.08$	&	$3.49\pm0.04$	&	$181.8\pm0.4$	&	1752	&	2313	&	958	&	HSS	\\
\hline
2211	&	2019-09-16T10:08:08	&	58742.53 	&	$12.63\pm0.09$	&	$3.87\pm0.04$	&	$179.0\pm0.3$	&	1437	&	2046	&	1482	&	HSS	\\
2212	&	2019-09-16T13:18:57	&	58742.63 	&	$7.18\pm0.05$	&	$3.25\pm0.04$	&	$178.0\pm0.3$	&	2928	&	1977	&	2274	&	HSS	\\
2213	&	2019-09-16T16:29:46	&	58742.76 	&	$10.70\pm0.06$	&	$4.23\pm0.05$	&	$180.6\pm0.3$	&	2645	&	1785	&	1556	&	HSS	\\
2214	&	2019-09-16T19:40:35	&	58742.89 	&	$1.51\pm0.05$	&	$3.52\pm0.08$	&	$179.7\pm0.7$	&	824	&	555	&	359	&	HSS	\\
2215	&	2019-09-16T22:51:24	&	58743.03 	&	$31.4\pm0.3$	&	$5.76\pm0.12$	&	$193.4\pm0.6$	&	286	&	387	&	619	&	HSS	\\
2301	&	2019-09-19T14:02:31	&	58745.71 	&	$20.40\pm0.07$	&	$7.91\pm0.06$	&	$181.3\pm0.3$	&	3672	&	2490	&	2693	&	HSS	\\
2401	&	2019-09-21T04:14:14	&	58747.21 	&	$28.32\pm0.12$	&	$8.06\pm0.05$	&	$172.1\pm0.2$	&	1873	&	3466	&	2858	&	HSS	\\
2601	&	2019-09-23T02:23:00	&	58749.19 	&	$6.18\pm0.04$	&	$2.32\pm0.03$	&	$159.8\pm0.3$	&	3664	&	3677	&	2274	&	HSS	\\
2701	&	2019-09-25T00:31:11	&	58751.14 	&	$20.13\pm0.07$	&	$4.30\pm0.03$	&	$160.8\pm0.3$	&	4617	&	3725	&	1855	&	HSS	\\
2801	&	2019-09-26T00:22:50	&	58752.13 	&	$11.91\pm0.06$	&	$3.07\pm0.03$	&	$152.3\pm0.3$	&	3701	&	3639	&	1556	&	HSS	\\
2901	&	2019-09-27T09:46:58	&	58753.51 	&	$4.60\pm0.03$	&	$1.80\pm0.02$	&	$140.6\pm0.3$	&	4988	&	3443	&	1377	&	HSS	\\
3001	&	2019-09-29T22:13:09	&	58756.03 	&	$6.34\pm0.04$	&	$1.95\pm0.02$	&	$136.1\pm0.3$	&	4537	&	3129	&	1317	&	HSS	\\
3102	&	2019-10-01T04:34:57	&	58757.28 	&	$8.58\pm0.05$	&	$1.63\pm0.03$	&	$127.9\pm0.4$	&	2925	&	1884	&	761	&	HSS	\\
3103	&	2019-10-01T08:01:33	&	58757.40 	&	$7.68\pm0.05$	&	$2.08\pm0.03$	&	$128.4\pm0.5$	&	2768	&	1874	&	479	&	HSS	\\
3203	&	2019-10-03T07:45:33	&	58759.39 	&	$3.66\pm0.04$	&	$1.38\pm0.03$	&	$125.1\pm0.8$	&	2531	&	1726	&	898	&	HSS	\\
3301	&	2019-10-05T05:26:59	&	58761.30 	&	$18.64\pm0.07$	&	$3.70\pm0.04$	&	$122.0\pm0.3$	&	4398	&	2966	&	1736	&	HSS	\\
3302	&	2019-10-05T09:04:46	&	58761.46 	&	$11.38\pm0.06$	&	$3.39\pm0.04$	&	$119.5\pm0.6$	&	3211	&	2074	&	388	&	HSS	\\
3303	&	2019-10-05T12:15:40	&	58761.56 	&	$24.76\pm0.15$	&	$5.03\pm0.08$	&	$120.3\pm0.7$	&	1143	&	712	&	239	&	HSS	\\
3401	&	2019-10-07T00:23:37	&	58763.14 	&	$26.6\pm0.2$	&	$5.21\pm0.06$	&	$124.7\pm0.5$	&	532	&	1240	&	562	&	HSS	\\
3402	&	2019-10-07T03:41:08	&	58763.23 	&	$28.17\pm0.09$	&	$5.73\pm0.05$	&	$121.4\pm0.5$	&	3444	&	2344	&	599	&	HSS	\\
3403	&	2019-10-07T07:12:40	&	58763.36 	&	$21.53\pm0.11$	&	$4.59\pm0.06$	&	$118.3\pm0.7$	&	1914	&	1217	&	239	&	HSS	\\
3701	&	2019-10-11T07:47:38	&	58767.40 	&	$27.95\pm0.09$	&	$4.59\pm0.05$	&	$106.7\pm0.3$	&	3206	&	2047	&	1240	&	HSS	\\
3702	&	2019-10-11T11:24:31	&	58767.53 	&	$29.07\pm0.14$	&	$5.21\pm0.05$	&	$109.4\pm0.3$	&	1534	&	2150	&	1017	&	HSS	\\
3802	&	2019-10-13T23:15:55	&	58770.05 	&	$21.49\pm0.09$	&	$1.95\pm0.03$	&	$97.1\pm0.2$	&	2602	&	2095	&	2028	&	HSS	\\
3803	&	2019-10-14T03:00:59	&	58770.21 	&	$14.85\pm0.09$	&	$1.49\pm0.03$	&	$96.6\pm0.3$	&	1841	&	1230	&	1017	&	HSS	\\
3901	&	2019-10-16T03:56:56	&	58772.26 	&	$11.25\pm0.07$	&	$0.93\pm0.02$	&	$88.9\pm0.3$	&	2611	&	1735	&	1197	&	HSS	\\
3902	&	2019-10-16T07:29:26	&	58772.40 	&	$10.75\pm0.06$	&	$1.09\pm0.02$	&	$89.4\pm0.2$	&	2920	&	1950	&	1676	&	HSS	\\
4602	&	2019-10-20T19:37:11	&	58776.89 	&	$24.60\pm0.12$	&	$2.63\pm0.04$	&	$85.3\pm0.3$	&	1764	&	1382	&	1317	&	HSS	\\
4603	&	2019-10-20T22:48:00	&	58777.03 	&	$19.57\pm0.09$	&	$2.53\pm0.04$	&	$85.4\pm0.2$	&	2316	&	1562	&	1436	&	HSS	\\
4701	&	2019-10-22T06:19:14	&	58778.35 	&	$15.1\pm0.1$	&	$2.48\pm0.03$	&	$81.9\pm0.2$	&	1516	&	2122	&	2095	&	HSS	\\
4702	&	2019-10-22T09:47:07	&	58778.52 	&	$11.1\pm0.1$	&	$2.77\pm0.04$	&	$83.6\pm0.2$	&	1097	&	2050	&	2214	&	HSS	\\
4703	&	2019-10-22T12:57:58	&	58778.62 	&	$17.6\pm0.1$	&	$2.40\pm0.04$	&	$83.9\pm0.2$	&	1659	&	1459	&	1356	&	HSS	\\
4801	&	2019-10-23T23:40:30	&	58780.07 	&	$13.29\pm0.07$	&	$4.24\pm0.05$	&	$81.3\pm0.2$	&	2534	&	1702	&	1616	&	HSS	\\
4803	&	2019-10-24T06:18:54	&	58780.31 	&	$20.95\pm0.12$	&	$4.80\pm0.05$	&	$81.4\pm0.2$	&	1503	&	1839	&	1637	&	HSS	\\
4901	&	2019-10-25T05:53:35	&	58781.30 	&	$21.25\pm0.13$	&	$4.08\pm0.04$	&	$78.3\pm0.2$	&	1310	&	2188	&	2274	&	HSS	\\
4902	&	2019-10-25T09:21:10	&	58781.48 	&	$14.80\pm0.09$	&	$4.59\pm0.05$	&	$78.0\pm0.2$	&	1655	&	1866	&	2211	&	HSS	\\
4903	&	2019-10-25T12:32:02	&	58781.57 	&	$22.15\pm0.14$	&	$5.24\pm0.06$	&	$79.7\pm0.3$	&	1204	&	1263	&	898	&	HSS	\\
\bottomrule
\end{longtable*}
\tablecomments{$^a$ Exposure ID, e.g., 0101: P0214050\emph{XXXX}, \emph{XXXX}=0101; $^b$ HE: 25--150~keV; $^c$ ME: 10--20~keV; $^d$ LE: 1--10~keV; $^e$ Exposure time.}
%\footnotesize{$^a$ Exposure ID, e.g., 0101:P0214050\emph{XXXX}, \emph{XXXX}=0101;} 
%\footnotesize{$^b$ HE: 30--150~keV;}
%\footnotesize{$^c$ ME: 10--20~keV;}
%\footnotesize{$^d$ LE: 1--10~keV;}
%\footnotesize{$^e$ Exposure time.}
\end{center}

\clearpage
%\begin{center}
%\onecolumn
\begin{longtable*}{cccccccccccc}
	\caption{Spectral fitting results of EXO~1846--031 using \texttt{TBabs*(cutoffpl+diskbb)} model, $N_{\rm H}$ fixed at $5.34 \times 10^{22}$ cm$^{-2}$.}\\
	\label{tab:fitting} \\
	\toprule
	\toprule
	 ExpID	&	Model	&	$\Gamma$	&	$E_{\rm cut}$	&	k$T_{\rm in}$	&	$N_{\rm dbb}$	&	$E_{\rm gau}^{c}$	&	$\sigma_{\rm gau}^{d}$ & $\chi^2$/dof  &  $F_{\rm dbb}^{e}$  &  $F_{\rm pl}^{e}$  &  $f_{\rm dbb}$  		\\
	  &	  &		&	(keV)	&	(keV)	&		&	(keV)	&	 (keV)	&   &   &  &    \\
	\hline
0101	&	CD$^{a}$	&	$1.39^{+0.04}_{-0.04}$	&	$50^{+3}_{-3}$	&	$1.65^{+0.13}_{-0.16}$	&	$6^{+2}_{-2}$	&	...	&	...	&	1280.8/1302	&	$0.08^{+0.03}_{-0.03}$	&	$0.705^{+0.007}_{-0.007}$	&	$0.12^{+0.04}_{-0.04}$	\\
0102	&	CD	&	$1.43^{+0.04}_{-0.04}$	&	$50^{+3}_{-3}$	&	$1.5^{+0.2}_{-0.3}$	&	$5^{+4}_{-2}$	&	...	&	...	&	1328.6/1443	&	$0.05^{+0.02}_{-0.02}$	&	$0.731^{+0.005}_{-0.005}$	&	$0.07^{+0.03}_{-0.03}$	\\
0103	&	CD	&	$1.46^{+0.03}_{-0.04}$	&	$50^{+3}_{-3}$	&	$1.7^{+0.2}_{-0.2}$	&	$4^{+2}_{-2}$	&	...	&	...	&	1336.9/1442	&	$0.06^{+0.03}_{-0.02}$	&	$0.66^{+0.04}_{-0.04}$	&	$0.08^{+0.03}_{-0.03}$	\\
0104	&	CD	&	$1.49^{+0.04}_{-0.04}$	&	$50^{+5}_{-4}$	&	$1.72^{+0.11}_{-0.12}$	&	$6^{+2}_{-2}$	&	...	&	...	&	1388.9/1401	&	$0.10^{+0.03}_{-0.03}$	&	$0.779^{+0.006}_{-0.006}$	&	$0.13^{+0.03}_{-0.03}$	\\
0105	&	CD	&	$1.56^{+0.04}_{-0.05}$	&	$50^{+8}_{-7}$	&	$1.6^{+0.3}_{-0.3}$	&	$5^{+5}_{-3}$	&	...	&	...	&	1364.9/1470	&	$0.04^{+0.03}_{-0.02}$	&	$0.810^{+0.006}_{-0.006}$	&	$0.05^{+0.03}_{-0.03}$	\\
0106	&	CD	&	$1.58^{+0.04}_{-0.04}$	&	$60^{+6}_{-5}$	&	$1.2^{+0.2}_{-0.2}$	&	$17^{+14}_{-7}$	&	...	&	...	&	1231.5/1224	&	$0.07^{+0.03}_{-0.03}$	&	$0.74^{+0.04}_{-0.05}$	&	$0.09^{+0.04}_{-0.03}$	\\
0107	&	CD	&	$1.65^{+0.06}_{-0.06}$	&	$60^{+14}_{-11}$	&	$1.31^{+0.15}_{-0.16}$	&	$18^{+12}_{-8}$	&	...	&	...	&	1222.4/1281	&	$0.11^{+0.04}_{-0.04}$	&	$0.73^{+0.05}_{-0.05}$	&	$0.13^{+0.04}_{-0.04}$	\\
0201	&	CD	&	$1.72^{+0.04}_{-0.04}$	&	$50^{+5}_{-4}$	&	$1.10^{+0.09}_{-0.09}$	&	$60^{+20}_{-17}$	&	...	&	...	&	1197.2/1342	&	$0.20^{+0.03}_{-0.03}$	&	$1.230^{+0.012}_{-0.012}$	&	$0.16^{+0.03}_{-0.03}$	\\
0301	&	CDG$^{b}$	&	$1.86^{+0.04}_{-0.04}$	&	$70^{+9}_{-7}$	&	$1.19^{+0.05}_{-0.05}$	&	$90^{+17}_{-15}$	&	$6.4^{+0.2}_{-p}$	&	$0.5^{+0.3}_{-0.2}$	&	1250.3/1325	&	$0.37^{+0.04}_{-0.04}$	&	$0.99^{+0.06}_{-0.06}$	&	$0.27^{+0.03}_{-0.03}$	\\
0302	&	CDG	&	$1.72^{+0.06}_{-0.07}$	&	$40^{+6}_{-5}$	&	$1.16^{+0.04}_{-0.05}$	&	$130^{+40}_{-30}$	&	$6.5^{+0.2}_{-0.1 }$	&	$1.2^{+0.3}_{-0.4}$	&	1232.1/1378	&	$0.45^{+0.04}_{-0.05}$	&	$0.80^{+0.09}_{-0.09}$	&	$0.36^{+0.04}_{-0.04}$	\\
0303	&	CDG	&	$1.8^{+0.10}_{-0.06}$	&	$50^{+14}_{-9}$	&	$1.06^{+0.08}_{-0.03}$	&	$200^{+60}_{-70}$	&	$6.4^{+0.4}_{-p}$	&	$1.6^{+0.3}_{-0.4}$	&	995.1/1137	&	$0.47^{+0.07}_{-0.08}$	&	$0.81^{+0.15}_{-0.13}$	&	$0.36^{+0.06}_{-0.06}$	\\
0502	&	CDG	&	$1.85^{+0.05}_{-0.06}$	&	$50^{+9}_{-7}$	&	$1.15^{+0.04}_{-0.04}$	&	$160^{+40}_{-30}$	&	$6.4^{+0.3}_{-p}$	&	$1.1^{+0.3}_{-0.4}$	&	1235.5/1323	&	$0.56^{+0.05}_{-0.05}$	&	$0.90^{+0.10}_{-0.11}$	&	$0.38^{+0.04}_{-0.03}$	\\
0503	&	CDG	&	$1.72^{+0.08}_{-0.10}$	&	$40^{+9}_{-7}$	&	$1.12^{+0.04}_{-0.04}$	&	$200^{+50}_{-40}$	&	$6.6^{+0.3}_{-0.1 }$	&	$1.2^{+0.4}_{-0.4}$	&	1128.0/1264	&	$0.62^{+0.06}_{-0.06}$	&	$0.77^{+0.11}_{-0.10}$	&	$0.45^{+0.05}_{-0.05}$	\\
0601	&	CDG	&	$1.95^{+0.08}_{-0.10}$	&	$60^{+30}_{-16}$	&	$1.12^{+0.04}_{-0.05}$	&	$200^{+70}_{-50}$	&	$6.8^{+0.1}_{-0.4 }$	&	$1.1^{+0.6}_{-0.6}$	&	1268.5/1394	&	$0.68^{+0.06}_{-0.06}$	&	$0.93^{+0.13}_{-0.12}$	&	$0.41^{+0.04}_{-0.04}$	\\
0701	&	CDG	&	$1.92^{+0.06}_{-0.06}$	&	$50^{+9}_{-7}$	&	$1.18^{+0.01}_{-0.01}$	&	$300^{+30}_{-20}$	&	$6.7^{+0.1}_{-0.1 }$	&	$0.4^{+0.5}_{-0.2}$	&	1324.9/1435	&	$1.28^{+0.04}_{-0.04}$	&	$0.83^{+0.08}_{-0.10}$	&	$0.62^{+0.02}_{-0.02}$	\\
0703	&	CDG	&	$1.83^{+0.07}_{-0.08}$	&	$40^{+10}_{-7}$	&	$1.17^{+0.02}_{-0.02}$	&	$300^{+30}_{-20}$	&	$6.6^{+0.2}_{-0.2 }$	&	$0.6^{+0.4}_{-0.2}$	&	1222.0/1351	&	$1.12^{+0.07}_{-0.07}$	&	$0.77^{+0.11}_{-0.11}$	&	$0.62^{+0.04}_{-0.04}$	\\
0802	&	CDG	&	$1.71^{+0.14}_{-0.17}$	&	$40^{+7}_{-8}$	&	$1.183^{+0.009}_{-0.009}$	&	$500^{+17}_{-19}$	&	$6.8^{+0.2}_{-0.4 }$	&	$0.9^{+0.3}_{-0.4}$	&	1328.3/1417	&	$1.67^{+0.04}_{-0.03}$	&	$0.62^{+0.05}_{-0.05}$	&	$0.76^{+0.02}_{-0.02}$	\\
0803	&	CDG	&	$1.79^{+0.09}_{-0.23}$	&	$50^{+20}_{-12}$	&	$1.153^{+0.013}_{-0.013}$	&	$500^{+30}_{-30}$	&	$6.4^{+0.4}_{-p}$	&	$...^{u}$	&	1214.4/1299	&	$1.77^{+0.04}_{-0.09}$	&	$0.33^{+0.10}_{-0.07}$	&	$0.90^{+0.02}_{-0.06}$	\\
0901	&	CDG	&	$1.91^{+0.09}_{-0.11}$	&	$40^{+14}_{-10}$	&	$1.04^{+0.02}_{-0.02}$	&	$500^{+60}_{-50}$	&	$6.6^{+0.3}_{-0.1 }$	&	$0.8^{+0.5}_{-0.4}$	&	1234.3/1329	&	$1.12^{+0.06}_{-0.06}$	&	$0.81^{+0.11}_{-0.19}$	&	$0.62^{+0.04}_{-0.04}$	\\
0902	&	CDG	&	$1.95^{+0.07}_{-0.08}$	&	$66^{+17}_{-12}$	&	$1.05^{+0.02}_{-0.02}$	&	$400^{+50}_{-40}$	&	$6.8^{+0.2}_{-0.2 }$	&	$0.6^{+0.7}_{-0.3}$	&	1285.6/1423	&	$0.88^{+0.04}_{-0.04}$	&	$0.77^{+0.10}_{-0.09}$	&	$0.53^{+0.03}_{-0.03}$	\\
1001	&	CDG	&	$2.08^{+0.05}_{-0.06}$	&	$170^{+170}_{-60}$	&	$0.999^{+0.018}_{-0.014}$	&	$300^{+40}_{-40}$	&	$6.4^{+0.1}_{-p}$	&	$1.1^{+0.2}_{-0.2}$	&	1323.7/1469	&	$0.65^{+0.04}_{-0.03}$	&	$0.78^{+0.10}_{-0.08}$	&	$0.45^{+0.03}_{-0.03}$	\\
1002	&	CDG	&	$2.15^{+0.05}_{-0.06}$	&	$120^{+60}_{-30}$	&	$0.98^{+0.02}_{-0.02}$	&	$300^{+40}_{-40}$	&	$6.4^{+0.1}_{-p}$	&	$1.0^{+0.2}_{-0.2}$	&	1205.3/1424	&	$0.53^{+0.04}_{-0.04}$	&	$0.94^{+0.08}_{-0.09}$	&	$0.37^{+0.03}_{-0.03}$	\\
1003	&	CDG	&	$2.09^{+0.10}_{-0.12}$	&	$110^{+200}_{-50}$	&	$0.94^{+0.03}_{-0.03}$	&	$400^{+80}_{-70}$	&	$6.4^{+0.1}_{-p}$	&	$1.2^{+0.3}_{-0.3}$	&	1084.8/1256	&	$0.58^{+0.07}_{-0.07}$	&	$0.81^{+0.14}_{-0.14}$	&	$0.42^{+0.06}_{-0.06}$	\\
1101	&	CDG	&	$1.95^{+0.08}_{-0.09}$	&	$57^{+17}_{-12}$	&	$1.043^{+0.011}_{-0.012}$	&	$500^{+30}_{-30}$	&	$6.6^{+0.2}_{-0.2 }$	&	$1.2^{+0.1}_{-0.3}$	&	1281.5/1461	&	$1.15^{+0.03}_{-0.03}$	&	$0.46^{+0.07}_{-0.06}$	&	$0.74^{+0.02}_{-0.03}$	\\
1102	&	CDG	&	$1.95^{+0.09}_{-0.10}$	&	$50^{+18}_{-11}$	&	$1.053^{+0.012}_{-0.012}$	&	$500^{+30}_{-30}$	&	$6.5^{+0.2}_{-0.1 }$	&	$0.8^{+0.4}_{-0.3}$	&	1195.0/1443	&	$1.23^{+0.04}_{-0.04}$	&	$0.57^{+0.07}_{-0.15}$	&	$0.75^{+0.03}_{-0.03}$	\\
1103	&	CDG	&	$2.16^{+0.05}_{-0.08}$	&	$500^{+p}_{-300}$	&	$1.082^{+0.012}_{-0.016}$	&	$400^{+40}_{-30}$	&	$6.4^{+0.4}_{-p}$	&	$1.1^{+0.3}_{-0.5}$	&	1123.4/1352	&	$1.14^{+0.04}_{-0.04}$	&	$0.56^{+0.09}_{-0.08}$	&	$0.69^{+0.03}_{-0.03}$	\\
1301	&	CD	&	$2.0^{+0.09}_{-0.09}$	&	$500^{f}$	&	$0.906^{+0.007}_{-0.007}$	&	$900^{+30}_{-30}$	&	...	&	...	&	1134.7/1280	&	$1.188^{+0.011}_{-0.012}$	&	$0.19^{+0.03}_{-0.03}$	&	$0.891^{+0.014}_{-0.013}$	\\
1303	&	CD	&	$2.2^{+0.2}_{-0.2}$	&	$500^{f}$	&	$0.91^{+0.01}_{-0.01}$	&	$900^{+50}_{-50}$	&	...	&	...	&	944.3/1125	&	$1.17^{+0.02}_{-0.02}$	&	$0.25^{+0.07}_{-0.06}$	&	$0.87^{+0.03}_{-0.02}$	\\
1501	&	CD	&	$1.95^{+0.08}_{-0.08}$	&	$500^{f}$	&	$0.900^{+0.006}_{-0.006}$	&	$1000^{+30}_{-30}$	&	...	&	...	&	1213.4/1285	&	$1.273^{+0.009}_{-0.010}$	&	$0.11^{+0.02}_{-0.02}$	&	$0.939^{+0.011}_{-0.011}$	\\
1601	&	CD	&	$2.05^{+0.06}_{-0.06}$	&	$500^{f}$	&	$0.904^{+0.005}_{-0.005}$	&	$1000^{+30}_{-30}$	&	...	&	...	&	1322.7/1347	&	$1.282^{+0.009}_{-0.010}$	&	$0.20^{+0.02}_{-0.02}$	&	$0.897^{+0.010}_{-0.010}$	\\
1701	&	CD	&	$1.81^{+0.14}_{-0.14}$	&	$500^{f}$	&	$0.987^{+0.009}_{-0.010}$	&	$900^{+40}_{-40}$	&	...	&	...	&	1047.1/1120	&	$1.70^{+0.02}_{-0.02}$	&	$0.15^{+0.04}_{-0.03}$	&	$0.94^{+0.02}_{-0.02}$	\\
1702	&	CD	&	$1.7^{+0.2}_{-0.2}$	&	$500^{f}$	&	$0.976^{+0.005}_{-0.006}$	&	$1000^{+30}_{-30}$	&	...	&	...	&	1188.9/1242	&	$1.800^{+0.011}_{-0.011}$	&	$0.10^{+0.02}_{-0.02}$	&	$0.981^{+0.009}_{-0.009}$	\\
1703	&	CD	&	$1.74^{+0.09}_{-0.20}$	&	$500^{f}$	&	$0.981^{+0.004}_{-0.005}$	&	$1000^{+20}_{-20}$	&	...	&	...	&	1252.6/1334	&	$1.803^{+0.009}_{-0.009}$	&	$0.113^{+0.016}_{-0.013}$	&	$0.971^{+0.007}_{-0.007}$	\\
1801	&	CD	&	$1.87^{+0.06}_{-0.06}$	&	$500^{f}$	&	$0.972^{+0.004}_{-0.004}$	&	$1100^{+20}_{-20}$	&	...	&	...	&	1615.0/1478	&	$1.928^{+0.009}_{-0.009}$	&	$0.24^{+0.03}_{-0.03}$	&	$0.906^{+0.007}_{-0.007}$	\\
1901	&	CD	&	$1.67^{+0.21}_{-0.07}$	&	$500^{f}$	&	$1.009^{+0.003}_{-0.003}$	&	$1000^{+14}_{-14}$	&	...	&	...	&	1400.2/1433	&	$2.031^{+0.007}_{-0.007}$	&	$0.140^{+0.017}_{-0.014}$	&	$0.970^{+0.005}_{-0.005}$	\\
2201	&	CD	&	$1.7^{+0.2}_{-0.3}$	&	$500^{f}$	&	$0.991^{+0.005}_{-0.005}$	&	$1100^{+20}_{-20}$	&	...	&	...	&	1121.2/1261	&	$2.014^{+0.011}_{-0.011}$	&	$0.069^{+0.024}_{-0.015}$	&	$0.990^{+0.008}_{-0.008}$	\\
2202	&	CD	&	$1.59^{+0.27}_{-0.09}$	&	$500^{f}$	&	$0.985^{+0.005}_{-0.005}$	&	$1100^{+30}_{-30}$	&	...	&	...	&	1129.2/1212	&	$1.995^{+0.012}_{-0.012}$	&	$0.073^{+0.020}_{-0.013}$	&	$0.992^{+0.008}_{-0.008}$	\\
2204	&	CD	&	$1.6^{+0.3}_{-0.2}$	&	$500^{f}$	&	$0.980^{+0.004}_{-0.004}$	&	$1100^{+20}_{-20}$	&	...	&	...	&	1332.5/1241	&	$1.944^{+0.010}_{-0.010}$	&	$0.07^{+0.02}_{-0.02}$	&	$0.994^{+0.008}_{-0.007}$	\\
2205	&	CD	&	$1.7^{+0.2}_{-0.2}$	&	$500^{f}$	&	$0.986^{+0.003}_{-0.003}$	&	$1100^{+16}_{-16}$	&	...	&	...	&	1361.1/1355	&	$1.935^{+0.007}_{-0.007}$	&	$0.08^{+0.02}_{-0.02}$	&	$0.994^{+0.005}_{-0.005}$	\\
2206	&	CD	&	$1.7^{+0.3}_{-0.2}$	&	$500^{f}$	&	$0.989^{+0.005}_{-0.005}$	&	$1100^{+20}_{-20}$	&	...	&	...	&	1219.1/1225	&	$1.960^{+0.011}_{-0.011}$	&	$0.09^{+0.03}_{-0.02}$	&	$0.990^{+0.008}_{-0.008}$	\\
2208	&	CD	&	$1.7^{+0.3}_{-0.2}$	&	$500^{f}$	&	$0.999^{+0.006}_{-0.006}$	&	$1000^{+30}_{-30}$	&	...	&	...	&	1091.1/1192	&	$1.908^{+0.013}_{-0.013}$	&	$0.051^{+0.019}_{-0.011}$	&	$0.994^{+0.010}_{-0.010}$	\\
2209	&	CD	&	$1.6^{+0.3}_{-0.2}$	&	$500^{f}$	&	$0.987^{+0.004}_{-0.004}$	&	$1100^{+20}_{-20}$	&	...	&	...	&	1180.3/1259	&	$1.931^{+0.010}_{-0.010}$	&	$0.074^{+0.025}_{-0.013}$	&	$0.991^{+0.007}_{-0.007}$	\\
2210	&	CD	&	$1.8^{+0.2}_{-0.2}$	&	$500^{f}$	&	$0.986^{+0.006}_{-0.006}$	&	$1100^{+30}_{-30}$	&	...	&	...	&	1151.6/1149	&	$1.936^{+0.014}_{-0.014}$	&	$0.08^{+0.03}_{-0.02}$	&	$0.989^{+0.011}_{-0.010}$	\\
2211	&	CD	&	$1.6^{+0.2}_{-0.2}$	&	$500^{f}$	&	$0.981^{+0.005}_{-0.005}$	&	$1100^{+30}_{-30}$	&	...	&	...	&	1093.9/1196	&	$1.926^{+0.012}_{-0.012}$	&	$0.08^{+0.03}_{-0.02}$	&	$0.990^{+0.009}_{-0.009}$	\\
\hline
2212	&	CD	&	$1.6^{+0.4}_{-0.2}$	&	$500^{f}$	&	$0.982^{+0.003}_{-0.004}$	&	$1100^{+19}_{-18}$	&	...	&	...	&	1278.0/1315	&	$1.925^{+0.008}_{-0.008}$	&	$0.061^{+0.020}_{-0.013}$	&	$0.994^{+0.006}_{-0.006}$	\\
2213	&	CD	&	$1.6^{+0.3}_{-0.2}$	&	$500^{f}$	&	$0.981^{+0.004}_{-0.004}$	&	$1100^{+20}_{-20}$	&	...	&	...	&	1189.9/1284	&	$1.941^{+0.009}_{-0.009}$	&	$0.08^{+0.02}_{-0.02}$	&	$0.990^{+0.007}_{-0.007}$	\\
2214	&	CD	&	$1.6^{+0.2}_{-0.3}$	&	$500^{f}$	&	$0.981^{+0.006}_{-0.006}$	&	$1100^{+30}_{-30}$	&	...	&	...	&	989.4/1095	&	$1.939^{+0.013}_{-0.013}$	&	$0.06^{+0.03}_{-0.02}$	&	$0.995^{+0.010}_{-0.010}$	\\
2215	&	CD	&	$1.6^{+0.2}_{-0.2}$	&	$500^{f}$	&	$0.945^{+0.009}_{-0.009}$	&	$1300^{+50}_{-50}$	&	...	&	...	&	1312.9/1159	&	$1.92^{+0.02}_{-0.03}$	&	$0.154^{+0.013}_{-0.010}$	&	$0.902^{+0.013}_{-0.014}$	\\
2301	&	CD	&	$1.7^{+0.2}_{-0.1}$	&	$500^{f}$	&	$0.990^{+0.004}_{-0.004}$	&	$100^{+17}_{-16}$	&	...	&	...	&	1327.1/1407	&	$1.892^{+0.008}_{-0.008}$	&	$0.134^{+0.017}_{-0.014}$	&	$0.965^{+0.006}_{-0.006}$	\\
2401	&	CD	&	$1.99^{+0.07}_{-0.07}$	&	$500^{f}$	&	$0.965^{+0.004}_{-0.004}$	&	$1100^{+20}_{-19}$	&	...	&	...	&	1257.3/1377	&	$1.797^{+0.009}_{-0.009}$	&	$0.19^{+0.03}_{-0.03}$	&	$0.940^{+0.008}_{-0.008}$	\\
2601	&	CD	&	$1.6^{+0.3}_{-0.2}$	&	$500^{f}$	&	$0.951^{+0.003}_{-0.003}$	&	$1100^{+16}_{-16}$	&	...	&	...	&	1422.5/1423	&	$1.775^{+0.006}_{-0.006}$	&	$0.08^{+0.06}_{-0.03}$	&	$0.995^{+0.005}_{-0.005}$	\\
2701	&	CD	&	$1.76^{+0.12}_{-0.12}$	&	$500^{f}$	&	$0.949^{+0.003}_{-0.003}$	&	$1100^{+17}_{-16}$	&	...	&	...	&	1393.7/1464	&	$1.758^{+0.006}_{-0.006}$	&	$0.10^{+0.03}_{-0.02}$	&	$0.981^{+0.006}_{-0.006}$	\\
2801	&	CD	&	$1.8^{+0.2}_{-0.3}$	&	$500^{f}$	&	$0.938^{+0.003}_{-0.003}$	&	$1200^{+19}_{-19}$	&	...	&	...	&	1361.1/1376	&	$1.703^{+0.007}_{-0.007}$	&	$0.07^{+0.02}_{-0.02}$	&	$0.991^{+0.006}_{-0.006}$	\\
2901	&	CD	&	$1.7^{+0.3}_{-0.3}$	&	$500^{f}$	&	$0.920^{+0.005}_{-0.005}$	&	$1200^{+30}_{-30}$	&	...	&	...	&	1120.2/1197	&	$1.597^{+0.011}_{-0.011}$	&	$0.045^{+0.028}_{-0.014}$	&	$0.993^{+0.010}_{-0.010}$	\\
3001	&	CD	&	$1.7^{+0.3}_{-0.3}$	&	$500^{f}$	&	$0.917^{+0.004}_{-0.005}$	&	$1200^{+30}_{-30}$	&	...	&	...	&	1194.6/1238	&	$1.565^{+0.009}_{-0.009}$	&	$0.08^{+0.07}_{-0.04}$	&	$0.993^{+0.009}_{-0.009}$	\\
3102	&	CD	&	$1.6^{+0.4}_{-0.3}$	&	$500^{f}$	&	$0.900^{+0.004}_{-0.004}$	&	$1200^{+30}_{-30}$	&	...	&	...	&	1098.9/1234	&	$1.532^{+0.01}_{-0.01}$	&	$0.05^{+0.06}_{-0.02}$	&	$0.995^{+0.009}_{-0.009}$	\\
3103	&	CD	&	$1.9^{+0.3}_{-0.3}$	&	$500^{f}$	&	$0.890^{+0.006}_{-0.006}$	&	$1300^{+40}_{-40}$	&	...	&	...	&	1094.3/1195	&	$1.537^{+0.011}_{-0.011}$	&	$0.07^{+0.08}_{-0.03}$	&	$0.986^{+0.014}_{-0.012}$	\\
3203	&	CD	&	$1.8^{+0.5}_{-0.5}$	&	$500^{f}$	&	$0.884^{+0.006}_{-0.006}$	&	$1300^{+40}_{-40}$	&	...	&	...	&	1084.7/1177	&	$1.499^{+0.011}_{-0.011}$	&	$0.08^{+0.18}_{-0.05}$	&	$0.993^{+0.013}_{-0.011}$	\\
3301	&	CD	&	$1.94^{+0.15}_{-0.15}$	&	$500^{f}$	&	$0.872^{+0.005}_{-0.005}$	&	$1300^{+30}_{-30}$	&	...	&	...	&	1287.7/1304	&	$1.427^{+0.009}_{-0.009}$	&	$0.11^{+0.03}_{-0.02}$	&	$0.957^{+0.011}_{-0.010}$	\\
3302	&	CD	&	$1.8^{+0.2}_{-0.2}$	&	$500^{f}$	&	$0.887^{+0.005}_{-0.005}$	&	$1200^{+30}_{-30}$	&	...	&	...	&	1150.9/1277	&	$1.415^{+0.008}_{-0.008}$	&	$0.07^{+0.04}_{-0.02}$	&	$0.981^{+0.010}_{-0.009}$	\\
3303	&	CD	&	$1.9^{+0.2}_{-0.2}$	&	$500^{f}$	&	$0.882^{+0.007}_{-0.007}$	&	$1200^{+40}_{-40}$	&	...	&	...	&	1079.1/1150	&	$1.430^{+0.012}_{-0.012}$	&	$0.09^{+0.05}_{-0.03}$	&	$0.974^{+0.015}_{-0.013}$	\\
3401	&	CD	&	$1.9^{+0.2}_{-0.2}$	&	$500^{f}$	&	$0.879^{+0.009}_{-0.009}$	&	$1200^{+60}_{-50}$	&	...	&	...	&	999.4/1150	&	$1.411^{+0.015}_{-0.015}$	&	$0.15^{+0.05}_{-0.04}$	&	$0.93^{+0.02}_{-0.02}$	\\
3402	&	CD	&	$1.9^{+0.11}_{-0.11}$	&	$500^{f}$	&	$0.868^{+0.008}_{-0.009}$	&	$1300^{+60}_{-50}$	&	...	&	...	&	940.5/1127	&	$1.386^{+0.015}_{-0.015}$	&	$0.13^{+0.03}_{-0.03}$	&	$0.94^{+0.02}_{-0.02}$	\\
3403	&	CD	&	$1.7^{+0.2}_{-0.2}$	&	$500^{f}$	&	$0.878^{+0.008}_{-0.009}$	&	$1200^{+60}_{-50}$	&	...	&	...	&	973.1/1094	&	$1.398^{+0.014}_{-0.014}$	&	$0.060^{+0.019}_{-0.012}$	&	$0.980^{+0.015}_{-0.015}$	\\
3701	&	CD	&	$1.88^{+0.14}_{-0.14}$	&	$500^{f}$	&	$0.867^{+0.006}_{-0.006}$	&	$1200^{+40}_{-40}$	&	...	&	...	&	1155.2/1211	&	$1.253^{+0.010}_{-0.010}$	&	$0.10^{+0.03}_{-0.02}$	&	$0.957^{+0.012}_{-0.012}$	\\
3702	&	CD	&	$1.92^{+0.13}_{-0.12}$	&	$500^{f}$	&	$0.867^{+0.005}_{-0.005}$	&	$1200^{+30}_{-30}$	&	...	&	...	&	1135.6/1309	&	$1.271^{+0.008}_{-0.008}$	&	$0.11^{+0.03}_{-0.02}$	&	$0.949^{+0.01}_{-0.01}$	\\
3802	&	CD	&	$2.4^{+0.4}_{-0.4}$	&	$500^{f}$	&	$0.843^{+0.007}_{-0.007}$	&	$1200^{+50}_{-60}$	&	...	&	...	&	1091.6/1155	&	$1.197^{+0.010}_{-0.011}$	&	$0.17^{+0.14}_{-0.10}$	&	$0.96^{+0.04}_{-0.02}$	\\
3803	&	CD	&	$1.9^{+0.5}_{-0.4}$	&	$500^{f}$	&	$0.848^{+0.005}_{-0.006}$	&	$1200^{+20}_{-40}$	&	...	&	...	&	985.5/1151	&	$1.20^{+0.01}_{-0.01}$	&	$0.06^{+0.18}_{-0.04}$	&	$0.992^{+0.015}_{-0.013}$	\\
3901	&	CD	&	$3.0^{+0.5}_{-0.6}$	&	$500^{f}$	&	$0.845^{+0.007}_{-0.006}$	&	$1100^{+50}_{-60}$	&	...	&	...	&	1008.9/1181	&	$1.10^{+0.02}_{-0.05}$	&	$0.3^{+0.5}_{-0.2}$	&	$0.94^{+0.07}_{-0.05}$	\\
3902	&	CD	&	$2.2^{+0.7}_{-0.5}$	&	$500^{f}$	&	$0.838^{+0.005}_{-0.005}$	&	$1200^{+30}_{-30}$	&	...	&	...	&	1033.9/1217	&	$1.132^{+0.008}_{-0.008}$	&	$0.21^{+0.23}_{-0.15}$	&	$0.990^{+0.032}_{-0.012}$	\\
4602	&	CD	&	$1.6^{+0.3}_{-0.3}$	&	$500^{f}$	&	$0.821^{+0.006}_{-0.006}$	&	$1300^{+50}_{-40}$	&	...	&	...	&	1026.0/1209	&	$1.084^{+0.009}_{-0.009}$	&	$0.039^{+0.027}_{-0.012}$	&	$0.983^{+0.013}_{-0.012}$	\\
4603	&	CD	&	$1.6^{+0.3}_{-0.2}$	&	$500^{f}$	&	$0.825^{+0.006}_{-0.006}$	&	$1300^{+40}_{-40}$	&	...	&	...	&	1124.0/1195	&	$1.084^{+0.009}_{-0.009}$	&	$0.033^{+0.016}_{-0.009}$	&	$0.987^{+0.013}_{-0.012}$	\\
4701	&	CD	&	$1.9^{+0.3}_{-0.2}$	&	$500^{f}$	&	$0.818^{+0.006}_{-0.006}$	&	$1200^{+40}_{-40}$	&	...	&	...	&	1009.0/1200	&	$1.032^{+0.008}_{-0.009}$	&	$0.06^{+0.04}_{-0.02}$	&	$0.967^{+0.015}_{-0.014}$	\\
4702	&	CD	&	$2.3^{+0.3}_{-0.3}$	&	$500^{f}$	&	$0.816^{+0.005}_{-0.005}$	&	$1200^{+40}_{-50}$	&	...	&	...	&	1160.7/1320	&	$1.043^{+0.007}_{-0.008}$	&	$0.16^{+0.11}_{-0.07}$	&	$0.94^{+0.03}_{-0.02}$	\\
4703	&	CD	&	$2.8^{+0.2}_{-0.3}$	&	$500^{f}$	&	$0.822^{+0.009}_{-0.008}$	&	$1100^{+80}_{-80}$	&	...	&	...	&	1110.5/1255	&	$1.021^{+0.013}_{-0.019}$	&	$0.29^{+0.11}_{-0.10}$	&	$0.86^{+0.06}_{-0.05}$	\\
4801	&	CD	&	$2.1^{+0.2}_{-0.2}$	&	$500^{f}$	&	$0.811^{+0.007}_{-0.007}$	&	$1200^{+50}_{-50}$	&	...	&	...	&	1027.2/1241	&	$0.979^{+0.010}_{-0.010}$	&	$0.13^{+0.04}_{-0.03}$	&	$0.91^{+0.02}_{-0.02}$	\\
4803	&	CD	&	$2.1^{+0.2}_{-0.2}$	&	$500^{f}$	&	$0.796^{+0.006}_{-0.006}$	&	$1300^{+50}_{-50}$	&	...	&	...	&	1121.5/1261	&	$0.998^{+0.010}_{-0.010}$	&	$0.14^{+0.04}_{-0.03}$	&	$0.91^{+0.02}_{-0.02}$	\\
4901	&	CD	&	$2.2^{+0.2}_{-0.2}$	&	$500^{f}$	&	$0.807^{+0.005}_{-0.005}$	&	$1200^{+40}_{-40}$	&	...	&	...	&	1180.2/1316	&	$0.970^{+0.007}_{-0.008}$	&	$0.13^{+0.05}_{-0.03}$	&	$0.93^{+0.02}_{-0.02}$	\\
4902	&	CD	&	$2.5^{+0.2}_{-0.2}$	&	$500^{f}$	&	$0.802^{+0.006}_{-0.006}$	&	$1100^{+50}_{-60}$	&	...	&	...	&	1099.1/1327	&	$0.932^{+0.011}_{-0.013}$	&	$0.26^{+0.07}_{-0.06}$	&	$0.85^{+0.03}_{-0.03}$	\\
4903	&	CD	&	$2.1^{+0.2}_{-0.2}$	&	$500^{f}$	&	$0.789^{+0.008}_{-0.008}$	&	$1300^{+60}_{-60}$	&	...	&	...	&	1062.9/1197	&	$0.945^{+0.013}_{-0.015}$	&	$0.17^{+0.05}_{-0.04}$	&	$0.88^{+0.02}_{-0.02}$	\\
\bottomrule
\end{longtable*}
\tablecomments{$^a$ CD: \texttt{cutoffpl+diskbb} model; $^b$ CDG: \texttt{cutoffpl+diskbb+gaussian} model;  $^c$ The central energy of the \texttt{gaussian} model; $^d$ Line width of the \texttt{gaussian} model; $^e$ Units: ($10^{-8}\ {\rm erg\ cm^{-2}\ s^{-1}}$); $^f$ $E_{\rm cut}$ fixed at 500~keV in the SIMS and HSS; $^p$ The parameter pegs at its lower or upper limit; $^u$ This value is unconstrained.}

\clearpage
%\onecolumn
\setlength{\tabcolsep}{2pt}
\begin{longtable*}{ccccccccccccc}
	\caption{Spectral fitting results of EXO~1846--031 using \texttt{TBabs*(simplcut*diskbb)} model, $N_{\rm H}$ fixed at $5.25 \times 10^{22}$ cm$^{-2}$.}\\
	\label{tab:fitting_1sc1d} \\
	\toprule
	\toprule
	 ExpID	&	Model	&	$\Gamma$	&	$E_{\rm cut}$  &	k$T_{\rm in}$	&	$N_{\rm dbb}$	&  $f^{g}_{\rm sc}$ &	$E_{\rm gau}^{c}$	& $\sigma_{\rm gau}^{d}$ &  $\chi^2$/dof  &  $F_{\rm dbb}^{e}$  &  $F_{\rm pl}^{e}$  &  $f_{\rm dbb}$  		\\
	  &	  &		&	(keV)	  &	(keV)	&	&	&	(keV)	&  (keV) & &   &  &    \\
\hline
0101	&	SD$^{a}$	&	$1.45^{+0.04}_{-0.04}$	&	$60^{+4}_{-3}$	&	$1.02^{+0.15}_{-0.16}$	&	$150^{+120}_{-60}$	&	$0.74^{+0.06}_{-0.06}$	&	...	&	...	&	1300.6/1302	&	$0.36^{+0.03}_{-0.03}$	&	$1.25^{+0.06}_{-0.06}$	&	$0.22^{+0.02}_{-0.02}$	\\
0102	&	SD	&	$1.47^{+0.04}_{-0.04}$	&	$50^{+4}_{-4}$	&	$0.7^{+0.2}_{-0.3}$	&	$600^{+600}_{-300}$	&	$0.85^{+0.06}_{-0.11}$	&	...	&	...	&	1350.8/1443	&	$0.30^{+0.03}_{-0.08}$	&	$1.35^{+0.05}_{-0.09}$	&	$0.18^{+0.02}_{-0.05}$	\\
0103	&	SD	&	$1.53^{+0.02}_{-0.04}$	&	$50^{+3}_{-3}$	&	$0.67^{+0.10}_{-0.06}$	&	$600^{+200}_{-300}$	&	$0.91^{+0.05}_{-0.07}$	&	...	&	...	&	1358.3/1442	&	$0.27^{+0.06}_{-0.04}$	&	$1.32^{+0.07}_{-0.06}$	&	$0.17^{+0.04}_{-0.03}$	\\
0104	&	SD	&	$1.59^{+0.05}_{-0.05}$	&	$60^{+7}_{-6}$	&	$0.8^{+0.2}_{-0.3}$	&	$300^{+400}_{-150}$	&	$0.83^{+0.09}_{-0.09}$	&	...	&	...	&	1431.5/1401	&	$0.36^{+0.03}_{-0.06}$	&	$1.19^{+0.05}_{-0.07}$	&	$0.23^{+0.02}_{-0.04}$	\\
0105	&	SD	&	$1.61^{+0.02}_{-0.02}$	&	$60^{+6}_{-5}$	&	$0.75^{+0.05}_{-0.06}$	&	$500^{+100}_{-100}$	&	$0.75^{+0.06}_{-0.04}$	&	...	&	...	&	1411.3/1470	&	$0.25^{+0.12}_{-0.05}$	&	$1.37^{+0.12}_{-0.07}$	&	$0.16^{+0.07}_{-0.03}$	\\
0106	&	SD	&	$1.60^{+0.04}_{-0.04}$	&	$60^{+6}_{-5}$	&	$0.81^{+0.12}_{-0.12}$	&	$400^{+300}_{-170}$	&	$0.72^{+0.06}_{-0.05}$	&	...	&	...	&	1238.3/1224	&	$0.42^{+0.03}_{-0.03}$	&	$1.25^{+0.05}_{-0.06}$	&	$0.25^{+0.02}_{-0.02}$	\\
0107	&	SD	&	$1.70^{+0.07}_{-0.07}$	&	$70^{+19}_{-13}$	&	$0.76^{+0.11}_{-0.16}$	&	$600^{+400}_{-200}$	&	$0.72^{+0.11}_{-0.07}$	&	...	&	...	&	1245.7/1281	&	$0.43^{+0.03}_{-0.04}$	&	$1.16^{+0.06}_{-0.07}$	&	$0.27^{+0.02}_{-0.03}$	\\
0201	&	SD	&	$1.75^{+0.04}_{-0.04}$	&	$50^{+6}_{-5}$	&	$0.78^{+0.05}_{-0.05}$	&	$800^{+140}_{-170}$	&	$0.62^{+0.04}_{-0.04}$	&	...	&	...	&	1218.3/1342	&	$0.64^{+0.02}_{-0.02}$	&	$1.15^{+0.04}_{-0.04}$	&	$0.356^{+0.011}_{-0.012}$	\\
0301	&	SDG$^{b}$	&	$1.93^{+0.04}_{-0.05}$	&	$80^{+13}_{-10}$	&	$0.93^{+0.04}_{-0.04}$	&	$600^{+100}_{-80}$	&	$0.50^{+0.04}_{-0.04}$	&	$6.4^{+0.2}_{-p}$	&	$0.6^{+0.4}_{-0.2}$	&	1256.9/1325	&	$0.864^{+0.015}_{-0.016}$	&	$1.07^{+0.04}_{-0.04}$	&	$0.446^{+0.012}_{-0.012}$	\\
0302	&	SDG	&	$1.76^{+0.07}_{-0.08}$	&	$50^{+8}_{-6}$	&	$1.00^{+0.03}_{-0.03}$	&	$400^{+60}_{-50}$	&	$0.37^{+0.03}_{-0.05}$	&	$6.5^{+0.2}_{-0.1}$	&	$1.3^{+0.2}_{-0.3}$	&	1257.8/1378	&	$0.857^{+0.013}_{-0.014}$	&	$1.03^{+0.04}_{-0.04}$	&	$0.455^{+0.011}_{-0.011}$	\\
0303	&	SDG	&	$1.81^{+0.12}_{-0.12}$	&	$50^{+15}_{-10}$	&	$0.96^{+0.04}_{-0.05}$	&	$600^{+120}_{-80}$	&	$0.32^{+0.07}_{-0.06}$	&	$6.4^{+0.4}_{-p}$	&	$1.6^{+0.2}_{-0.3}$	&	1006.9/1137	&	$0.94^{+0.02}_{-0.02}$	&	$1.06^{+0.06}_{-0.06}$	&	$0.47^{+0.02}_{-0.02}$	\\
0502	&	SDG	&	$1.94^{+0.06}_{-0.07}$	&	$60^{+14}_{-10}$	&	$0.99^{+0.03}_{-0.03}$	&	$500^{+70}_{-60}$	&	$0.38^{+0.05}_{-0.05}$	&	$6.6^{+0.3}_{-0.1}$	&	$1.1^{+0.3}_{-0.3}$	&	1236.3/1323	&	$1.056^{+0.014}_{-0.014}$	&	$0.99^{+0.04}_{-0.04}$	&	$0.516^{+0.012}_{-0.012}$	\\
0503	&	SDG	&	$1.81^{+0.10}_{-0.11}$	&	$50^{+12}_{-9}$	&	$1.00^{+0.04}_{-0.03}$	&	$500^{+60}_{-70}$	&	$0.33^{+0.06}_{-0.06}$	&	$6.6^{+0.3}_{-0.1}$	&	$1.2^{+0.3}_{-0.4}$	&	1130.7/1264	&	$1.03^{+0.02}_{-0.02}$	&	$0.96^{+0.05}_{-0.05}$	&	$0.517^{+0.014}_{-0.014}$	\\
0601	&	SDG	&	$2.06^{+0.12}_{-0.14}$	&	$90^{+80}_{-30}$	&	$0.97^{+0.03}_{-0.03}$	&	$700^{+90}_{-70}$	&	$0.34^{+0.06}_{-0.06}$	&	$6.8^{+0.1}_{-0.4}$	&	$1.5^{+0.2}_{-0.3}$	&	1301.4/1394	&	$1.217^{+0.013}_{-0.013}$	&	$1.04^{+0.05}_{-0.05}$	&	$0.540^{+0.013}_{-0.013}$	\\
0701	&	SDG	&	$2.21^{+0.07}_{-0.08}$	&	$90^{+40}_{-20}$	&	$1.07^{+0.014}_{-0.014}$	&	$600^{+30}_{-30}$	&	$0.28^{+0.02}_{-0.02}$	&	$6.7^{+0.1}_{-0.1}$	&	$0.4^{+0.2}_{-0.4}$	&	1294.0/1435	&	$1.758^{+0.010}_{-0.010}$	&	$0.75^{+0.03}_{-0.03}$	&	$0.701^{+0.009}_{-0.009}$	\\
0703	&	SDG	&	$2.06^{+0.09}_{-0.10}$	&	$70^{+30}_{-16}$	&	$1.07^{+0.02}_{-0.02}$	&	$600^{+40}_{-40}$	&	$0.28^{+0.03}_{-0.03}$	&	$6.6^{+0.2}_{-0.2}$	&	$0.1^{+0.1}_{-0.6}$	&	1216.5/1351	&	$1.534^{+0.012}_{-0.012}$	&	$0.78^{+0.04}_{-0.04}$	&	$0.663^{+0.012}_{-0.012}$	\\
0802	&	SDG	&	$2.09^{+0.08}_{-0.09}$	&	$90^{+40}_{-20}$	&	$1.136^{+0.011}_{-0.011}$	&	$600^{+20}_{-20}$	&	$0.16^{+0.02}_{-0.02}$	&	$6.8^{+0.2}_{-0.4}$	&	$0.5^{+0.4}_{-0.5}$	&	1320.0/1417	&	$2.038^{+0.010}_{-0.010}$	&	$0.58^{+0.03}_{-0.03}$	&	$0.778^{+0.009}_{-0.008}$	\\
0803	&	SDG	&	$2.12^{+0.11}_{-0.12}$	&	$120^{+190}_{-50}$	&	$1.11^{+0.02}_{-0.02}$	&	$600^{+40}_{-30}$	&	$0.17^{+0.02}_{-0.02}$	&	$6.4^{+0.4}_{-p}$	&	$0.1^{+4.2}_{-0.1}$	&	1209.3/1299	&	$1.933^{+0.013}_{-0.013}$	&	$0.59^{+0.04}_{-0.04}$	&	$0.766^{+0.014}_{-0.013}$	\\
0901	&	SDG	&	$1.93^{+0.13}_{-0.16}$	&	$50^{+20}_{-12}$	&	$0.97^{+0.02}_{-0.02}$	&	$800^{+60}_{-60}$	&	$0.20^{+0.04}_{-0.05}$	&	$6.6^{+0.3}_{-0.1}$	&	$1.1^{+0.4}_{-0.5}$	&	1262.7/1329	&	$1.515^{+0.012}_{-0.012}$	&	$0.69^{+0.04}_{-0.04}$	&	$0.687^{+0.013}_{-0.012}$	\\
0902	&	SDG	&	$2.10^{+0.07}_{-0.07}$	&	$90^{+40}_{-20}$	&	$0.94^{+0.04}_{-0.02}$	&	$800^{+60}_{-100}$	&	$0.31^{+0.03}_{-0.12}$	&	$6.8^{+0.2}_{-0.2}$	&	$0.4^{+1.2}_{-0.4}$	&	1325.2/1423	&	$1.356^{+0.009}_{-0.009}$	&	$0.79^{+0.03}_{-0.03}$	&	$0.632^{+0.010}_{-0.010}$	\\
1001	&	SDG	&	$2.18^{+0.04}_{-0.07}$	&	$400^{+p}_{-200}$	&	$0.90^{+0.02}_{-0.02}$	&	$800^{+60}_{-60}$	&	$0.31^{+0.03}_{-0.04}$	&	$6.4^{+0.1}_{-p}$	&	$1.1^{+0.2}_{-0.2}$	&	1373.2/1469	&	$1.109^{+0.007}_{-0.007}$	&	$0.79^{+0.03}_{-0.03}$	&	$0.584^{+0.008}_{-0.008}$	\\
1002	&	SDG	&	$2.19^{+0.08}_{-0.08}$	&	$140^{+120}_{-50}$	&	$0.85^{+0.02}_{-0.02}$	&	$1000^{+120}_{-90}$	&	$0.35^{+0.05}_{-0.04}$	&	$6.6^{+0.2}_{-0.1}$	&	$1.1^{+0.2}_{-0.2}$	&	1303.8/1424	&	$1.033^{+0.009}_{-0.010}$	&	$0.76^{+0.03}_{-0.03}$	&	$0.577^{+0.011}_{-0.011}$	\\
1003	&	SDG	&	$2.03^{+0.16}_{-0.19}$	&	$80^{+150}_{-40}$	&	$0.87^{+0.03}_{-0.04}$	&	$900^{+160}_{-120}$	&	$0.26^{+0.07}_{-0.06}$	&	$6.4^{+0.1}_{-p}$	&	$1.3^{+0.2}_{-0.3}$	&	1117.2/1256	&	$1.044^{+0.014}_{-0.014}$	&	$0.75^{+0.06}_{-0.05}$	&	$0.58^{+0.02}_{-0.02}$	\\
1101	&	SDG	&	$2.03^{+0.10}_{-0.11}$	&	$70^{+30}_{-17}$	&	$1.001^{+0.011}_{-0.010}$	&	$700^{+30}_{-30}$	&	$0.14^{+0.02}_{-0.02}$	&	$6.6^{+0.2}_{-0.2}$	&	$1.2^{+0.2}_{-0.2}$	&	1289.1/1461	&	$1.433^{+0.007}_{-0.007}$	&	$0.46^{+0.02}_{-0.02}$	&	$0.756^{+0.009}_{-0.009}$	\\
1102	&	SDG	&	$2.09^{+0.12}_{-0.14}$	&	$70^{+40}_{-20}$	&	$1.001^{+0.012}_{-0.011}$	&	$700^{+30}_{-30}$	&	$0.17^{+0.03}_{-0.03}$	&	$6.4^{+0.2}_{-p}$	&	$0.8^{+0.4}_{-0.3}$	&	1222.9/1443	&	$1.482^{+0.008}_{-0.008}$	&	$0.47^{+0.03}_{-0.03}$	&	$0.760^{+0.011}_{-0.010}$	\\
1103	&	SDG	&	$2.20^{+0.07}_{-0.08}$	&	$500^{+p}_{-300}$	&	$1.021^{+0.012}_{-0.013}$	&	$700^{+30}_{-30}$	&	$0.15^{+0.02}_{-0.02}$	&	$6.4^{+0.4}_{-p}$	&	$1.2^{+0.2}_{-0.3}$	&	1149.3/1352	&	$1.489^{+0.009}_{-0.009}$	&	$0.48^{+0.03}_{-0.03}$	&	$0.755^{+0.012}_{-0.012}$	\\
1301	&	SD	&	$2.01^{+0.09}_{-0.09}$	&	$500^{f}$	&	$0.900^{+0.008}_{-0.008}$	&	$1000^{+40}_{-40}$	&	$0.054^{+0.008}_{-0.007}$	&	...	&	...	&	1147.3/1280	&	$1.384^{+0.01}_{-0.01}$	&	$0.20^{+0.03}_{-0.03}$	&	$0.87^{+0.02}_{-0.02}$	\\
1303	&	SD	&	$2.3^{+0.2}_{-0.2}$	&	$500^{f}$	&	$0.898^{+0.012}_{-0.012}$	&	$1000^{+60}_{-60}$	&	$0.066^{+0.018}_{-0.014}$	&	...	&	...	&	947.5/1125	&	$1.363^{+0.014}_{-0.014}$	&	$0.15^{+0.04}_{-0.03}$	&	$0.90^{+0.02}_{-0.02}$	\\
1501	&	SD	&	$1.95^{+0.09}_{-0.08}$	&	$500^{f}$	&	$0.898^{+0.007}_{-0.007}$	&	$1000^{+40}_{-40}$	&	$0.029^{+0.005}_{-0.005}$	&	...	&	...	&	1229.0/1285	&	$1.457^{+0.01}_{-0.01}$	&	$0.13^{+0.02}_{-0.02}$	&	$0.920^{+0.013}_{-0.012}$	\\
1601	&	SD	&	$2.06^{+0.07}_{-0.06}$	&	$500^{f}$	&	$0.899^{+0.006}_{-0.006}$	&	$1100^{+30}_{-30}$	&	$0.050^{+0.005}_{-0.005}$	&	...	&	...	&	1338.5/1347	&	$1.492^{+0.009}_{-0.009}$	&	$0.18^{+0.02}_{-0.02}$	&	$0.893^{+0.011}_{-0.011}$	\\
1701	&	SD	&	$1.82^{+0.15}_{-0.14}$	&	$500^{f}$	&	$0.985^{+0.010}_{-0.011}$	&	$1000^{+50}_{-40}$	&	$0.027^{+0.008}_{-0.007}$	&	...	&	...	&	1048.9/1120	&	$1.92^{+0.02}_{-0.02}$	&	$0.20^{+0.05}_{-0.04}$	&	$0.91^{+0.02}_{-0.02}$	\\
1702	&	SD	&	$1.65^{+0.15}_{-0.15}$	&	$500^{f}$	&	$0.977^{+0.006}_{-0.006}$	&	$1000^{+30}_{-30}$	&	$0.011^{+0.003}_{-0.003}$	&	...	&	...	&	1210.6/1242	&	$2.001^{+0.012}_{-0.012}$	&	$0.15^{+0.03}_{-0.03}$	&	$0.930^{+0.014}_{-0.012}$	\\
1703	&	SD	&	$1.73^{+0.10}_{-0.10}$	&	$500^{f}$	&	$0.982^{+0.005}_{-0.005}$	&	$1000^{+20}_{-20}$	&	$0.015^{+0.003}_{-0.002}$	&	...	&	...	&	1283.4/1334	&	$2.011^{+0.010}_{-0.010}$	&	$0.18^{+0.03}_{-0.02}$	&	$0.920^{+0.011}_{-0.010}$	\\
1801	&	SD	&	$1.90^{+0.07}_{-0.07}$	&	$500^{f}$	&	$0.967^{+0.005}_{-0.005}$	&	$1200^{+30}_{-20}$	&	$0.055^{+0.005}_{-0.004}$	&	...	&	...	&	1650.3/1478	&	$2.197^{+0.008}_{-0.008}$	&	$0.39^{+0.04}_{-0.03}$	&	$0.851^{+0.012}_{-0.011}$	\\
1901	&	SD	&	$1.76^{+0.18}_{-0.08}$	&	$500^{f}$	&	$1.010^{+0.004}_{-0.004}$	&	$1000^{+17}_{-16}$	&	$0.016^{+0.002}_{-0.002}$	&	...	&	...	&	1466.7/1433	&	$2.260^{+0.008}_{-0.008}$	&	$0.19^{+0.02}_{-0.02}$	&	$0.922^{+0.008}_{-0.008}$	\\
2201	&	SD	&	$1.62^{+0.24}_{-0.24}$	&	$500^{f}$	&	$0.993^{+0.005}_{-0.005}$	&	$1100^{+30}_{-20}$	&	$0.006^{+0.003}_{-0.002}$	&	...	&	...	&	1134.4/1261	&	$2.231^{+0.012}_{-0.012}$	&	$0.10^{+0.03}_{-0.03}$	&	$0.957^{+0.014}_{-0.011}$	\\
2202	&	SD	&	$1.6^{+0.2}_{-0.3}$	&	$500^{f}$	&	$0.987^{+0.005}_{-0.005}$	&	$1100^{+30}_{-30}$	&	$0.005^{+0.002}_{-0.001}$	&	...	&	...	&	1140.3/1212	&	$2.210^{+0.013}_{-0.013}$	&	$0.11^{+0.03}_{-0.03}$	&	$0.951^{+0.013}_{-0.012}$	\\
2204	&	SD	&	$1.6^{+0.2}_{-0.2}$	&	$500^{f}$	&	$0.981^{+0.004}_{-0.004}$	&	$1100^{+20}_{-20}$	&	$0.003^{+0.001}_{-0.001}$	&	...	&	...	&	1365.4/1241	&	$2.154^{+0.012}_{-0.012}$	&	$0.09^{+0.03}_{-0.03}$	&	$0.958^{+0.015}_{-0.012}$	\\
2205	&	SD	&	$1.7^{+0.3}_{-0.3}$	&	$500^{f}$	&	$0.988^{+0.003}_{-0.003}$	&	$1100^{+20}_{-20}$	&	$0.004^{+0.001}_{-0.001}$	&	...	&	...	&	1413.3/1355	&	$2.142^{+0.008}_{-0.008}$	&	$0.13^{+0.02}_{-0.03}$	&	$0.941^{+0.007}_{-0.013}$	\\
2206	&	SD	&	$1.6^{+0.2}_{-0.2}$	&	$500^{f}$	&	$0.990^{+0.005}_{-0.005}$	&	$1100^{+30}_{-20}$	&	$0.006^{+0.002}_{-0.001}$	&	...	&	...	&	1241.5/1225	&	$2.171^{+0.013}_{-0.013}$	&	$0.13^{+0.03}_{-0.03}$	&	$0.945^{+0.014}_{-0.012}$	\\
2208	&	SD	&	$1.58^{+0.14}_{-0.26}$	&	$500^{f}$	&	$1.002^{+0.006}_{-0.006}$	&	$1000^{+30}_{-30}$	&	$0.004^{+0.002}_{-0.001}$	&	...	&	...	&	1106.1/1192	&	$2.109^{+0.015}_{-0.015}$	&	$0.13^{+0.03}_{-0.05}$	&	$0.942^{+0.013}_{-0.021}$	\\
2209	&	SD	&	$1.7^{+0.2}_{-0.2}$	&	$500^{f}$	&	$0.989^{+0.004}_{-0.004}$	&	$1100^{+20}_{-20}$	&	$0.005^{+0.002}_{-0.001}$	&	...	&	...	&	1202.7/1259	&	$2.139^{+0.012}_{-0.011}$	&	$0.10^{+0.03}_{-0.02}$	&	$0.956^{+0.012}_{-0.011}$	\\
2210	&	SD	&	$1.6^{+0.2}_{-0.2}$	&	$500^{f}$	&	$0.987^{+0.006}_{-0.006}$	&	$1100^{+30}_{-30}$	&	$0.006^{+0.003}_{-0.002}$	&	...	&	...	&	1163.5/1149	&	$2.15^{+0.02}_{-0.02}$	&	$0.10^{+0.03}_{-0.03}$	&	$0.957^{+0.014}_{-0.013}$	\\
2211	&	SD	&	$1.7^{+0.2}_{-0.2}$	&	$500^{f}$	&	$0.982^{+0.005}_{-0.005}$	&	$1100^{+30}_{-30}$	&	$0.006^{+0.002}_{-0.002}$	&	...	&	...	&	1109.4/1196	&	$2.136^{+0.014}_{-0.014}$	&	$0.10^{+0.03}_{-0.03}$	&	$0.955^{+0.013}_{-0.012}$	\\
\hline
2212	&	SD	&	$1.7^{+0.2}_{-0.3}$	&	$500^{f}$	&	$0.984^{+0.004}_{-0.004}$	&	$1100^{+20}_{-20}$	&	$0.004^{+0.001}_{-0.001}$	&	...	&	...	&	1301.9/1315	&	$2.133^{+0.010}_{-0.010}$	&	$0.09^{+0.03}_{-0.02}$	&	$0.961^{+0.012}_{-0.010}$	\\
2213	&	SD	&	$1.7^{+0.3}_{-0.2}$	&	$500^{f}$	&	$0.983^{+0.004}_{-0.004}$	&	$1100^{+20}_{-20}$	&	$0.006^{+0.002}_{-0.001}$	&	...	&	...	&	1221.2/1284	&	$2.151^{+0.011}_{-0.011}$	&	$0.11^{+0.03}_{-0.02}$	&	$0.952^{+0.012}_{-0.010}$	\\
2214	&	SD	&	$1.6^{+0.2}_{-0.2}$	&	$500^{f}$	&	$0.982^{+0.005}_{-0.005}$	&	$1100^{+30}_{-30}$	&	$0.004^{+0.001}_{-0.001}$	&	...	&	...	&	994.4/1095	&	$2.15^{+0.02}_{-0.02}$	&	$0.16^{+0.02}_{-0.06}$	&	$0.932^{+0.010}_{-0.024}$	\\
2215	&	SD	&	$1.6^{+0.3}_{-0.2}$	&	$500^{f}$	&	$0.946^{+0.009}_{-0.010}$	&	$1300^{+60}_{-50}$	&	$0.047^{+0.009}_{-0.004}$	&	...	&	...	&	1323.5/1159	&	$2.22^{+0.02}_{-0.02}$	&	$0.05^{+0.05}_{-0.11}$	&	$0.98^{+0.02}_{-0.05}$	\\
2301	&	SD	&	$1.71^{+0.18}_{-0.08}$	&	$500^{f}$	&	$0.991^{+0.004}_{-0.004}$	&	$1000^{+20}_{-20}$	&	$0.018^{+0.003}_{-0.002}$	&	...	&	...	&	1357.7/1407	&	$2.114^{+0.008}_{-0.008}$	&	$0.18^{+0.02}_{-0.02}$	&	$0.922^{+0.009}_{-0.008}$	\\
2401	&	SD	&	$1.99^{+0.08}_{-0.08}$	&	$500^{f}$	&	$0.963^{+0.005}_{-0.005}$	&	$1100^{+30}_{-30}$	&	$0.032^{+0.004}_{-0.004}$	&	...	&	...	&	1284.8/1377	&	$2.028^{+0.009}_{-0.009}$	&	$0.17^{+0.02}_{-0.02}$	&	$0.921^{+0.009}_{-0.008}$	\\
2601	&	SD	&	$1.8^{+0.2}_{-0.3}$	&	$500^{f}$	&	$0.953^{+0.003}_{-0.003}$	&	$1100^{+20}_{-20}$	&	$0.003^{+0.001}_{-0.001}$	&	...	&	...	&	1467.1/1423	&	$1.974^{+0.008}_{-0.008}$	&	$0.06^{+0.02}_{-0.02}$	&	$0.969^{+0.012}_{-0.009}$	\\
2701	&	SD	&	$1.71^{+0.23}_{-0.12}$	&	$500^{f}$	&	$0.951^{+0.003}_{-0.004}$	&	$1100^{+20}_{-20}$	&	$0.010^{+0.002}_{-0.002}$	&	...	&	...	&	1433.3/1464	&	$1.960^{+0.007}_{-0.007}$	&	$0.10^{+0.02}_{-0.02}$	&	$0.951^{+0.009}_{-0.008}$	\\
2801	&	SD	&	$1.72^{+0.18}_{-0.18}$	&	$500^{f}$	&	$0.940^{+0.003}_{-0.003}$	&	$1100^{+20}_{-20}$	&	$0.005^{+0.002}_{-0.001}$	&	...	&	...	&	1393.1/1376	&	$1.898^{+0.008}_{-0.008}$	&	$0.08^{+0.02}_{-0.02}$	&	$0.960^{+0.010}_{-0.009}$	\\
2901	&	SD	&	$1.7^{+0.3}_{-0.3}$	&	$500^{f}$	&	$0.922^{+0.005}_{-0.005}$	&	$1200^{+30}_{-30}$	&	$0.004^{+0.002}_{-0.001}$	&	...	&	...	&	1144.1/1197	&	$1.787^{+0.013}_{-0.013}$	&	$0.04^{+0.02}_{-0.02}$	&	$0.979^{+0.012}_{-0.011}$	\\
3001	&	SD	&	$1.7^{+0.2}_{-0.4}$	&	$500^{f}$	&	$0.918^{+0.004}_{-0.005}$	&	$1200^{+30}_{-30}$	&	$0.004^{+0.003}_{-0.001}$	&	...	&	...	&	1214.8/1238	&	$1.750^{+0.011}_{-0.011}$	&	$0.05^{+0.02}_{-0.02}$	&	$0.974^{+0.012}_{-0.011}$	\\
3102	&	SD	&	$1.7^{+0.3}_{-0.4}$	&	$500^{f}$	&	$0.901^{+0.004}_{-0.004}$	&	$1200^{+30}_{-30}$	&	$0.003^{+0.002}_{-0.001}$	&	...	&	...	&	1110.8/1234	&	$1.718^{+0.012}_{-0.012}$	&	$0.03^{+0.02}_{-0.02}$	&	$0.983^{+0.013}_{-0.011}$	\\
3103	&	SD	&	$1.9^{+0.3}_{-0.3}$	&	$500^{f}$	&	$0.891^{+0.006}_{-0.007}$	&	$1300^{+50}_{-40}$	&	$0.009^{+0.006}_{-0.004}$	&	...	&	...	&	1093.6/1195	&	$1.727^{+0.014}_{-0.013}$	&	$0.05^{+0.03}_{-0.02}$	&	$0.972^{+0.016}_{-0.014}$	\\
3203	&	SD	&	$1.7^{+0.5}_{-0.3}$	&	$500^{f}$	&	$0.885^{+0.006}_{-0.006}$	&	$1300^{+40}_{-40}$	&	$0.004^{+0.005}_{-0.002}$	&	...	&	...	&	1095.1/1177	&	$1.685^{+0.013}_{-0.013}$	&	$0.04^{+0.04}_{-0.03}$	&	$0.978^{+0.020}_{-0.014}$	\\
3301	&	SD	&	$1.92^{+0.15}_{-0.15}$	&	$500^{f}$	&	$0.872^{+0.006}_{-0.006}$	&	$1300^{+40}_{-40}$	&	$0.023^{+0.005}_{-0.004}$	&	...	&	...	&	1290.4/1304	&	$1.62^{+0.01}_{-0.01}$	&	$0.12^{+0.03}_{-0.03}$	&	$0.931^{+0.016}_{-0.014}$	\\
3302	&	SD	&	$1.8^{+0.2}_{-0.2}$	&	$500^{f}$	&	$0.889^{+0.005}_{-0.005}$	&	$1200^{+40}_{-30}$	&	$0.011^{+0.004}_{-0.003}$	&	...	&	...	&	1152.6/1277	&	$1.59^{+0.01}_{-0.01}$	&	$0.07^{+0.03}_{-0.02}$	&	$0.955^{+0.015}_{-0.012}$	\\
3303	&	SD	&	$1.9^{+0.2}_{-0.2}$	&	$500^{f}$	&	$0.883^{+0.007}_{-0.007}$	&	$1200^{+50}_{-50}$	&	$0.014^{+0.006}_{-0.004}$	&	...	&	...	&	1084.2/1150	&	$1.615^{+0.014}_{-0.014}$	&	$0.07^{+0.03}_{-0.03}$	&	$0.96^{+0.02}_{-0.02}$	\\
3401	&	SD	&	$1.8^{+0.2}_{-0.2}$	&	$500^{f}$	&	$0.879^{+0.011}_{-0.011}$	&	$1300^{+70}_{-60}$	&	$0.036^{+0.009}_{-0.007}$	&	...	&	...	&	1005.8/1150	&	$1.61^{+0.02}_{-0.02}$	&	$0.22^{+0.07}_{-0.05}$	&	$0.88^{+0.03}_{-0.03}$	\\
3402	&	SD	&	$1.9^{+0.12}_{-0.11}$	&	$500^{f}$	&	$0.868^{+0.009}_{-0.009}$	&	$1300^{+70}_{-60}$	&	$0.031^{+0.006}_{-0.006}$	&	...	&	...	&	952.9/1127	&	$1.59^{+0.02}_{-0.02}$	&	$0.17^{+0.04}_{-0.04}$	&	$0.91^{+0.02}_{-0.02}$	\\
3403	&	SD	&	$1.7^{+0.2}_{-0.2}$	&	$500^{f}$	&	$0.879^{+0.009}_{-0.009}$	&	$1200^{+60}_{-60}$	&	$0.011^{+0.004}_{-0.003}$	&	...	&	...	&	981.4/1094	&	$1.58^{+0.02}_{-0.02}$	&	$0.14^{+0.06}_{-0.04}$	&	$0.92^{+0.03}_{-0.02}$	\\
3701	&	SD	&	$1.87^{+0.14}_{-0.14}$	&	$500^{f}$	&	$0.867^{+0.007}_{-0.007}$	&	$1200^{+50}_{-40}$	&	$0.023^{+0.005}_{-0.005}$	&	...	&	...	&	1169.9/1211	&	$1.426^{+0.011}_{-0.011}$	&	$0.12^{+0.03}_{-0.03}$	&	$0.93^{+0.02}_{-0.02}$	\\
3702	&	SD	&	$1.89^{+0.13}_{-0.12}$	&	$500^{f}$	&	$0.867^{+0.006}_{-0.006}$	&	$1200^{+40}_{-40}$	&	$0.027^{+0.005}_{-0.004}$	&	...	&	...	&	1155.1/1309	&	$1.448^{+0.009}_{-0.009}$	&	$0.13^{+0.03}_{-0.02}$	&	$0.917^{+0.016}_{-0.014}$	\\
3802	&	SD	&	$2.2^{+0.4}_{-0.4}$	&	$500^{f}$	&	$0.843^{+0.008}_{-0.008}$	&	$1300^{+60}_{-60}$	&	$0.016^{+0.011}_{-0.007}$	&	...	&	...	&	1096.9/1155	&	$1.360^{+0.013}_{-0.013}$	&	$0.04^{+0.02}_{-0.02}$	&	$0.97^{+0.02}_{-0.02}$	\\
3803	&	SD	&	$1.9^{+0.5}_{-0.4}$	&	$500^{f}$	&	$0.849^{+0.006}_{-0.006}$	&	$1200^{+40}_{-40}$	&	$0.004^{+0.004}_{-0.002}$	&	...	&	...	&	995.1/1151	&	$1.360^{+0.012}_{-0.012}$	&	$0.02^{+0.02}_{-0.02}$	&	$0.985^{+0.016}_{-0.014}$	\\
3901	&	SD	&	$2.9^{+0.6}_{-0.7}$	&	$500^{f}$	&	$0.842^{+0.006}_{-0.008}$	&	$1200^{+50}_{-40}$	&	$0.012^{+0.021}_{-0.008}$	&	...	&	...	&	1014.5/1181	&	$1.262^{+0.011}_{-0.010}$	&	$0.01^{+0.02}_{-0.02}$	&	$0.992^{+0.013}_{-0.012}$	\\
3902	&	SD	&	$2.0^{+0.6}_{-0.5}$	&	$500^{f}$	&	$0.839^{+0.005}_{-0.005}$	&	$1200^{+40}_{-40}$	&	$0.005^{+0.006}_{-0.003}$	&	...	&	...	&	1047.0/1217	&	$1.285^{+0.010}_{-0.010}$	&	$0.02^{+0.02}_{-0.02}$	&	$0.988^{+0.013}_{-0.012}$	\\
4602	&	SD	&	$1.6^{+0.3}_{-0.3}$	&	$500^{f}$	&	$0.823^{+0.006}_{-0.006}$	&	$1300^{+50}_{-50}$	&	$0.008^{+0.004}_{-0.003}$	&	...	&	...	&	1028.6/1209	&	$1.235^{+0.011}_{-0.011}$	&	$0.07^{+0.04}_{-0.03}$	&	$0.95^{+0.03}_{-0.02}$	\\
4603	&	SD	&	$1.7^{+0.2}_{-0.3}$	&	$500^{f}$	&	$0.827^{+0.006}_{-0.006}$	&	$1200^{+50}_{-40}$	&	$0.007^{+0.003}_{-0.002}$	&	...	&	...	&	1127.4/1195	&	$1.233^{+0.011}_{-0.011}$	&	$0.06^{+0.03}_{-0.03}$	&	$0.95^{+0.03}_{-0.02}$	\\
4701	&	SD	&	$1.8^{+0.2}_{-0.2}$	&	$500^{f}$	&	$0.820^{+0.007}_{-0.007}$	&	$1200^{+50}_{-50}$	&	$0.015^{+0.006}_{-0.004}$	&	...	&	...	&	1010.6/1200	&	$1.182^{+0.011}_{-0.011}$	&	$0.07^{+0.03}_{-0.02}$	&	$0.94^{+0.02}_{-0.02}$	\\
4702	&	SD	&	$2.0^{+0.2}_{-0.2}$	&	$500^{f}$	&	$0.816^{+0.006}_{-0.006}$	&	$1300^{+40}_{-40}$	&	$0.023^{+0.007}_{-0.005}$	&	...	&	...	&	1189.3/1320	&	$1.201^{+0.008}_{-0.008}$	&	$0.07^{+0.02}_{-0.02}$	&	$0.943^{+0.018}_{-0.015}$	\\
4703	&	SD	&	$2.4^{+0.3}_{-0.3}$	&	$500^{f}$	&	$0.810^{+0.007}_{-0.008}$	&	$1300^{+60}_{-50}$	&	$0.038^{+0.016}_{-0.011}$	&	...	&	...	&	1130.8/1255	&	$1.195^{+0.010}_{-0.010}$	&	$0.06^{+0.02}_{-0.02}$	&	$0.955^{+0.016}_{-0.014}$	\\
4801	&	SD	&	$2.1^{+0.2}_{-0.2}$	&	$500^{f}$	&	$0.808^{+0.008}_{-0.008}$	&	$1300^{+60}_{-50}$	&	$0.040^{+0.009}_{-0.007}$	&	...	&	...	&	1039.6/1241	&	$1.144^{+0.010}_{-0.010}$	&	$0.11^{+0.03}_{-0.02}$	&	$0.91^{+0.02}_{-0.02}$	\\
4803	&	SD	&	$2.02^{+0.15}_{-0.15}$	&	$500^{f}$	&	$0.794^{+0.007}_{-0.007}$	&	$1400^{+60}_{-60}$	&	$0.040^{+0.008}_{-0.007}$	&	...	&	...	&	1132.0/1261	&	$1.173^{+0.011}_{-0.010}$	&	$0.13^{+0.03}_{-0.03}$	&	$0.90^{+0.02}_{-0.02}$	\\
4901	&	SD	&	$2.1^{+0.2}_{-0.2}$	&	$500^{f}$	&	$0.805^{+0.006}_{-0.006}$	&	$1300^{+50}_{-50}$	&	$0.033^{+0.008}_{-0.006}$	&	...	&	...	&	1192.4/1316	&	$1.125^{+0.008}_{-0.008}$	&	$0.08^{+0.02}_{-0.02}$	&	$0.934^{+0.016}_{-0.014}$	\\
4902	&	SD	&	$2.3^{+0.2}_{-0.2}$	&	$500^{f}$	&	$0.792^{+0.007}_{-0.007}$	&	$1300^{+60}_{-50}$	&	$0.058^{+0.012}_{-0.010}$	&	...	&	...	&	1123.0/1327	&	$1.111^{+0.009}_{-0.009}$	&	$0.10^{+0.02}_{-0.02}$	&	$0.920^{+0.015}_{-0.014}$	\\
4903	&	SD	&	$2.0^{+0.2}_{-0.2}$	&	$500^{f}$	&	$0.786^{+0.009}_{-0.009}$	&	$1400^{+80}_{-70}$	&	$0.052^{+0.012}_{-0.010}$	&	...	&	...	&	1074.9/1197	&	$1.126^{+0.012}_{-0.012}$	&	$0.16^{+0.04}_{-0.03}$	&	$0.88^{+0.03}_{-0.02}$	\\
\bottomrule
\end{longtable*}
\tablecomments{$^a$ SD: \texttt{simplcut*diskbb} model; $^b$ SDG: \texttt{simplcut*diskbb+gaussian} model;  $^c$ The central energy of the \texttt{gaussian} model; $^d$ Line width of the \texttt{gaussian} model; $^e$ Units: ($10^{-8}\ {\rm erg\ cm^{-2}\ s^{-1}}$); $^f$ $E_{\rm cut}$ fixed at 500~keV in the SIMS and HSS; $^p$ Refers to reaching the lower or upper limit; $^g$  $f_{\rm sc}$ is the scattered fraction, {\it i.e.}, the proportion of disk photons scattered by the corona.}

\end{document}